\documentclass{article}

\usepackage{PRIMEarxiv}

\usepackage[utf8]{inputenc} % allow utf-8 input
\usepackage[T1]{fontenc}    % use 8-bit T1 fonts
\usepackage{url}            % simple URL typesetting
\usepackage{booktabs}       % professional-quality tables
\usepackage{amsfonts}       % blackboard math symbols
\usepackage{nicefrac}       % compact symbols for 1/2, etc.
\usepackage{microtype}      % microtypography
\usepackage{lipsum}
\usepackage{fancyhdr}       % header
\usepackage{graphicx}       % graphics
\graphicspath{{media/}}     % organize your images and other figures under media/ folder

\usepackage{hyperref}
\hypersetup{colorlinks=true,allcolors = blue,citecolor=blue}

\title{Data-Driven Design for Metamaterials and Multiscale Systems: A Review

}

\author{
  Doksoo Lee \\
  Dept. of Mechanical Engineering \\
  Northwestern University \\
  Evanston, IL 60208\\
  \texttt{dslee@northwestern.edu} \\
   \And
  Wei (Wayne) Chen \\
  Dept. of Mechanical Engineering \\
  Texas A\&M University \\
  College Station, TX 77840 \\
  \texttt{w.chen@tamu.edu} \\   
    \And
 Liwei Wang \\
  Dept. of Mechanical Engineering \\
  Northwestern University \\
  Evanston, IL 60208 \\
  \texttt{liwei.wang@northwestern.edu} \\   
     \And
  Yu-Chin Chan \\
  Siemens Corporate Technology \\
  Princeton, New Jersey 08540 \\
  \texttt{yu-chin.chan@siemens.com} \\
    \And
  Wei Chen\thanks{Corresponding author: Wei Chen (weichen@northwestern.edu)} \\%\\ This manuscript has been submitted to \textit{Advanced Materials}.} \\
  Dept. of Mechanical Engineering \\
  Northwestern University \\
  Evanston, IL 60208\\
  \texttt{weichen@northwestern.edu} \\  
}

\begin{document}
\maketitle

\begin{abstract}
Metamaterials are artificial materials designed to exhibit effective material parameters that go beyond those found in nature. Composed of unit cells with rich designability that are assembled into multiscale systems, they hold great promise for realizing next-generation devices with exceptional, often exotic, functionalities. However, the vast design space and intricate structure-property relationships pose significant challenges in their design. A compelling paradigm that could bring the full potential of metamaterials to fruition is emerging: data-driven design. In this review, we provide a holistic overview of this rapidly evolving field, emphasizing the general methodology instead of specific domains and deployment contexts. We organize existing research into data-driven modules, encompassing data acquisition, machine learning-based unit cell design, and data-driven multiscale optimization. We further categorize the approaches within each module based on shared principles, analyze and compare strengths and applicability, explore connections between different modules, and identify open research questions and opportunities.
\end{abstract}

% keywords can be removed

\tableofcontents

\section{Introduction}
Engineered material structures generally benefit from some extreme or spatially varying material properties to achieve higher performance or complex functionalities \cite{skylar2019voxelated,wehner2016integrated, zhang2020seamless, kim2019ferromagnetic}. Typically, tuning material properties involves precise control of material composition and processing conditions, which can be both technically challenging and cost-prohibitive, particularly when aiming for spatially varying properties. In contrast, metamaterials are engineered architectural materials that can reach a broad range of properties by carefully designing their architectures or microstructures rather than altering the material composition itself. \cite{yu2018mechanical,zheludev2010road,kadic20193d,lumpe2021exploring}. Along with the recent enhancements in manufacturing capabilities, metamaterials are emerging as a new paradigmatic material system to enable unprecedented design flexibility in properties. They can advance applications in a wide range of fields, including optics ~\cite{leonhardt2006optical,leonhardt2009broadband,liu2008three}, electromagnetics ~\cite{ma2013first,pendry2006controlling}, thermology ~\cite{fan2008shaped,schittny2013experiments,han2013homogeneous,hu2020machine}, acoustics ~\cite{chen2014periodic,deymier2013acoustic,phani2017dynamics,ma2019valley,wang2020valley}, and mechanics~\cite{yu2018mechanical,florijn2014programmable,wang2021structured,jenett2020discretely,kochmann2017exploiting, frenzel2017three}.  

% \textbf{[Challenges associated with the conventional methods]}
Nevertheless, the design of metamaterials and their multiscale structures proves to be a complex process that involves navigating an infinite-dimensional topological design space, mapping microstructures to their effective properties across multiple scales, dealing with numerous local optima in design optimization, the absence of analytical gradient information, and expensive property or performance evaluation. Most existing metamaterial designs adopt trial-and-error and heuristic methods \cite{lumpe2021exploring,wang2021structured,frenzel2017three,jin2020guided,reid2018auxetic,kadic2012practicability,fernandes2021mechanically}, which rely heavily on the experience of a designer, or simple parameter optimization methods \cite{bessa2019bayesian,hedayati20213d}, confining the metamaterials to a restricted selection of properties. In some specific cases with relatively simple and differentiable physical models, gradient-based topology optimization (TO) has been utilized to facilitate the automatic design of metamaterials. Nonetheless, these methods are generally not scalable to the design of multiscale systems of metamaterials that feature nested optimization loops and require numerous microscale designs at different spatial locations. 

% \textbf{[the emerging of data-driven]}
The emergence of data-driven methods has provided solutions to these challenges by enabling high-throughput property or response prediction, reducing the dimensionality of complex problems, accelerating design space exploration and design optimization, and allowing fast solutions to ill-posed inverse design problems.
Data-driven metamaterials design typically includes three modules, 1) data acquisition: acquiring a precomputed
dataset of unit cells; 2) machine learning-based unit cell design: using machine learning to extract information from data and help unit cell designs; 3) multiscale design: utilizing unit-cell database and machine learning models for design synthesis at the system level. In practical applications, it is possible to integrate all of these modules into a unified framework or selectively focus on specific modules based on the design requirements. However, the core idea underlying data-driven design remains consistent across these modules, which is to extract meaningful patterns from data that are unavailable or difficult to obtain with physical models, and incorporate them into the design process to simplify the complexity in traditional design approaches.. Although these capabilities come with the cost of data collection and model training, deploying the trained model has the benefits of negligible inference time, which can significantly speed up the design process. Data-driven design methods are especially useful in scenarios where the design problems are high-dimensional or the governing physics is unknown. They can also achieve unprecedented design performance owing to their capability of encapsulating higher design freedom (e.g., heterogeneous metamaterials system design) compared to conventional design methods.

% \textbf{[observations on existing reviews]}
As evidence of growing attention to data-driven multiscale design, with or without the use of data, a suite of literature reviews from different communities have been published in recent years. Each review is centered on particular topics, such as design~\cite{Wu2021TopologyReview, Regenwetter2021DeepReview, woldseth2022use, Mukherjee2021AcceleratingChallenges, so2022revisiting}, manufacturing~\cite{Wang2020MachinePerspectives, Mozaffar2022MechanisticPerspectives}, mechanics~\cite{So2020DeepNanophotonics, Liu2021HowStructures, Kumar2021WhatMechanics, so2022revisiting, liu2021tackling, jin2022intelligent, liu2023deep}, and specific machine learning methods~\cite{Regenwetter2021DeepReview}. Despite the meaningful contributions of each, and of the aggregate, we recognize lack of discussion on some points that may impede researchers from unlocking the full potential of data-driven design for multiscale architectures. First, in the corpus, we observe a disconnect between two primary lines of approaches, one being the data-driven camp that harnesses pre-existing machine learning tools with minimal customization, and the other being the physics-based camp that specializes in physics with limited awareness to recent advancements of data-driven methods. Second, when covering data-driven design, the prior reviews are typically dedicated to specific aspects, e.g., machine learning methods~\cite{Regenwetter2021DeepReview}, physical mechanisms~\cite{Kumar2021WhatMechanics, So2020DeepNanophotonics, Liu2021HowStructures}, individual components of data-driven design~\cite{Wu2021TopologyReview, woldseth2022use}. In the current state of the field, it is difficult to find a singular review that provides a holistic overview of data-driven design, a summary of the archetypal design framework, and the critical inter-dependencies between design components. Third, while some existing reviews~\cite{Regenwetter2021DeepReview, Mozaffar2022MechanisticPerspectives} discussed data preparation as a core module, i.e., a step in data-driven design, none have systematically compared how existing datasets were created and how data quality could be ensured to better support the design of multiscale architectures.

The key contributions of our review include the following:
\begin{enumerate}
    \item We adopt a \textit{design}-centered perspective to examine a wide range of papers, categorizing existing methods from various domains into three modules within the data-driven design framework. This synthesis of different studies allows us to present a clear and cohesive picture of \textit {how} data-driven design has been practiced from data acquisition to single-scale and multiscale optimization. Our key focus is on the methodological aspects of design without  specific deployment goals, i.e., without filtering based on the underlying physical mechanisms, geometric families of unit cells, and fabrication methods. This holistic approach allows us to uncover the common threads and overarching principles that drive the field of data-driven design. %(holistic contribution)
    \item We review the common practices of current data generation strategies for metamaterial design through a standardized taxonomy, discuss key concerns in depth, and attempt to raise awareness on certain issues that are crucial yet underestimated, or even overlooked, in data acquisition. %(Section III)
    \item We review prior data-driven design methods that can be broadly applied to metamterials design under different physics (i.e., optical, acoustic, mechanical, thermal, etc.) and raise critical concerns that are generalizable to all types of metamaterial design problems addressed by data-driven methods. %(Section IV)
    \item  We investigate the role of data-driven models and methods in the context of multiscale system design. Unlike previous studies that treat data-driven models as isolated solutions, we highlight their integration into a complete design workflow and their scalability in handling large databases, multiple scales, and combinatorial design spaces. By examining these tools within the broader design process, we aim to offer insights into how data-driven approaches can be effectively utilized to enhance the efficiency and effectiveness of metamaterial design.%(Section V)  
\end{enumerate}

% \textbf{[scope (paper selected]}
We specify our scope as follows:
\begin{enumerate}
    \item We limit our scope to only the machine learning-based design methods for metamaterials and their multiscale systems.
    \item We consider a wide range of physics including optical, acoustic, mechanical, thermal, and magneto-mechanical, because the machine learning methods are usually applicable regardless of the physics that governs the problem.
    \item The reviewed machine learning methods do not necessarily require prior training data (i.e., we included past works using methods such as reinforcement learning or physics-informed neural networks).
    \item We encompass design for multiscale systems that are built on either a unit-cell database of metamaterials or machine learning models. It includes the use of descriptors of unit cells, surrogate models for constitutive laws, efficient simulation via machine learning, optimization, and assembly of unit cells based on the database. 
\end{enumerate}

% \textbf{[ the structure of this review]}
This paper is structured as follows: Section~\ref{2. Preliminaries} gives a brief definition of key concepts underpinning our review and major machine learning methods utilized in the works we will cover. In Section~\ref{3. Data Acquisition}, isolating data acquisition from the pipeline of data-driven multiscale design, we anatomize the common practices of data acquisition strategies, with particular attention to the methodological procedures of shape generation and property-aware sampling. Following this, we provide a brief review on the current practices of data assessment that ensure data quality and often shed light on how it can propagate into downstream tasks. %For all the individual themes, we raise issues of the current practice, share our perspectives, and suggest new research avenues.
Section~\ref{4 Learning and Generation: Data-Driven Metamaterials Design} reviews prior works using machine learning methods for single-scale metamaterials design, discusses some critical considerations when evaluating these methods, and proposes promising future directions. Section~\ref{5 Data-driven Multiscale TO} explores data-driven design methods for multiscale systems that are built on either a unit-cell database or machine learning models. It includes the use of descriptors of unit cells, surrogate models for constitutive laws, efficient simulation via machine learning, optimization, and assembly of unit cells based on the database.

\section{Preliminaries}
\label{2. Preliminaries}
This section defines key terminologies and concepts used throughout this paper as well as in other data-driven metamaterials design literature. We also briefly introduce common machine learning techniques and how these techniques were applied under the context of metamaterials design.

%In this section, we define key concepts to be frequently used throughout the review, and then walk you through individual modules and a common procedure of data-driven metamaterials design, which mirror our perspectives on the state-of-the-art.

%%%%%%%%%%%%%%%%%%%%%%%%%%%%%
\subsection{Key Concepts}
\label{key concepts}
Below, we list working definitions of key concepts to be frequently used throughout the paper.
%In this subsection, we define key concepts to be frequently used throughout the review, and then walk you through individual modules and a common procedure of data-driven metamaterials design, which mirror our perspectives on the state-of-the-art. 
%\subsubsection{Key Concepts}
%\subsection{Key Concepts}
% \newtcolorbox{boxH}{
%     colback = sub, 
%     colframe = main, 
%     boxrule = 0pt, 
%     leftrule = 6pt % left rule weight
% }

\begin{figure}[!htb]
\centering
\includegraphics[width=0.8\textwidth]{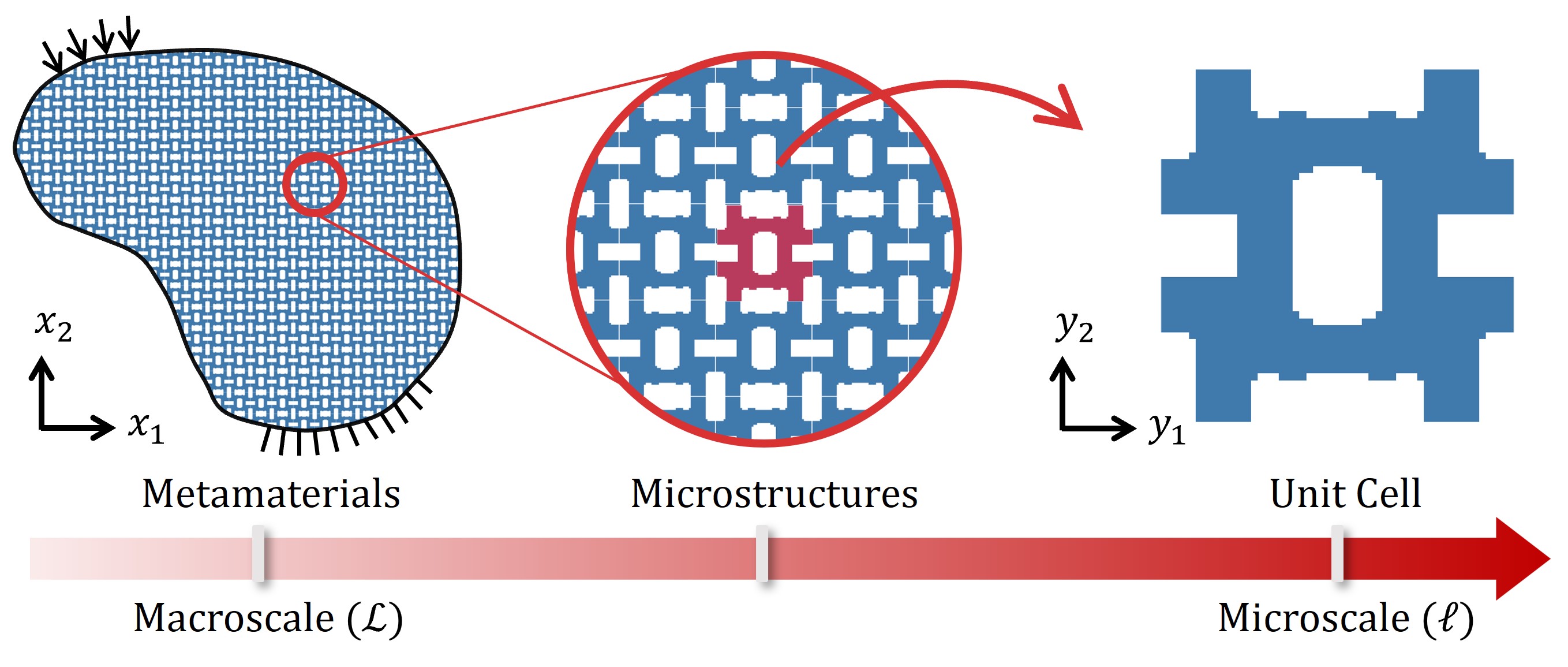}
\caption{Hierarchical terminology system in this review. Metamaterials are created by assembling multiple microstructures to achieve properties on the macroscale. A microstructure is an assembly of multiple unit cells. A unit cell is the smallest representative unit of a material in the microscale to control its properties.}
\label{fig:Hierarchical concepts}
\end{figure}

%\begin{boxH}
\paragraph{Unit Cell}
A unit cell is the smallest representative unit of a material used to control its properties, as shown in Figure~\ref{fig:Hierarchical concepts}. It is often referred to by interchangeable terms such as meta-atom, meta-molecule, building block, and cell \cite{kadic20193d}, depending on the physical mechanism, scale and geometry.

\paragraph{Microstructure}
A microstructure is an assembly of multiple unit cells arranged in a specific pattern to achieve more complex properties arising from their arrangement.

\paragraph{Metamaterials}
Metamaterials is the assembly of multiple microstructures, often in a periodic pattern, to achieve unique and tailored properties that cannot be found in natural materials. 

\paragraph{Multiscale design}
Multiscale design in the context of metamaterials refers to the process of designing metamaterials with desired properties at multiple length scales, from the macroscopic level of unit cell to the macroscopic level of the metamaterial structure.

\paragraph{Module}
Module refers to an independent and reusable unit or component within data-driven multiscale design. It provides a specific functionality that is required for design. An ordered sequence of modules forms a data-driven multiscale design framework. Modules of primary interest in this review include data acquisition (Section~\ref{3. Data Acquisition}), machine learning-based unit
cell design (Section~\ref{4 Learning and Generation: Data-Driven Metamaterials Design}), and data-driven mulitscale optimization (Section~\ref{5 Data-driven Multiscale TO}).

\paragraph{Effective Properties}
Effective property is the macroscopic property of a metamaterial that arises from the collective behavior of its microstructures \cite{li2008introduction}. These properties are often different from the intrinsic properties of the constituent materials, and can be tailored through design and optimization of the microstructure.

\paragraph{Class}
A group of unit cells that can be generated from the same geometric motif or design parameterization~\cite{Chan2022Yu-ChinDissertation}.

\paragraph{Representation}
A set of parameters, or models, used to directly characterize unit cells~\cite{Chan2022Yu-ChinDissertation}. Representations often involve the projection of high-dimensional instances into a lower-dimensional space.

\paragraph{Design Space}
The space of all possible design solutions. It contains all the combinations of design variables. In the context of metamaterials design, design variables usually refer to geometric or material design variables.

\paragraph{Shape Space}
The geometric design space of unit cells.
% unit cell
% representation
% class
% reparamterization
% tradeoff between expressivity and compactness
% high-D

\paragraph{Property Space}
The response space of unit cells.
% low-D

\paragraph{Evaluation}
To obtain system responses of concern given an architecture and loading conditions, typically through numerical analyses such as finite element methods. In machine learning literature, the evaluation process is similar to ``labeling", which refers to the process of adding attributes of interest (i.e., labels) to raw data. Throughout this review, we will assume that the term evaluation is interchangeable with labeling.
% evaluation or design evaluationhttps://www.overleaf.com/project/620e8169c8c9ffddcf758c79
% cf. assessment

\paragraph{Shape-Property Mapping}
A directional mapping from instances in shape space to those in property space. It is typically learned through machine learning using labeled data.

\paragraph{Compatibility}
Capability of neighboring unit cells to possess seamless geometric/mechanical connections, or lack of geometric frustration. It is often measured through geometric/mechanical similarities at the interface of neighboring unit cells~\cite{Wang2020, Wang2022IH-GAN:Structures}.
%It involves manufacturability, deviation from periodic boundary condition. 
% geometric compatibility
% mechanical compatibility
% homogenization
%\end{boxH}

%%%%%%%%%%%%%%%%%%%%%%%%%%%%%%%%%%%%%%
\subsection{Machine Learning}
\label{prelim: machine learning}

Data-driven metamaterial design processes usually include the use of machine learning models to extract useful information from data. The extracted information can then help the design process in different ways. Machine learning approaches used in this context were mainly from four categories~\textemdash~supervised learning, unsupervised learning, semi-supervised learning, and reinforcement learning. We briefly introduce these categories in this section.

\subsubsection{Supervised Learning}
% what is it?
In a supervised learning task, we train a machine learning model to predict attributes of interest (i.e., labels). The machine learning model is trained on a collection of data-label pairs.
% Supervised learning solves a function approximation problem where the training data forms a collection of input-output pairs. The learning goal is to predict the output of an unseen input not included in the training data. 
The data can take various forms, e.g., vectors, images, and graphs. Depending on the type of labels, supervised learning can be divided into a broad dichotomy: regression with continuous outputs and classification with discrete ones. Commonly-used machine learning models include neural networks, kernel machines, and decision trees. Selecting the type of model is generally done before training, and could be contingent upon the end task (e.g., regression/classification), data volume, need to predict uncertainty, and design applications.

% examples in MD
Within the context of data-driven metamaterial design (DMD), supervised learning is most widely for creating shape-property relations. A data pair typically describes a shape, where the input is its parameterization (e.g., parameterized lattice) and the output is quantities of interest (e.g., elasticity components; frequency dispersion). A key motivation for using supervised learning has been to replace the resource-intensive evaluation of unit cells with a surrogate model. The types of models have been dominantly based on neural networks~\cite{Liu2018, Wang2020, Chan2022RemixingBlending} and Gaussian processes~\cite{Wang2020}. Once trained on a large volume of data, a data-driven model offers on-the-fly predictions of the (effective) properties of unseen unit cells. Sometimes such a surrogate is incorporated into a larger network architecture and trained end-to-end together with other components~\cite{wang2020deep, Wang2022IH-GAN:Structures}.

% MLP, CNN, DT, GP, GNN, PCR, RNN, LR

\subsubsection{Unsupervised Learning}

In contrast to supervised learning, unsupervised learning extracts information from unlabeled data. More specifically, it addresses tasks such as clustering, anomaly detection, and dimensionality reduction. In metamaterials design, it is mainly applied to learning the representation or the distribution of complex metamaterial geometries. Unsupervised learning models commonly used in metamaterials design are autoencoders, variational autoencoders, and generative adversarial networks.

Autoencoder (AE)~\cite{kramer1991nonlinear} is a type of neural network that uses an encoder-decoder architecture to extract lower-dimensional latent variables of input data. In metamaterials design, AEs were used to reduce the dimensionality of either metamaterial geometries~\cite{li2020designing} or high-dimensional responses such as scattering parameters~\cite{qiu2019deep}. A variational autoencoder (VAE)~\cite{kingma2013auto} has a similar architecture with AEs, while being a type of deep generative model that learns the distribution of data. New data can be generated by sampling latent variables that are low-dimensional and follow a well-defined distribution. Therefore, the latent representation learned by a VAE is usually more efficient and interpretable than the original metamaterial design representation, especially when considering high design complexity and freedom~\cite{Wang2020DeepSystems,liu2020hybrid}. Same as VAEs, generative adversarial networks (GANs)~\cite{goodfellow2020generative} can also generate new metamaterial designs and learn efficient representations. A GAN models the generation as a game between its generator and discriminator. Compared to VAEs, GANs can generate higher-quality samples~\cite{radford2015unsupervised}.

While VAEs and GANs have been generally used for unsupervised learning, prior works have proposed variants of these models, such as conditional GANs (e.g., \cite{Liu2018,gurbuz2021generative,Wang2022IH-GAN:Structures}, conditional VAEs (e.g., \cite{ma2019probabilistic,ma2020data}), and VAE-regressor models (e.g., \cite{Wang2020DeepSystems}), that require supervised learning. These supervised learning models either enable the inverse design of metamaterials~\cite{Liu2018,gurbuz2021generative,Wang2022IH-GAN:Structures,ma2019probabilistic,ma2020data} or help construct a property-related metamaterial latent representation~\cite{Wang2020DeepSystems}.

% DGM

\subsubsection{Semi-Supervised Learning}
% what is it?
Semi-supervised learning trains a machine learning model on partially-labeled data so that the model can predict the labels of any given data. Typically it is assumed that the portion of labeled data is much smaller than the counterpart. This is a commonplace scenario in machine learning due to the high labeling cost. Marrying supervised and unsupervised learning, semi-supervised learning aims to improve the performances of either. It also inherits most of the machine learning tasks above, such as semi-supervised classification and regression, and semi-supervised generative modeling. 

% examples in DMD
Within DMD, the efficacy of semi-supervised learning has been demonstrated in some works. In designing photonic metasurfaces, Ma et al. used both labeled and unlabeled data when training the prediction network to improve the regression with a less computational overhead of data preparation~\cite{ma2019probabilistic}. Although the idea of semi-supervised learning could sound compelling, the efficacy comes under some conditions; in-depth discussion on this point can be found in some reviews dedicated to semi-supervised learning~\cite{zhu2005semi, van2020survey}.

\subsubsection{Reinforcement Learning}
\label{sec:pre_rl}

Reinforcement learning (RL) is another category of machine learning methods used in metamaterials design. It is modeled as a Markov decision process~\cite{bellman1957markovian}, where an agent takes actions in an environment to maximize a reward. The goal is to learn an optimal policy that guides the action-taking strategy. Different from supervised, unsupervised, and semi-supervised learning, RL does not require an initial dataset to learn from. Instead, the agent in RL learns from its experience of exploring the action space and receiving rewards. There have been successful applications of RL in areas such as gaming and robotics. RL is usually applied to solving sequential decision-making processes. But with a proper definition of the action space and sequential decision-making setting, the design of metamaterials can also be formulated as RL problems, as shown in prior works~\cite{sajedian2019double,liu2021reinforcement,sui2021deep}.

%%%%%%%%%%%%%%%%%%%%%%%%%%%%%%%%%%%%%%%%%%%

\section{Data Acquisition}
\label{3. Data Acquisition}
\subsection{Overview} % new outline
%III-1-1 A working definition of data acquisition for DMD
\label{3.1 Overview}
Creating and leveraging datasets of unit cells has been a core enabler of the recent success of DMD. Data acquisition in multiscale design is a decision-making activity that determines a ﬁnite collection of structure-property pairs, which delegates the space to be explored in downstream tasks. When one strives to exploit the power of DMD, data acquisition presents multifaceted open challenges, such as exploration of high-dimensional design space of unit cells, resource-intensive design evaluation in particular for large datasets, many-to-one mapping from shape to property, distributional bias in data, and opaque compounding eﬀects of data quality to downstream tasks. Generally taken as the first module of DMD, data acquisition was tackled through diverse strategies in past works. Despite the diversity of strategies seen across different communities, it is difficult to draw connections among them due to the absence of attempts to build a common context that bridges different strategies.

To this end, we present a review on data acquisition for DMD with a particular focus on the methodological perspectives of prior works. We propose a standardized taxonomy which which to organize the literature in a relatable and easy-to-compare manner. Based on our observations of the current research trend, our review of data acquisition consists of two parts: Shape-Centric Data Generation Method (Section~\ref{3.2 Shape Generation Heuristics}) and Property-Aware Data Acquisition Strategy (Section~\ref{3.5.2 Acquisition Strategy .}). In the corpus, we recognize that many demonstrations of data acquisition adopted shape generation heuristics, which do not necessarily consider property, in order to incorporate domain knowledge into datasets and avoid handcrafting a large number of shape instances one by one. Section~\ref{3.2 Shape Generation Heuristics} offers a detailed review of these shape-centric data acquisition methods. In contrast, Section~\ref{3.5.2 Acquisition Strategy .} reviews acquisition strategies that take property into account. These methods could boost sampling efficiency and facilitate data customization for specific design tasks. 

Following the current research trend, the scope of this section is centered on data acquisition, but we also brieﬂy introduce exemplar demonstrations of data assessment in Section~\ref{3.5.1 Data Assessment .} that are key to underpinning data quality assurance and data sharing practice. Each subsection includes a discussion specific to its topic, while Section~\ref{3.5 Discussion . } offers a more general and comprehensive discussion that covers multiple themes in data acquisition.

    \subsection{Shape-Centric Data Generation Method}               
\label{3.2 Shape Generation Heuristics}
Data acquisition for DMD often entails a large collection of unit cell shapes. Handcrafting individual shape instances one by one is intractable for large data, as is running inverse optimization to obtain all unit cells corresponding to a massive set of pre-specified target properties. To create large data in an effective, systematic manner, past works presented a diverse array of methodological procedures. We recognize that individual approaches commonly involved two research questions in shape generation: 1) How to specify a group of unit cells? and 2) How to grow sparse data to large data? Answering the first essentially entails a representation of unit cells (Section~\ref{3.2.1 Representation . .}), which was usually pre-selected at the early stages of data acquisition and often justified based on criteria such as domain knowledge and fabrication methods. On the other hand, answering the second question involves the reproduction of unit cells (Section~\ref{3.2.2 reproduction . .}), which facilitates the collection of a large enough dataset from sparse data without both handcrafting and optimizing a large volume of individual samples. We remark that in literature the two questions were addressed primarily based on generation strategies driven by shape rather than property; we call these types of approaches shape-centric generation methods. Given this taxonomy, we can bridge diverse data generation approaches scattered over different communities in order to offer a comparative review.

        \subsubsection{Representation of Unit Cells} %%%%%
\label{3.2.1 Representation . .}
A representation refers to a set of parameters, or models, used to directly characterize unit cells, typically by projecting high-dimensional instances into a lower-dimensional space (Section~\ref{key concepts}). Determining what representation to use, specifically for the unit cells in this subsection, is a key decision that should be made at the early stages of data acquisition, as it dictates the nature of resulting data distributions.

\begin{figure}[t!]
\centering
\includegraphics[width=0.9\textwidth]{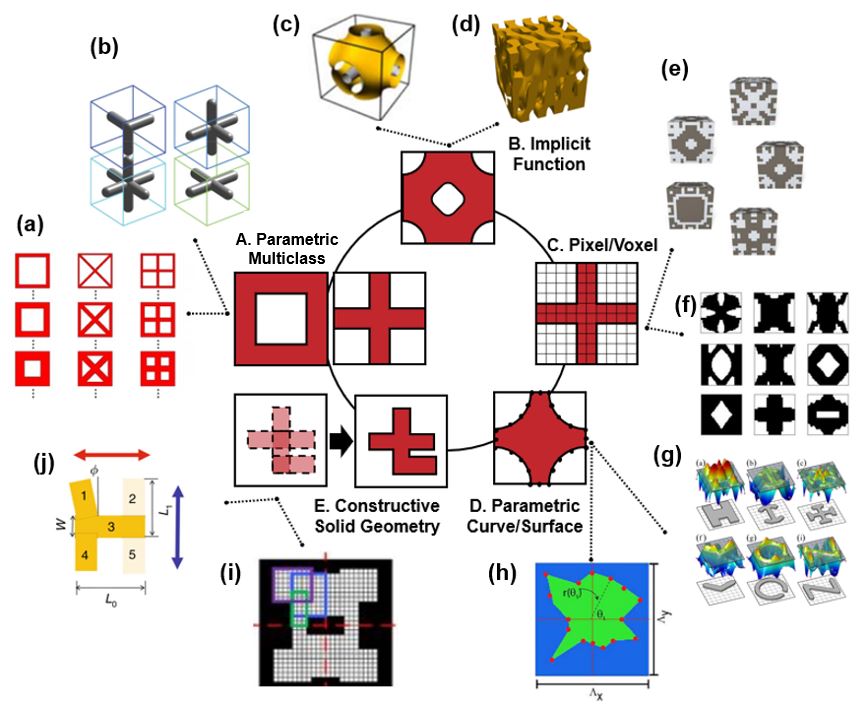}
\caption{Representations of building blocks. a) and b): Parametric Multiclass. a) The mixed-variable multiclass lattice representation of mechanical metamaterials. Reproduced from Ref.~\cite{wang2021data} with permission from ASME. b) The parametric representation of 3D multiclass building blocks of mechanical metamaterials. Reproduced from Ref.~\cite{liu2022growth} with permission from American Association for the Advancement of Science.   
c) and d): Implicit Function. c) The representation based on Triply Minimal Periodic Surfaces of mechanical metamaterials. Reproduced from Ref.~\cite{Li2019DesignShapes} with permission from ASME. d) The spinodoid representation of mechanical metamaterials. Reproduced from Ref.~\cite{kumar2020inverse} with permission from Creative Commons CC BY. e) and f): Pixel/Voxel. e) 3D voxelated representation of mechanical metamaterials. Reproduced from Ref.~\cite{Zhu2017} with permission from Association for Computing Machinery. f) 2D pixelated representation. Reproduced from Ref.~\cite{wang2020deep} with permission from Elsevier B.V. g) and h): Parametric Curve/Surface. g) The parametric surface representation of photonic metasurfaces. Reproduced from Ref.~\cite{Whiting2020Meta-atomMethod} with permission from Optica Publishing Group. h) The parametric curve representation of photonic metasurfaces. Reproduced from Ref.~\cite{inampudi2018neural} with permission from AIP Publishing. i) and j): Constructive Solid Geometry. i) The representation based on primitive superposition and four-fold symmetry of dielectric metasurfaces. Reproduced from Ref.~\cite{An2021MultifunctionalNetwork} with permission from Wiley-VCH GmbH. j) The union-based representation of plasmonic metasurfaces. Reproduced from Ref.~\cite{Malkiel2018PlasmonicLearning} with permission from Creative Commons Attribution 4.0.
}
\label{fig:representation}
\end{figure}

Figure~\ref{fig:representation} depicts widely-used representations reported to date, organized based on our literature survey. The list might not be exhaustive, but we believe that it provides a sufficient overview of common practices in data generation. For each category, we discuss the definition, hallmarks, and relevant works. A comparative, multifaceted discussion across all the representations covered herein can be found in Section~\ref{3.2.3 Discussion . .}.

\paragraph{A. Parametric Multiclass}
An intuitive way to encode domain knowledge into a dataset is to directly include a set of geometric \textit{classes} (Section~\ref{key concepts}) intensely studied in the literature. Herein the seed classes are expected to serve as ``pivots" of the shape generation. Each class is typically endowed with some low-dimensional explicit parameterization, e.g., the length/thickness of a bar entity~\cite{ma2018deep, an2019deep, wang2021data, wang2022generalizedOptimization}, volume fraction~\cite{wang2021data}, angle between geometric entities~\cite{wilt2020accelerating}, or rotational angle~\cite{liu2022growth, wang2022generalizedOptimization}, which supports direct design exploration either within a class~\cite{wang2021data, wilt2020accelerating, liu2022growth} or across classes~\cite{wang2022generalizedOptimization}.

In literature, some advocated for a mixed-variable representation where qualitative (e.g., building block type) and quantitative (e.g., scaling factor) variables were used together as an explicit multiclass representation for TO (Figure~\ref{fig:representation}(b))~\cite{Wang2022ScalableFactors}. Conceptual generalization was proposed by Liu et al., showcased by lattice-like building blocks, spinodal pattern-like ones, and multimaterial composites (Figure~\ref{fig:representation}(a)), each of which involved an explicit, mixed-variable representation with ensured compatibility~\cite{liu2022growth}. Meanwhile, others conceived a versatile parametric representation able to generate multiple classes without explicitly defining classes~\cite{wang2022generalizedOptimization}. To avoid geometric frustration when assembling building blocks in the multiscale design of mechanical metamaterials, the compatibility among neighboring unit cells based on geometric and/or mechanical factors often served as a primary criterion for choosing classes~\cite{wang2021data, liu2022growth, wang2022generalizedOptimization, choi2019design}.

\paragraph{B. Implicit Function}
% formulation?
It has been widely adopted to represent unit cells using implicit functions that can generate geometric families. Therein, a shape instance is implicitly represented through a surface function. 

Demonstrations in literature have been centered on functions that enjoy smooth topological variations, as opposed to lattice representations, and which are tunable by a handful of parameters. % In addition, one can directly encourage the instances to have preferred shape-related attributes, such as symmetry and periodicity.
The most widely used families in DMD are Triply Periodic Minimal Surfaces (TPMS) that feature zero mean curvature and large surface areas (Figure~\ref{fig:representation}(c))~\cite{Li2019DesignShapes, Wang2022IH-GAN:Structures}. Another isosurface representation based on linear combinations of analytical crystallographic symmetry functions was implemented by Chan et al~\cite{Chan2020METASET:Design}. Boddapati et al. proposed another representation that can synthesize diverse classes of quasi-free unit cells of mechanical metamaterials by a linear superposition of periodic cosine functions~\cite{boddapati2023single}.  Inspired by the phase separation process described by the Cahn-Hilliard equation, Kumar et al.~\cite{kumar2020inverse} reported a spinodiod representation (Figure~\ref{fig:representation}(d)), which features smooth, aperiodic variations of complex topologies and tunable anisotropy.

A variant under this branch harnesses spectral decomposition. A manifestation of this for photonic metasurfaces was shown by Liu et al.~\cite{LiuADesign}, where Fourier transform and level-set function of shapes served as key pillars for the new representation. The spectral representation enjoys topologically rich unit cells, reconstruction capability supported by inverse Fourier Transform, and efficient symmetry handling. Another example in this line was used by Wang et al.~\cite{Wang2020}, where the Laplace-Beltrami operation was employed for dimension reduction of the freeform unit cells.

\paragraph{C. Pixel/Voxel} % need to incorporate TO-based and image-based (photonics)
Pixel/voxel representation builds on the assumption that any shape instance can be viewed as a spatial aggregate of solid/void elements. They are typically freeform. Distinct from other representations, this approach offers a direct connection with inverse topology optimization. 

As an early demonstration in DMD, Zhu et al. employed the voxelated representation with TO (Figure~\ref{fig:representation}(e))~\cite{Zhu2017}. Wang et al. implemented inverse TO to find hundreds of freeform unit cells to start with, each of which closely matches the target effective property specified \textit{a priori} (Figure~\ref{fig:representation}(f))~\cite{Wang2020, wang2020deep}. Li et al. used multimaterial TO to systemically construct a library of freeform unit cells, each of whose response was programmed to exhibit a prescribed target force-displacement behavior~\cite{li2022digital}. TO was also used for a thermal emitter design that aims for frequency selective reflectivity when generating seed instances~\cite{Kudyshev2020Machine-learning-assistedOptimization}. Harnessing interpretable machine learning for band gap engineering, Chen et al. adopted the 2D pixel representation to generate unit cell templates of phononic metasurfaces~\cite{chen2022see}.

In the literature, we also observe another subcategory that advocated the pixel-/voxelated representation while considering user-defined classes. For design of photonic metasurfaces, many built datasets spanning from a group of canonical classes, or meta-atoms, such as cross, bow-tie, V-shape, I-beam, split ring resonator, and others~\cite{Ma2020AStructures, Liu2018, ma2019probabilistic, an2019deep, so2019designing}. This allows one to encode the data generation procedure with domain knowledge in contrast to the optimization-based pixel representation introduced above.% Inter-class pixelated instances can be produced upon the combination with a proper reproduction strategies (Section~\ref{3.2.2 reproduction . .} ); but this multiclass pixel representation shares the drawbacks mentioned above, namely high-computational overhead and needs for post-processing~\cite{liu2018generative, ma2019probabilistic}. 

\paragraph{D. Parametric Curve/Surface}
Boundary-based representations, also referred to as contour-based shape descriptors~\cite{kazmi2013survey}, have been commonly used to describe shapes as well. In these approaches, a shape is represented by an ordered sequence of control points on curves or surfaces. 

Within DMD, this approach has been primarily favored in the design of wave-based metamaterials that pursues design exploration beyond canonical families. For metagrating design, Inampudi et al. employed a boundary representation that specified shape instances with 16 boundary Parametric Curve/Surfaces (Figure~\ref{fig:representation}(h))~\cite{inampudi2018neural}. Li et al. used the 4-order formulation of trigonometric functions with tunable parameters to explicitly represent a boundary curve of scattering inclusions of phononic crystals~\cite{li2020designing}. Tanriover et al. also harnessed such a representation to construct a shape dataset not restricted to the canonical meta-atoms in the literature under curvature constraints~\cite{tanriover2022deep}. As an extension, it was also shown that a higher dimensional embedding of parametric curves/surfaces can be used as an implicit representation of 2D boundaries. For example, in photonic metasurfaces, Whiting et al. conceived a representation that offers topologically diverse instances in order to generate quasi-free building blocks (Figure~\ref{fig:representation}(g))~\cite{Whiting2020Meta-atomMethod}. A key distinction between this and the Implicit Function method above is that the 3D embedding here is fully governed by control points, which are explicitly defined. As an example in mechanical metamaterials, Wang et al. adopted the Cassini oval curve to represent the proposed auxetic planar metasurfaces with oval holes~\cite{wang2021novel}.

\paragraph{E. Constructive Solid Geometry}
Constructive Solid Geometry (CGS) is a geometric modeling method to create a solid object in a syntactic manner~\cite{requicha1977constructive}. Its underlying concept is to compose an instance by following a sequence of set-theoretic operations (e.g., union; intersection) acting on primitives (e.g., rectangle, cylinder, sphere). The semantic nature makes instances of CGS highly interpretable, and offers a seamless connection with Computer-Aided Design (CAD). Capitalizing on explicit parameterization, a similar approach, the so-called moving morphable components~\cite{Guo2014DoingFramework,Lei2019MachineFramework}, has been developed in the TO community.

Within the context of DMD, CGS has been utilized in some works, in particular for design of photonic metasurfaces. Malkie et al.~\cite{Malkiel2018PlasmonicLearning} applied the primitive rectangle, whose presence, length, and angle were parameterized, along with the union operation to synthesize plasmonic nanostructures (Figure~\ref{fig:representation}(j)). A recent work that builds on further design freedom was reported by An et al.~\cite{an2019deep, An2021MultifunctionalNetwork}, where a heuristic shape composition strategy, named the needle-drop approach by the authors, was employed to produce a large volume of quasi-freeform unit cells as unions of rectangle primitives (Figure~\ref{fig:representation}(i)).

        \subsubsection{Reproduction from Sparse Data to Large Data}
\label{3.2.2 reproduction . .}
\begin{figure}[t!]
\centering
\includegraphics[width=0.95\textwidth]{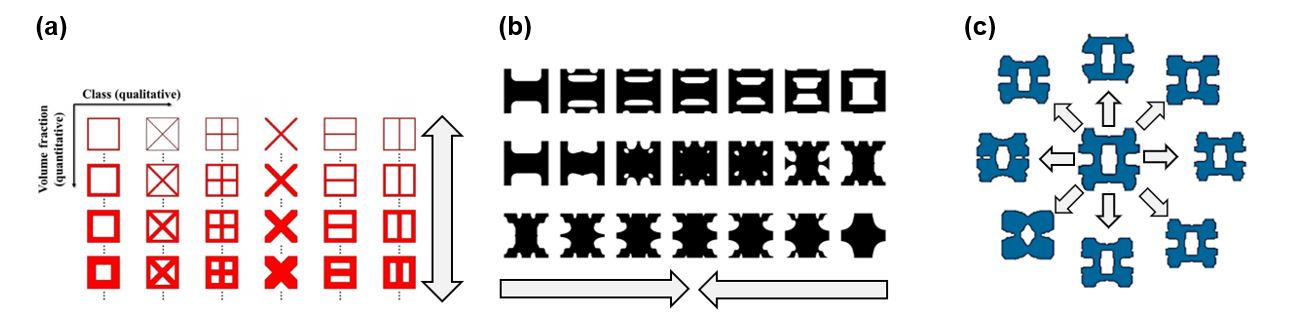}
\caption{
Reproduction strategies applied to sparse data to generate large data. White arrows indicate the direction from existing shapes to new ones. a) Parametric Sweep applied to Wang et al., where six lattice classes were explored by varying their volume fraction. Reproduced from Ref.~\cite{wang2021data} with permission from ASME. b) Multiclass Blending employed in Chan et al. where the unit cells at the leftmost and rightmost columns serve as seed classes. The proposed class remixing generates inter-class instances with ensured connectivity. Reproduced from Ref.~\cite{Chan2022RemixingBlending} with permission from Springer-Verlag GmbH Germany. c) Perturbation implemented in Wang et al~\cite{Wang2020}, where radial distortion was harnessed as the perturbation method to recursively expand the coverage in property space while preserving the topology as well as axial symmetry. Reproduced from Ref.~\cite{lee2023t} with permission from ASME.
}
\label{fig:reproduction}
\end{figure}

Once a representation is determined, a typical next step that follows is to use it to grow a large-scale shape collection, with an optional target dataset size. We will call this task reproduction throughout this review. A reproduction strategy dictates the way of producing generic instances in a shape set and the distributional nature of resulting data, thus significantly affecting the quality of downstream tasks of DMD. Through effective reproduction, DMD can enjoy a quality dataset that represents the property distribution with space-filling samples and wide coverage. Based on our survey, we observe three primary lines of reproduction methods: (i) Parametric Sweep, (ii) Multiclass Blending, and (iii) Perturbation. Figure~\ref{fig:reproduction} illustrates each with an example in literature.

\paragraph{A. Parametric Sweep}
\label{3.2.2.1 A. Parametric sweep}
Given a representation of unit cells that spans a descriptor space with low/moderate dimensionality, Parametric Sweep is often applied to explore the descriptor space as uniformly as possible with a finite number of samples. Space-ﬁlling sampling that is eﬀective in low-dimensional spaces has been intensely studied for a long time. Readers interested in the topic are referred to reviews~\cite{lin2001sampling, jin2002sequential}.

We observe that, perhaps not surprisingly, Parametric Sweep is the most widely-used reproduction strategy~\cite{an2019deep, wang2021data, Wang2022IH-GAN:Structures, so2019simultaneous, liu2022growth, kumar2020inverse, inampudi2018neural, wilt2020accelerating, wang2022generalizedOptimization}. It has been combined with diverse types of unit cell representations, especially low-dimensional ones such as the mixed-variable lattice representation (Figure~\ref{fig:reproduction}(a))~\cite{Wang2022ScalableFactors}, the six-bar lattice representation~\cite{wang2022generalizedOptimization}, the H-shape meta-atom parameterized with six variables~\cite{an2019deep}, the CGS-based representation~\cite{An2021MultifunctionalNetwork}, and the pixel/voxel dataset with canonical classes~\cite{Ma2020AStructures}, to mention a few.

Despite its wide use, three key drawbacks are that (i) the approaches offer little design freedom, as the sweep takes place only within a selected pivot and hence cannot bridge multiple unit cell classes; (ii) as the dimensionality gets larger, the density of space-filing sampling drastically drops due to \textit{the curse of dimensionality}~\cite{bishop2006pattern}; (iii) sampling only in shape space, even if done well, typically leads to huge bias in property space (discussed in detail in Section~\ref{3.3.3 perspectives on data assessment}).

\paragraph{B. Multiclass Blending}
\label{3.2.2.1 B. Muilticlass Blending}
Blending, or interpolating, across classes offers an avenue to grow a large shape library from a few initial seed classes. This approach differs from Parametric Sweep in that multiple classes jointly contribute to a new instance. Hence, this line of reproduction approaches could be powerful for merging different classes into a unified landscape that includes unseen inter-class instances.

%For example in mechanical metamaterials, Wang et al. created a shape library containing diverse cellular instances based on three TPMS families with cubic symmetry through tuning the weights and level-set parameters of each class~\cite{Wang2022IH-GAN:Structures}.
In DMD, some works addressing photonic metasurfaces also utilized image transformations based on Boolean operations (i.e., union, intersection, complement) among canonical classes (e.g., cross, split ring resonator, I-beam) to synthesize freeform inter-class instances~\cite{Liu2018, ma2019probabilistic}. In the corpus, such class blending for DMD is often followed by 
deep generative modeling~\cite{goodfellow2020generative, kingma2013auto}, which distills a continuous shape manifold of multiclass unit cells~\cite{Wang2022IH-GAN:Structures, Liu2018, ma2019probabilistic}. For a recent method developed in this branch, Chan et al. proposed a versatile Multiclass Blending scheme for functionally graded structures~\cite{Chan2022RemixingBlending, Chan2022Yu-ChinDissertation}. The entire scheme combines a weighted sum of seed classes and an activated soft union with lower feasible bounds. This enables the method to be directly integrated into multiscale TO with assurances on both structural integrity (i.e., connectivity within a unit cell) and feasibility (i.e., connectivity among neighbors) simultaneously, while avoiding the restriction of the unit cell to predefined shapes (Figure~\ref{fig:reproduction}(b)).

While feasible blending operations are not tied only to set-theoretic Boolean operations (e.g., union, complement), relevant works seem to have reported either simple unions~\cite{Wang2022IH-GAN:Structures} or its variants~\cite{Chan2022RemixingBlending}. For implementation one needs to specify (i) the type of operations (e.g., union; intersection); (ii) the number of classes (iii) rules for weighting factors. We point out that either justifications or analyses on the impact of the choices have been rarely reported in the literature, with a possible exception of Chan et al.~\cite{Chan2022RemixingBlending}.

\paragraph{C. Perturbation} 
\label{3.2.2.3 C. Perturbation}
Perturbation-based reproduction utilizes heuristics that also allow data expansion. The core idea is to (1) either look up near-boundary or on-boundary instances in the property space of a given dataset at the current iteration, (2) apply geometric perturbations on them in the ambient shape space, often beyond the given representation space, and (3) iterate the perturbation to drive the data acquisition for on-demand purposes (e.g., coverage expansion). 

An example of this approach is the iterative database expansion proposed by Wang et al.~\cite{Wang2020, wang2020deep}, where the radial distortion was recursively applied to sampled freeform building blocks for progressive growth of the property coverage. This reproduction strategy enables extensible data acquisition, which can possibly go beyond user-defined seed instances and representation, contrary to the two aforementioned strategies. At the heart of its implementation are sampling strategies that enable efficient, property-aware exploration of the shape space~\cite{Zhu2017, Wang2020}.

Perturbation has been mostly applied to the Pixel/Voxel representations; yet it could be even more effective for other lower-dimensional representations, e.g., a lattice representation using the Parametric Multiclass method (Section~\ref{3.2.1 Representation . .}), to explore new instances beyond them, as depicted in Figure~\ref{fig:reproduction}(c).

% \paragraph{Optimization}
% (** still debating whether this is eligible) All instances observed during optimization are saved in the dataset without considering optimality~\cite{Chan2022Yu-ChinDissertation, Whiting2020Meta-atomMethod}.

% Is it valid/eligible as a dataset generation method?
% good for task-awareness (as objective function). but typically does not consider diversity, with possible 1 of BO
% good for reproduction-based optimizers (e.g., metaheuristic optimizations; ...)

%    \subsection{Acquisition Strategy}    

        \subsubsection{Perspectives on Shape Generation}
\label{3.2.3 Discussion . .}        
Hinged on the taxonomy presented, we relate individual representations and reproduction strategies, and share our perspectives in multiple aspects.

\paragraph{Shape Dataset as a Design Element}
It is generally affordable to produce a bounty of unit cell shapes for DMD without obtaining their properties. Nevertheless, the shape collection needs to be judiciously prepared since (i) it primarily determines the landscape to be explored by the downstream tasks; and (ii) its utility (which we assess later in Section~\ref{3.5.1 Data Assessment .}) is related to the resulting property distribution, which is, in general, initially unknown and resource intensive to obtain. Thus, we argue shape data for DMD is a critical design element.

\paragraph{A Trade-Off between Dimensional Compactness and Expressivity}
Any representation is subject to the trade-off between dimensional compactness and expressivity. For example, both the Parametric Multiclass and Implicit Function methods enjoy dimensional compactness. Combined with Parametric Sweep, performing data generation with these is relatively straightforward. However, design exploration could be restricted unless supported by effective reproduction strategies. In contrast, the Pixel/Voxel representation supports freeform topologies without restrictions; yet the advantage comes with the challenges of efficient exploration in the huge design space and enforcing desirable design attributes, such as manufacturability~\cite{Kudyshev2020Machine-learning-assistedOptimization, ma2019probabilistic}. Parametric Curve/Surface allows for free boundary variations but suffers from topological restriction. As another moderate-dimensional representation, Constructive Solid Geometry offers topologically quasi-free instances, yet its coverage of possible shapes highly depends on the shape generation heuristics that are agnostic to property; hence it is prone to distributional bias in property space. This property bias could be a hurdle for a data-driven model to accurately learn and perform inference, and trigger the compounding effects of data quality issues in downstream tasks~\cite{Chan2020METASET:Design, lee2023t}, or, as it is known in the machine learning community, \textit{Data Cascade}~\cite{sambasivan2021everyone}. Relevant works that attempted to tackle the property bias are reviewed in Section~\ref{sequential acquisition}.

\paragraph{Class-Centric vs. Class-Free}
Depending on the presence of user-defined classes, the representations introduced above can be divided into two groups: class-centric approaches that typically include predefined classes (such as Parametric Multiclass and Implicit Function), and class-free (such as Pixel/Voxel, which is generally the case). % The key advantages of class-based approaches are (i) ease of imposing on-demand geometric properties (e.g., cubic symmetry and periodicity in TPMS~\cite{Wang2022IH-GAN:Structures}); (ii) a parsimonious number of parameters that represents complex topologies and facilitate functional grading~\cite{Li2019DesignShapes, Wang2022IH-GAN:Structures, Chan2020METASET:Design, Chan2022RemixingBlending, kumar2020inverse}. 
By pre-specifying seed classes, class-centric approaches enjoy a database that can take advantage of desirable features inherited from the user-defined unit cell templates, i.e., it can include domain expertise. However, resorting to a particular set of user-defined classes tends to restrict the design freedom early in the procedure of DMD and bias the resulting data distribution in undesirable ways. A workaround applicable during reproduction is to consider Multiclass Blending (Section~\ref{3.2.2 reproduction . .}) introduced earlier, which offers seamless connection across seed classes~\cite{Chan2022RemixingBlending}.

\paragraph{Choosing Seed Classes}
A crucial step for approaches under the class-centric umbrella is to choose, and justify the choice of, the classes with which to start. They can be chosen based on attributes related to shapes (e.g., topological features, mass/volume, smoothness, manufacturability) or their properties (e.g., elastic anisotropy, performance-to-mass ratio, broadband response). It is also important to ensure shape diversity among the classes, since it secures broad coverage of shape space. Last not but least, it has been pointed out that diversity in the shapes of unit cell data barely contributes to property diversity or task-awareness~\cite{Chan2020METASET:Design, lee2023t}. When wider coverage and better uniformity are sought, property diversity, in addition to shape diversity, could serve as a selection criterion of seed classes~\cite{wang2021data}.

% single-class vs mulitclass; why multiclass?
%\paragraph{Multiclass Representation}

\paragraph{Inspiration-Based Data Acquisition}
Natural materials that exhibit outstanding properties supported by complicated structures evolved over a long time have been a great inspiration for innovation in design-by-analogy~\cite{fu2014bio, jiang2022data}. Some works dedicated to engineering metamaterials have espoused motifs from biosystems%, such as spider web inspired acoustic metamaterials used for wave attenuation
~\cite{miniaci2016spider, liu2018fractal, hamzehei20223d, li20214d}. Although biologically inspired design provides a compelling avenue to concept generation, relevant works seem to mainly focus on proof-of-concept demonstrations with little design exploration due to grand design challenges such as scalability and repeatability~\cite{nagel2010function, goel2015biologically}. We believe that DMD can tackle the challenges by marrying the inspiration from bio-systems and data-driven exploration, especially in combination with effective reproduction strategies (Section~\ref{3.2.2 reproduction . .}).

\paragraph{Deterministic vs. Stochastic}
To date, most datasets prepared for multiscale architectural systems were created based on deterministic representations. % They are assumed to deal with an ideal situation where a multiscale system of interest is completely free from any randomness or sources of uncertainty.
On the other hand, a huge line of work addresses multiscale systems whose microstructures are either intrinsically random or subject to irreducible uncertainties associated with system deployment, e.g., material, operating conditions, and fabrications. Such scenarios can be better addressed via stochastic representations. Examples in the literature include spectral density function~\cite{farooq2018spectral, iyer2020designing} proposed for quasi-random nanophotonic structures and photovoltaic cells, and the spinodoid representation~\cite{kumar2020inverse}, which was claimed to be more robust to fabrication imperfection compared to deterministic counterparts. We believe that a comparative study between deterministic and stochastic representations is yet to be explored. Relevant discussion with more focus on the trustworthiness of DMD can be found in Section~\ref{trustworthiness}.

\paragraph{Handling of On-Demand Attributes}
%(e.g., periodicity, symmetry, manufacturability, connectivity)
In choosing a representation, the capability of handling desirable attributes could be decisive. Attributes of potential concern in DMD include symmetry, periodicity, invariance, volume constraint, manufacturability, connectivity among unit cells, and others. In general, the handling of those attributes is easier for explicit, low-dimensional representations, namely the Parametric Multiclass, Implicit Function, and Constructive Solid Geometry types under the proposed taxonomy. On the other hand, the Pixel/Voxel representations tend to need special techniques to enforce those attributes, typically with the aid of constraints~\cite{Zhu2017, Wang2020, tanriover2022deep} or data augmentation~\cite{Kudyshev2020Machine-learning-assistedOptimization, tanriover2022deep}.% We give a set of examples regarding each attribute that could be of interest for DMD. 

\subsection{Property-Aware Data Acquisition Strategy}
\label{3.5.2 Acquisition Strategy .}
Once a shape collection is prepared, design evaluation of all or some of the individual shapes usually follows in order to build training data for semi-/supervised learning (Section~\ref{prelim: machine learning}). In literature, the exhaustive evaluation of a large amount of data has been widely employed by means of space-filling sampling, such as Latin hypercube sampling~\cite{loh1996latin}. Examples in DMD include the 6-D lattice representation~\cite{wang2022generalizedOptimization}, the mixed-variable multiclass representation~\cite{wang2021data}, and the Fourier transform based representation~\cite{LiuADesign}. It has been widely used in a parametric space associated with reproduction (Section~\ref{3.2.2 reproduction . .}) as well, such as for exploration of the weight space of some isosurface representations~\cite{Chan2020METASET:Design, Wang2022IH-GAN:Structures}. The popular use of space-filing design is perhaps attributable to its simplicity and generality of implementation.

However, exhaustive sampling could become intractable when (i) the relevant simulation is time-consuming (e.g., high resolution; 3D simulation), (ii) the on-demand data size is too large (e.g., more than 100k~\cite{wang2020deep}), or (iii) the sampling space is too high dimensional (e.g., more than 50-D). It could also be the case that one wishes to acquire a data distribution with particular characteristics related to downstream tasks (e.g., negative Poisson's ratio or strong elastic anisotropy). Under such scenarios, it is warranted to exploit the acquisition strategies that take (estimated) property into account as a complement to the aforementioned shape generation heuristics (Section~\ref{3.2 Shape Generation Heuristics}). Compared to a plethora of works on sampling in the small data regime, not many methods dedicated to DMD with large data have been reported. We introduce some within the context of data acquisition for DMD.

% \subsubsection{Metaheuristic Optimization}
% \label{metaheuristic optimization}
% % whitening's work -> 
% % issue:
% % property-aware?
% % GA; PSO; ...
% % population-based

\subsubsection{Sequential Acquisition with Active Learning}
\label{sequential acquisition}

\begin{figure}[t!]
\centering
\includegraphics[width=0.9\textwidth]{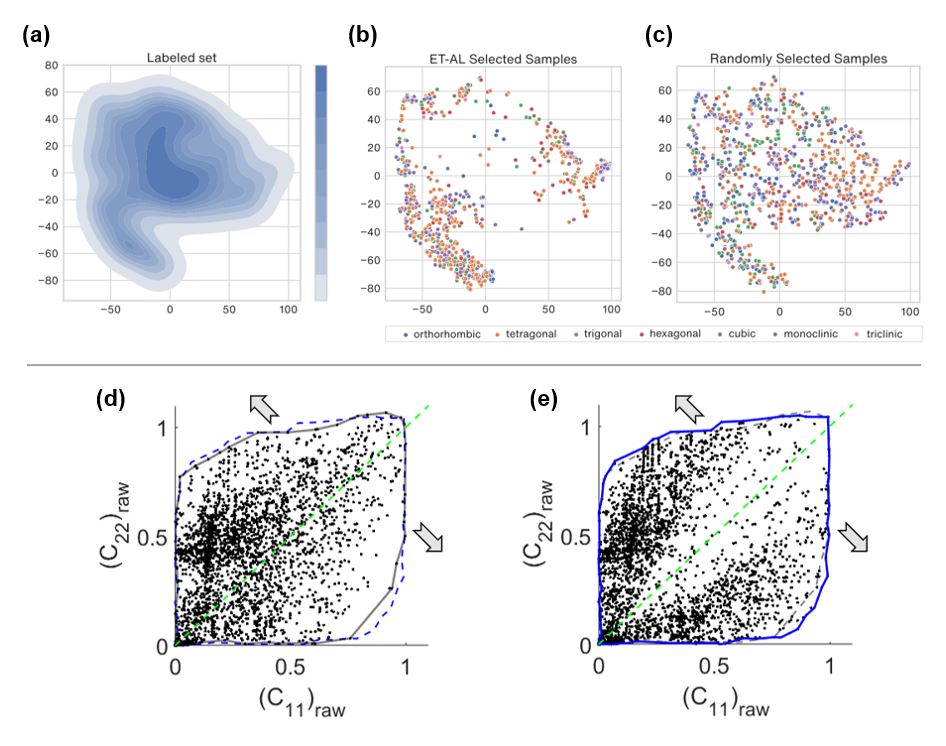}
\caption{Data acquisition for DMD through active learning. a)-c) Entropy-based active learning (ET-AL)~\cite{zhang2023entropy} demonstrated by the J-CFID dataset~\cite{choudhary2020joint}. The dataset includes 10,898 instances with seven types of crystal symmetries. a) Kernel density estimation plot of t-distributed stochastic embeddings (t-SNE), where regions with light colors are covered by sparse data. b) and c) t-SNE plots of graph embeddings of the materials selected by ET-AL and random sampling, respectively. The proposed ET-AL better covers the sparse regions, hence mitigating bias in the multiclass crystal dataset. Reproduced from Ref.~\cite{zhang2023entropy} with permission from AIP Publishing. d)-e) Task-aware diversity-based active learning demonstrated by purposefully preferring data with strong anisotropic elasticity~\cite{lee2023t}. The test dataset includes 88,180 instances of freeform pixelated unit cells of orthotropic mechanical metamaterials~\cite{Wang2020}. The resulting property distributions of 3k datapoints in the $C_{11}$-$C_{22}$ space obtained by random sampling and the proposed task-aware active learning, respectively. Reproduced from Ref.~\cite{lee2023t} with permission from ASME. %The active learning framework leads to a property distribution tailored with respect to user-defined preferences.
}
\label{fig:distributional bias}
\end{figure}

Active learning~\cite{settles2011theories, gal2017deep, ren2021survey, monarch2021human} refers to machine learning approaches that iteratively guide the locations of the next samples. In DMD, it is common to obtain a large pool of shape instances, but have labels for none or only a small portion of it. Under such a case, active learning offers a systematic, efficient, and general route to acquire evaluated samples, and thus could help prepare on-demand data.

In DMD literature, a few works specifically harnessed sequential, heuristic sampling as part of their data acquisition. In an early demonstration, Zhu et al. used a sequential sampling score that aims at property boundary expansion through randomly flipping voxels of near-boundary instances~\cite{Zhu2017}. The estimated data density and distance to the boundary were two key criteria that constitute the sampling score. Inspired by that work, Wang et al. developed an iterative stochastic data expansion scheme that builds on a sampling rule accommodating both infilling and gamut growth in the property space~\cite{Wang2020}.

With the aim to develop a method that is widely applicable to data acquisition in DMD, Lee et al. proposed a diversity-based active learning framework specialized in customizing metamaterial datasets with respect to design tasks (Figure~\ref{fig:distributional bias}(d) and (e)) ~\cite{lee2023t}. As opposed to one-shot sampling where the whole samples are collected in a single iteration, sequential data acquisition powered by active learning uses metrics to monitor the growth of a dataset, thus offering a potential answer to a pressing research question in DMD: \textit{``How much data?"}~\cite{lee2023t}. In addition to progressive dataset generation, active learning can serve as a key component for other tasks of data management, such as domain adaptation~\cite{farahani2021brief} and bias mitigation. An example was shown by Zhang et al.~\cite{zhang2023entropy}, where the proposed entropy-based active learning was demonstrated to substantially reduce the structure-stability bias of two public crystal datasets (Figure~\ref{fig:distributional bias}(a)-(c)).

\subsubsection{Downsampling Representative Subsets}
All our discussion above can be summarized as how to grow a sparse, existing dataset into a large one. Some downstream tasks of data acquisition, however, do not always benefit from using an entire, massive dataset. A prevalent issue of large datasets in DMD is distributional biases. They typically present as containing more of certain shapes or properties, typically in undesirable ways, hosting the issue known as \textit{learning under data imbalance} during the downstream tasks~\cite{Branco2016ADomains}. To this end, Chan et al. proposed a diversity-based subset selection framework built on Determinantal Point Processes~\cite{kulesza2012determinantal}, a probabilistic way of modeling diversity in relation to the determinant of a similarity matrix. The key idea is to find small yet representative subsets, whose diversity in terms of shape, property, or a joint of them is tunable~\cite{Chan2020METASET:Design}.

Such downsampling could also be useful for training the models that do not scale gracefully to large datasets, such as vanilla Gaussian processes~\cite{rasmussen2004gaussian} that scale cubically with respect to data size for training and inference. When large databases (e.g., more than 20k) are to be used as a ground set of downsampling based on pairwise metrics, e.g., diversity based on Euclidean distance, scalability of the downsampling algorithms becomes critical. It is often resolved through special schemes related to large-scale kernel learning~\cite{rahimi2007random, gillenwater2012near, affandi2013approximate}. Lastly, downsampling can be harnessed for determining a set of initial shapes to serve as seeds during shape generation, as shown by Chan et al., who leveraged Multiclass Blending (Section~\ref{3.2.2.1 B. Muilticlass Blending}) as their data reproduction strategy~\cite{Chan2022RemixingBlending}.% The selection of seed classes in Wang et al. can also be regarded as a similar example in that the the diversity of shape as well as property (elastic modulus surface) was what qualitatively justified the selection.

% An example is combinatorial tiling optimization of mechanical metamaterials, where the compatibility among neighboring microstructures must be ensured~\cite{Wang2020, Chan2020METASET:Design, wang2020deep, Wang2022IH-GAN:Structures}. Under top-down approaches of Multscale synthesis (Section~\ref{5.3 Top-Down Framework}) where all the local properties are optimized first and then set as targets, it is necessary to look up a precomputed library to place building blocks at the right place. Due to one-to-many mapping between building block and property, typically multiple instances can achieve a property of interest. Considering only 10 candidate building blocks at each location in a 50 by 50 array leads to $(50\times50)^10$ , considering all the instances in large data frequently hosts combinatorial explosion. A workaround is to prepare a small subset at each location~\cite{Wang2020}.

\subsubsection{Perspectives on Acquisition Strategy}
\paragraph{Sequential Acquisition for Generic Use: Uncertainty vs. Diversity}
Sequential acquisition can be thought of as designing the rules with which to query an existing pool of unlabeled data, which, in our review, is typically a large number of unit cells with unknown properties. Here, we discuss and compare two key approaches to acquisition for DMD: uncertainty-based sampling~\cite{lee2021dynamic, zhang2022uncertainty}, and diversity-based sampling~\cite{Chan2020METASET:Design, Chan2022RemixingBlending, lee2023t, zhang2023entropy}. Uncertainty-based sampling is centered on improving the prediction confidence of a model, typically resulting in a distributional imbalance that poorly represents the distribution of unlabeled data. Meanwhile, diversity-based sampling focuses on identifying a finite number of landmark data points to combat the distributional bias, and could include samples with little information for query. Practical implementation of either approach entails deciding the input dimensionality of data and the computational cost of the sampling algorithms.

Uncertainty is typically formulated as a model-specific, point-wise function that takes a query point as the input. Including uncertainty within sampling is effective for directly improving the predictive performance of models, whereas the computational complexity escalates as the input dimensionality increases. Within DMD, acquisition methods utilizing uncertainty could be useful when fitting machine learning models that offer uncertainty quantification, e.g., Gaussian processes or Bayesian linear regression, with a low-dimensional representation relative to a large number of data (say $ > \mathcal{O}(10^4)$). In general, uncertainty-based acquisition can be conducted based on either frequentist approaches, e.g., random forests and deep neural networks, or Bayesian approaches, e.g., Gaussian processes and generalized linear models. For practical guidance on which to employ, readers are referred to Zhang et al.~\cite{zhang2022uncertainty}.

Meanwhile, diversity is frequently modeled as a pair-wise, model-agnostic metric that involves a mapping from a pair of instances to a scalar similarity~\cite{kulesza2012determinantal}. By harnessing the pair-wise kernel trick, the diversity-based acquisition is capable of handling high-dimensional input instances. However, the acquisition does not gracefully scale with respect to data size due to large storage requirement ($\mathcal{O}(N^2)$ where $N$ is the data size) and matrix inversions that involve time complexity of $\mathcal{O}(N^3)$ unless any large-scale kernel approximations~\cite{rahimi2007random, affandi2013approximate} are employed.% In addition, it should be ensured whether the pairwise similarity computed from the representation is robust upon scaling (e.g. the bandwidth of the Gaussian similarity measure).

% \paragraph{Data Acquisition as a Dynamic Procedure}
% %- active learning -> modularity; multi-purpose; can be applied incrementally. Also %allows ones to dynamically maneuver the data acquisition based on monitoring metrics

% \paragraph{Downstream Impact of Data Quality}
% % downstream impact of data quality at the model level (RF, G,B, NN); system-level is still difficult

\paragraph{Tailoring Property Distributions}
%We remark that to date the practice of data acquisition in DMD has largely resorted to space-filling design in shape space, which is prone to property distributions that are highly imbalanced or incongruent with design tasks in concern.

\begin{figure}[t!]
\centering
\includegraphics[width=0.95\textwidth]{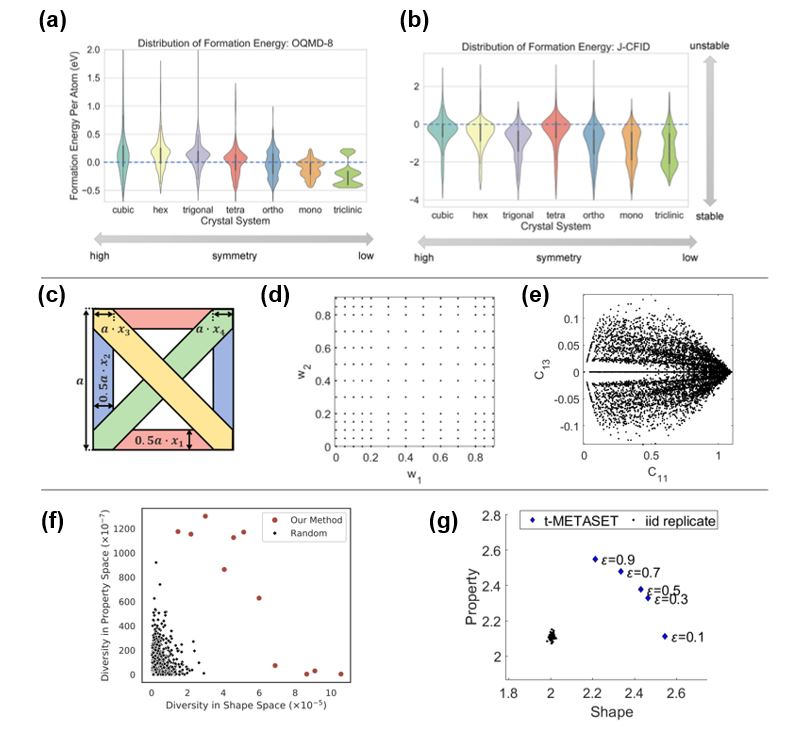}
\caption{Illustration on distributional bias in the property space of existing DMD datasets. a) and b) Stability distributions of two different crystal datasets, 2,953-size OQMD-8~\cite{saal2013materials} and 10,898-size CFID~\cite{choudhary2020joint}, whose instances are categorized based on symmetry. Reproduced from Ref.~\cite{zhang2023entropy} with permission from AIP Publishing. c)-e) Visualization of the six-bar parametric lattice dataset~\cite{wang2022generalizedOptimization, lee2023t}. c) Conceptual illustration of the unit cell shape generation based on Parametric Multiclass. d) The near-uniform space-filling design in the projected $w_1$-$w_2$ shape descriptor space. e) The resulting property distribution in the $C_{11}$-$C_{13}$ space. The near uniformity in the weight space leads to a strong bias in the $C_{11}$-${C13}$ space. f) and g) The near-zero correlation between shape diversity and property diversity observed during downsampling of the isosurface dataset and active learning applied to the orthotropic freeform dataset~\cite{Wang2020, lee2023t}, respectively. f) was reproduced from Ref.~\cite{Chan2020METASET:Design} with permission from ASME. g) was reproduced from~\cite{lee2023t} with permission from ASME.
}
\label{fig:property bias}
\end{figure}

Data acquisition methods that are agnostic to property can lead to datasets that are prone to data bias. For example, Figure~\ref{fig:property bias}(a) and (b) show the highly biased distributions of formation energy in public crystal datasets. A plethora of sampling methods in low-dimensional input domains have secured datasets with decent coverage as well as uniformity~\cite{lin2001sampling, jin2003efficient, du2004sequential, pronzato2012design, shang2021fully}. The quality in the input domain, however, has been found to barely transfer to that in the output domain. Figure~\ref{fig:property bias}(c)-(e) shows such an example in a mechanical metamaterial dataset. Within DMD literature, this point was observed through the near-zero correlation between shape diversity and property diversity depicted in Figure~\ref{fig:property bias}(f) during downsampling and in Figure~\ref{fig:property bias}(g) under active learning~\cite{Chan2020METASET:Design, lee2023t}. The implication is that property distributions are likely to be highly imbalanced even though the design in shape space is space-filling, a problem that is currently overlooked in DMD.

Furthermore, for design purposes, addressing property imbalance is only part of data quality assurance. Not all data are equally useful because users frequently have certain preferences in terms of shape, property, or both. For example, a user might wish to collect unit cells of mechanical metamaterials with negative Poisson's ratios, with packaging or shock absorption applications in mind. During data acquisition of photonic metasurfaces, broadband reflectivity might be preferred over narrowband for some defense applications. Therefore, data acquisition for DMD should, ideally, be task-aware so that more resources can be invested in the region of central interest. Figure~\ref{fig:distributional bias}(d) and (e) illustrates a methodology where batch sequential data acquisition is encouraged to favor samples with high elastic anisotropy.

It is difficult to tailor the property distribution during data acquisition without supervising the properties of interest. This supervision is non-trivial in that (1) properties of unseen unit cells are unknown before evaluation; (2) the evaluation is typically resource-intensive, particularly for large datasets; and (3) distributional control in regression tasks is more challenging and has been under-explored compared to that in classification~\cite{Branco2016ADomains}. We believe this topic calls for more research attention.% Active learning (Section~\ref{sequential acquisition}) can possibly counter all the challenges, not only for data acquisition, but other tasks of data management (Section~\ref{Other Tasks on Data}) in general.

\subsection{Data Assessment}
\label{3.5.1 Data Assessment .}
Data assessment in DMD entails quality quantification across candidate sets with respect to either general use or specific design tasks. Large sizes of data render the close inspection of individual samples intractable. Thus, their assessment is often conducted through proxy measures such as a quantitative summary of distributional characteristics, or through data visualization for qualitative interpretation. Subjectivity perhaps cannot be totally excluded in data assessment, particularly for DMD applications that involve multiple assessment criteria, e.g., data size, distributional uniformity of data, property coverage, target design tasks, and manufacturability of unit cells. Nevertheless, protocols could help to compare datasets and make decisions on which set is best suited for the intended goal.

Below, we share feasible ideas and scenarios for data assessment that have been drawn from our survey of a large volume of existing metamaterial datasets, and a sparse number of exemplar quantitative/qualitative assessment methods. As reviewers, we hope our discussion will contribute to the establishment of assessment protocols on metamaterial datasets.

\subsubsection{Qunatitative Assessment with Metrics}
\paragraph{Space-Filling Metrics} In general, data acquisition aims to make the data distribution as uniform as possible to ensure any local region of potential interest is equally covered by the dataset. In doing so, some works in DMD employed sequential sampling that included the density, or concentration, of data in certain regions of the design space as part of the sampling utility function~\cite{Zhu2017, Wang2020, wang2020deep}. The density in these works was not used to assess dataset quality, but can possibly be used to do so~\cite{lee2023t}. As an alternative to point-wise density, the alternative of set-wise diversity, or pairwise dissimilarity~\cite{kulesza2012determinantal}, can also serve as a sampling criterion to suppress distributional biases in both shape and property space, as shown in some DMD works~\cite{Chan2020METASET:Design, lee2023t, Chan2022RemixingBlending}. Point-wise diversity can be measured through information entropy~\cite{cover2006elements}. A recent demonstration in literature was performed by Zhang et al.~\cite{zhang2023entropy}, where the coverage imbalance of formation energy across seven crystal systems within the dataset was quantified through point-wise entropy and mitigated by the proposed entropy-based active learning (see Figure~\ref{fig:property bias}(a) and (b)). 

\paragraph{Task-Related Metrics}
\label{task-related metrics}
In DMD, a dataset often ends up being used for a particular design scenario. In such cases, distributional metrics alone may not ensure the on-demand deployment of DMD. Even with a dataset that exhibits perfect uniformity, it could be the case that the region associated with a given design task (e.g., high performance-to-mass ratio, high stiffness anisotropy, broadband reflectivity) happens to be covered by only a few, or even none, of the datapoints. This implies, if design tasks of interest are given, the assessment on a given dataset must vary.
% Liwei's two tasks
%This implies that (i) designers could consider utility of individual data points for the given task, on top of diversity, during both data acquisition and evaluation; (ii) it could be rather desirable to promote artificial data bias to a certain direction/area, as an on-demand inductive bias~\cite{} towards design tasks of interest.
The task-specificity of data assessment indicates that if a new task is given, data assessment must realign according to it. In a relevant work, Lee et al. specified a couple of design scenarios involving different design tasks given a shape-only dataset and showed that the resulting property distribution for each case can be tailored through diversity-driven active learning~\cite{lee2023t}.

\paragraph{Within-Dataset Assessment} 
In case multiple metamaterial datasets share the same input space, i.e., the representation space of building blocks, the quality of each can be comparatively measured based on metrics. Such comparisons could be useful for assessing across multiple data acquisition methods.

For example, Chan et al.~\cite{Chan2020METASET:Design} validated the proposed diversity-driven subset selection method for the isosurface representation by showing larger shape diversity of subsets against that of random, independent sampling. Assuming a similar setting but with more focus on sequential acquisition, Lee et al.~\cite{lee2023t} conceived a measure of diversity gain, which quantifies a relative ratio between the diversity of selected subsets and that of independent and identically distributed (\textit{i.i.d.}) samples with the same data size, to quantify the increase of shape diversity enabled by the proposed sequential sampling strategy. The authors also demonstrated better task-awareness in the two representation spaces, both of which were latent spaces distilled by training generative models.

% \paragraph{Cross-Dataset Assessment} 
% When multiple datasets prepared for a similar/equivalent design problem are available, what one might essentially wish to identify the best among them, by measuring a relative utility of each.  %Given a set of relevant but different datasets, the decision making for DMD could host questions: (i) how to compare different representations? (e.g., freeform vs isosurface) (ii) how to compare datasets with different sizes and reprensetations? (e.g., a 50k-size freeform dataset built with a freeform pixel representation versus a 1k-size Parametric Multiclass representation (Section )) 

% When the assessment is allowed to involve downstream tasks of DMD, e.g., model training, cross-dataset generalization~\cite{Torralba2011} offers a way to quantify relative utility of a dataset against another. Given two datasets (A, B) to be compared, it involves training a regression model on one (A) and test it on the other (B), and vice versa. This gauges the relative generality of one dataset against the other. % If the two model performances are more or less the same, it implies that the two datasets have a large overlap in the representation spaces in concern.
% A caveat is that the assessment involving models is subject to the choice of models (e.g., Gaussian processes; fully connected neural network) and their hyperparameters (e.g., scale parameters of Gaussian processes).

\subsubsection{Qualitative Assessment Through Visualization}
%We also introduce some scenarios where visual comparisons across datasets can be made. 
\paragraph{Property Coverage} 
Property coverage offers an intuitive, relative criterion to comparatively gauge utility across datasets, similar to how the Ashby chart visualizes a modulus-density space for disparate materials~\cite{ashby2013designing}. Upon valid normalization across datasets, data assessment in property space is usually less subjective compared to that in shape space due to lower dimensionality. Intuitive examples in literature include elasticity components~\cite{wang2022data, wang2022generalizedOptimization, Wang2020}, transmission-phase delay at a single frequency~\cite{An2021MultifunctionalNetwork}, and formation energy of crystal structures~\cite{zhang2023entropy}.

\begin{figure}[t!]
\centering
\includegraphics[width=0.8\textwidth]{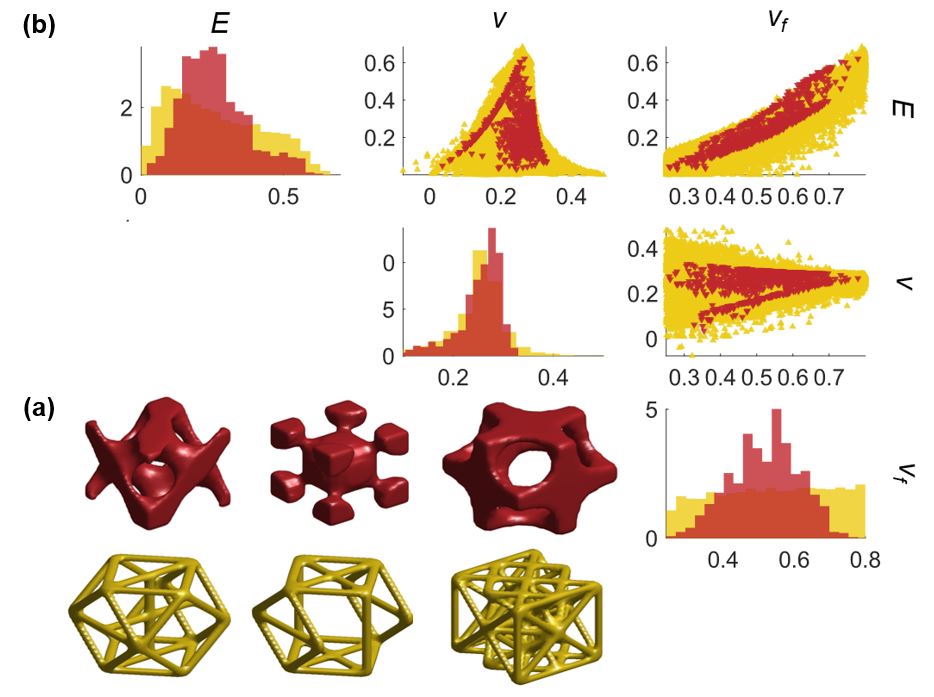}
\caption{An example visual comparison between two 3D mechanical metamaterial datasets. a) Examples of unit cells in the 924-size TPMS dataset.~\cite{Wang2022IH-GAN:Structures} (red, top) and the 21,684-size multiclass lattice dataset.~\cite{Chan2022Yu-ChinDissertation} (yellow, bottom). b) Pairwise plots of the properties of interest: Young’s modulus ($E$), Poisson’s ratio ($\nu$), and volume fraction ($v_f$). The effective properties are computed by energy-based homogenization~\cite{xia2015design}. The plots located at the diagonal depict a histogram of each component. All the off-diagonal components show scatter plots of two different properties. The upper triangle plots are shown considering the symmetry. The volume fraction ranges from 0.25 to 0.8.}
\label{fig:Fig3-5}
%\caption{Illustration on fig: task specificity of data assessment. A hypothetical data distribution of mechanical metamaterials in the $E$ (Young's modulus)-$\nu$ (Possion's ratio) space is depicted. (i) (blue) an exemplar region; (ii) (green) negative Poisson's ratio; (iii) (yellow) another exemplar region that is not covered by the given dataset; (iv) (gray) region with near-zero Young's modulus, which might not be useful for design that pursues high stiffness. The core idea is that the utility of a given dataset is contingent upon design tasks in concern.}
\end{figure}

Figure~\ref{fig:Fig3-5} shows an example of a visual comparison between two datasets in a low-dimensional property space. Both carry 3D mechanical metamaterial instances under linear elasticity. Considering the geometrical symmetry, we only consider three components, Young’s modulus ($E$), Poisson’s ratio ($\nu$), and volume fraction ($v_f$). The red one denotes the 924-size TPMS dataset generated by Wang et al.~\cite{Wang2022IH-GAN:Structures} using Implicit Function and Parametric Sweep. The yellow one denotes the 21,684-size multiclass lattice dataset presented by Chan et al.~\cite{Chan2022Yu-ChinDissertation}, created with Parametric Multiclass and Multiclass Blending. The pairwise plots of projected properties in Figure~\ref{fig:Fig3-5}(b) shed light on some insightful information in a comparative sense including: (i) Overall, the multiclass dataset has better data uniformity in terms of Young's modulus ($E$) and volume fraction ($v_f$); (ii) In the $E-\nu$ space, the TPMS dataset has some regions that are covered only by sparse data, arguably attributed to the limitation of Parametric Sweep with a few classes; (iii) Although the data size of the multiclass dataset is more than 20 times larger, some property values in the $E-v_f$ space are only available in the TPMS dataset.

\begin{figure}[t!]
\centering
\includegraphics[width=0.8\linewidth]{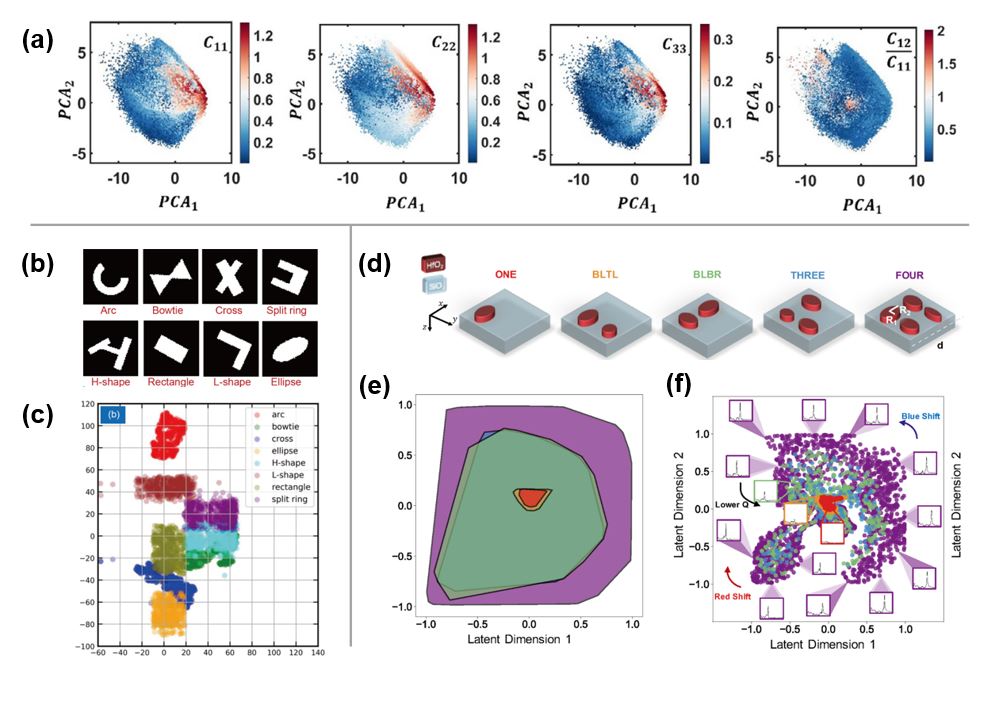}
\caption{
Visualization of high-dimensional data. a) Visualization of the latent space in the orthotropic mechanical metamaterials dataset with the property distributions included~\cite{Wang2020}. The 16-D latent representation distilled by variational autoencoder is visualized in 2D space using principal component analysis~\cite{abdi2010principal}. Reproduced from Ref.~\cite{Wang2020} with permission from Elsevier B.V. b)-c) Visualization of the latent space in plasmonic metasurfaces~\cite{ma2020data}. b) The representative images of the seed classes included in the dataset. c) The resulting data distributions in the 2D latent space using t-distributed stochastic neighbor embedding~\cite{van2008visualizing}. Each class forms separate clusters in the latent space, projected from 20D into 2D. Reproduced from Ref.~\cite{ma2020data} with permission from Science China Press and Springer-Verlag GmbH Germany. d)-f) The latent space representation of the resonant reflection spectra in dielectric metasurfaces~\cite{zandehshahvar2022manifold}. d) Metasurface unit cells with five different levels of geometric complexities. e) The corresponding convex hulls of shape manifolds estimated through the one-class support vector machine~\cite{scholkopf2001estimating}. f) The corresponding property distribution of high-dimensional optical spectra in the property manifold. It encodes the shift of resonance frequency with respect to the traversal along counter-/clockwise directions. Reproduced from Ref.~\cite{zandehshahvar2022manifold} with permission from American Chemical Society.
}
\label{cross-dataset comparison}
% \caption{
% Examples of qualitative cross-data comparison based on visualization. (a) and (b): Cross-data comparisons between the multiclass lattice dataset~\cite{Chan2022Yu-ChinDissertation} and the TPMS dataset~\cite{Wang2022IH-GAN:Structures}. (a) A schematic of the data distributions in the image space. Due to the high-dimensionality, it is difficult to grasp how each relates to the other. (b) Pairwise plots of the corresponding data distribution in property space. The plots at the diagonal locations, while those at off-diagonal ones show the property distributions in each projected space. Upon proper normalization, this gives qualitative insight on how each dataset compares to another. (c) and (d): Cross-data comparisons in photonic metasurfaces~\cite{zandehshahvar2022manifold}. (c) A visualized shape manifold whose boundaries depend on design freedom. (d) the corresponding property distribution of high-dimensional optical spectra in the property manifold. It encodes the shift of resonance frequency with respect to the traversal along counter-/clockwise directions. Technically, the visualization in ths work was not used for comparing different datasets, but can be extended to do so.
% }
\end{figure}

In the case where the dimensionality of property is high-dimensional, e.g., optical spectra of transmission, dimensional reductions are necessary to visualize the data in two-dimensional space. Zandehshahvar et al.~\cite{zandehshahvar2022manifold} shows such an example built on an autoencoder, where the latent space of optical spectra (1) visualizes the coverage dependent on design complexity of unit cells (Figure~\ref{cross-dataset comparison}(d) and (e)) and (2) automatically encodes the shift of resonance frequency along the circumferential directions (Figure~\ref{cross-dataset comparison}(f)).
%Distributional metrics also can be directly computed and compared across datasets.

% Wayne's example (TPMS vs lattice)

\paragraph{Shape Manifolds} 
The distribution of data in shape space could give another insight into data assessment in DMD. Despite high dimensionality, ranging from several (e.g., Parametric Multiclass) to millions (e.g., Pixel/Voxel), there is an array of dimension reduction schemes developed for exploratory data analysis, such as principal component analysis (PCA)~\cite{abdi2010principal}, t-distributed stochastic neighbor embedding (t-SNE)~\cite{van2008visualizing}, and uniform manifold approximation and projection (UMAP)~\cite{mcinnes2018umap}. The projection is conducted preferably into 2D spaces for straightforward visualization. This can uncover the underlying characteristics of the data distribution, e.g., clusters formed by data acquisition, often dictated by reproduction strategies of unit cells (Section~\ref{3.2.2 reproduction . .}).

For example, in DMD, Ma et al.~\cite{ma2020data} employed a VAE~\cite{kingma2013auto} and then visualized the latent space using t-SNE. The visualization projected in a 2D space reveals the clustering across the seed eight classes learned in an unsupervised manner (Figure~\ref{cross-dataset comparison}). Meanwhile, Wang et al.~\cite{wang2020deep} employed PCA to claim the versatility of the proposed latent representation learned by a conditional VAE. The visualization was used to delineate that the continuous, interpretable latent space offers simple interpolation across building blocks, a shape similarity measure, and intrinsic clustering of associated properties.  

Even more interesting use of such high-dimensional data visualization is for comparisons across datasets. Employing the one-class support vector machine~\cite{scholkopf2001estimating}, Zandehshahvar et al. visually demonstrated the impact of geometric freedom in building blocks by visualizing different coverage in both shape and property space (see Figure~\ref{cross-dataset comparison}(d)-(f) for illustration).% It was also observed that the learned latent representation of property space exhibits a shift of resonance frequencies according to counter-/clockwise directions, which was encoded without human supervision.

\subsubsection{Perspectives on Data Assessment}
\label{3.3.3 perspectives on data assessment}
\paragraph{Task Specificity of Data Acquisition and Assessment}
In Section~\ref{task-related metrics} we covered the task-specificity of data assessment with a particular focus on metrics. We now summarize our general view on data acquisition and data assessment of DMD for commonplace scenarios as follows:
\begin{itemize}
    \item Provided that no target tasks have been specified, data acquisition can aim to create a dataset for generic use, i.e., focus on uniformity and wide coverage, in both the shape and property spaces of building blocks. The data assessment can also follow the same criteria without preferring a certain region in those spaces. Ref.~\cite{Chan2020METASET:Design} shows an example in literature.
    \item Even without being given any specific on-demand properties \textit{\textit{a priori}}, instance-wise preferences related to shape (e.g., fabrication feasibility), property (e.g., high physical anisotropy), or both (e.g., performance-to-mass ratio) can be enforced during data acquisition to tailor the data distribution as desired with minimal trial-and-error. The data assessment approach must address both distributional metrics and task-related metrics. Ref.~\cite{lee2023t} is an example.
    \item If a target task is required at downstream, or a set of target tasks is given, the data acquisition and assessment can be aligned to the specified task(s), in addition to data uniformity. In these cases, data uniformity is useful only within the domains associated with the tasks. Moreover, the assessment is subject to the definition of target tasks. A concrete example where the task of matching target displacements at the system level is addressed in Ref.~\cite{Wang2020} and discussed in Section~\ref{5.4.5 Task Specificity of Data Acquisition for Multiscale Design}.
%    \item When a new task comes in addition to existing ones, additional data acquisition could follow to address it.
\end{itemize}

% \begin{figure}[t!]
% \centering
% \includegraphics[width=.9\textwidth]{figure/Fig3-5-4.jpg}
% \caption{Task specificity of data acquisition and assessment based on the case study in Wang et al.~\cite{wang2020deep}. (a) Illustration of three system-level design tasks associated with different target displacements: smiley face (left), mouth (center), and bridge (right); (b) On-demand distributions of homogenized elasticity components $\{C_{11}$, $C_{12}$, $C_{22}$, $C_{33}\}$; (c) The distributions of required property for all the individual tasks plotted with respect to that of the orthotropic dataset (gray), created through Pixel/Voxel and Perturbation. All the results are reproduced from Wang et al.~\cite{wang2020deep} with permission.
% }
% \label{fig: task specificity}
% \end{figure}

\paragraph{Assessment Protocols}
Data assessment is essential for either diagnosing a dataset or choosing the best among competing ones. How to fairly measure the quality of metamaterial datasets is key to decision-making over competing datasets, as well as to minimization of iterations among the modules of data-driven metamaterials design. A general guideline for the assessment of synthetic datasets for engineering purposes was recently proposed~\cite{picard2023dated}. However, not many attempts that are dedicated to data assessment for DMD have been reported, compared to the rapidly growing volume of the corpus. Without agreement upon standard protocols of data assessment, it is difficult to judge the quality of individual works that include data acquisition and to draw meaningful conclusions among them. Thus, we assert the need for more research efforts centered on data assessment.

\paragraph{Benchmark Datasets}
The easy access to public datasets has been an enabler of the recent surge of machine learning. Ideally, newly proposed methods on any module of DMD should be validated through diverse benchmark datasets that are suggested by the communities. In the corpus of data-driven design, however, such solid validation seems difficult to find. A profound reason that applies to general data-driven design is, arguably, a dearth of public datasets and benchmarks~\cite{Regenwetter2021DeepReview, picard2023dated}. Echoing this, we argue that securing more public datasets and setting some of them as benchmarks will be the first step towards the research practice that prompts quantitative, rigorous comparisons of relevant works and reproducible research, hence helping readers better appreciate individual works with relation to the field. In doing so, it is highly encouraged to make new datasets publicly available, preferably in online repositories that support consolidation across datasets.

\subsection{Discussion}  
\label{3.5 Discussion . }

\subsubsection{Reusability of Datasets}
\label{3.5.3 Reusability . .}
We observe that the current practice in data-driven multiscale architectural design typically starts with creating one's own dataset, rather than with existing ones; arguably, the practice is what has led to ``per-task" datasets involving a similar/equivalent end-use. To avoid the trial-and-error and computational resources that data creation from scratch demands, new works can be conducted by either (i) creating a versatile, high-quality dataset that can address multiple, disparate design tasks, or (ii) customizing a public dataset with respect to new design tasks. We believe that the reusability of datasets is by no means trivial to achieve, and thus deserves further research attention.
%\subsubsection{Generality}

\subsubsection{Limitations of Unit Cell Datasets}
\label{3.5.5 Beyond Unit Cell Datasets . . }
To date, most demonstrations of data-driven multiscale design have been built on building block datasets. Each data point, i.e., a structure-property pair, is generated based on periodic boundary conditions (PBC), which assume a given building block is surrounded by infinitely many identical neighbors. Although the assumption is valid only for periodic multiscale architectures (e.g., crystal), some recent works involving fully aperiodic design still utilized the PBC-based effective properties. They justify this choice, which can accelerate the design process, through, e.g., geometric/mechanical compatibility under linear elasticity~\cite{wang2020deep}, functional grading~\cite{Chan2022RemixingBlending, Chan2022Yu-ChinDissertation}, and uncoupled operations in photonic metasurfaces~\cite{An2021MultifunctionalNetwork, tanriover2022deep}.

We point out that while the periodicity assumption has been a backbone of the recent progress in DMD, it is also a hurdle that impedes researchers from exploring other problems where the assumption does not hold true. There are a variety of cases where the deviation of effective properties under PBC at the system level is more than acceptable~\cite{so2022revisiting}. Examples include the systems under: (i) large deformation~\cite{ning2020low, ma2022deep}, (ii) strong local coupling among neighbors~\cite{zhelyeznyakov2021deep, an2022deep}, (iii) long-range interactions~\cite{ghavanloo2019wave, zhu2020nonlocal}, and (iv) heterogeneous loading conditions~\cite{ertsgaard2014dynamic, buijs2021programming, lee2021dynamic}. Preparing the datasets that take a supercell (i.e., a collection of neighboring unit cells) instead of a single unit cell as a datapoint could offer a simple yet powerful extension to the unit-cell-based approaches, as shown by ~\cite{an2022deep, spagele2021multifunctional, lee2021dynamic}. The extension would involve increased computational cost and validating the boundary conditions applied to the supercell simulations. A relevant discussion at the system level can be found in Section~\ref{5.4.2 Assumptions of homogenization}.

\subsubsection{Learning Global Responses vs Local Responses}
Many works reported in DMD include a surrogate model that directly maps the parameterized unit cells to effective, homogenized properties, such as elasticity components~\cite{Wang2020, Wang2022IH-GAN:Structures, Chan2022RemixingBlending, Chan2022Yu-ChinDissertation} and scattering parameters~\cite{liu2018generative, ma2019probabilistic, An2021MultifunctionalNetwork}. %In essence, such ``global" properties are a proxy for local physical fields, typically formulated through surface/volume integration within the measuring domain of interest. % Examples include elasticity components in mechanical metamaterials and scattering parameters and derivatives thereof (e.g., transmission/phase delay) in electromagnetic metamaterials.
The associated mapping can be relatively easily learned provided that there is enough training data and the output dimensionality is low. However, this comes at a significant loss of full field information.% It could also be the case that one might wish to replace the quantity of interest with another in data while keeping the rest same. When such scenarios are of interest, the global surrogate based approaches are ineffective. 

A workaround that can boost the generality of the structure-property mapping is to directly learn output physical fields, e.g., displacement, electric fields, and temperature, as a function of parameterized unit cells. The goal can be achieved through either physics-informed machine learning or operator learning~\cite{brunton2016discovering, raissi2019physics, goswami2022physics}, where an underlying physics is either imposed as a constraint or discovered by field data. Such a mapping can capture underlying spatial correlations of fields, which could be subject to strong long-range interaction across unit cells in DMD. This approach features decent transparency, generality, and sample efficiency. These approaches need to address challenges that include (1) how to effectively regularize the learning with regard to high-dimensional output fields (e.g., adding a sparsity penalty to avoid overfitting~\cite{brunton2016discovering}) and (2) how to impose priors associated with physics (e.g., smoothness of fields~\cite{zhang2022metanor}). Some works in DMD proposed to learn global behaviors~\cite{zhelyeznyakov2021deep}; however, the literature to date is sparse. There is room to place more attention on this branch of approaches. Details of this topic are reviewed in Section.~\ref{sec:physics_based}.

% long-range interactions

% field learning vs secondary learning
% generality
% # of data

% PDE learning, operator learning

\subsubsection{Determining Data Size}
In one form or another, ``\textit{How much data?}" has been a key research question in data-driven approaches. Within the scope of DMD, this question affects the following aspects: (1) model complexity (e.g., neural networks vs Gaussian processes~\cite{rasmussen2004gaussian}), (2) unit cell representation (e.g., high-dimensional vs low-dimensional), (3) simulation cost/fidelity, among others. 
Due to the multifaceted nature of DMD problems, it is difficult to predict an ``optimal" data size \textit{\textit{a priori}}, especially via one-shot sampling. The data sizes reported in the literature could be a good starting point, provided that a new design task of interest shares some attributes, e.g., model complexity and unit cell representation, with those of the reported works. Integrating active learning with data acquisition (Section~\ref{sequential acquisition}) can be a more rigorous, general approach to determining data size since it offers metrics related to either the data themselves or model performance. Lee et al.~\cite{lee2023t} claimed that diversity metrics can be monitored to gauge the relative utility of incoming data, hence serving as a proxy to determine data size based on the gamut growth in property space. For generic scenarios, a guideline on data size of engineering datasets was proposed by Picard et al.~\cite{picard2023dated}. %Retraining of large models with respect to every new observation could be ineffective, due to minute influence of it with regard to large data, and computationally prohibitive~\cite{biyik2019batch}. We advocate batch sequential sampling~\cite{} that recommends a set of samples to be observed at once to reduce the overhead of large-scale model retraining.

In the corpus, we observe most prior efforts were hinged upon large data. It is equally worth investigating in the small data regime~\cite{rixner2022self}, as doing so will help tackle design problems that involve expensive simulations and limited computational resources. This call resonates with data-centric AI~\cite{motamedi2021data, whang2023data, mazumder2022dataperf}, an initiative that propels a paradigm shift of data acquisition from \textit{more data} to \textit{better data}~\cite{strickland2022andrew}.

% More data to better data

% how many data?
% time-consuming; dynamic; 
% sampling efficiency
% data-hungry (e.g., deep generative models, mentioned in )

% An et al.~\cite{an2022deep} observed the deviation of effective optical properties obtained under PBCs and. proposed a CNN-based framework that can predict inter-coupling effects among building blocks. The prediction network took 1-D arrays of aperiodic unit cells as the input data and used for inverse optimization of aperiodic metasurfaces. More detailed discussions dedicated to photonics metasurfaces can be found in So et al.~\cite{so2022revisiting}.

\subsubsection{Data Sharing Practice}
% refs: refs from Nanomine; HMS proposal; Brite proposal;

% Online repository: Metamine / Nanomine (nanocomposites, metallic material systems)
The surge of DMD has inspired the emergence of open-source data sharing platforms, such as \textit{NanoMine}~\cite{zhao2016perspective, mccusker2020nanomine, brinson2020polymer}, % dedicated to the development of a data curation system, extensible knowledge representation, and interactive visualization tools,
which pays special attention to polymer nanocomposites. \textit{MetaMine}, its sister platform, currently stores 300k structure-property data of metamaterials with a diverse array of unit cell representations. Building a common platform and knowledge representation presents immense challenges. The endeavor will facilitate consolidating datasets that were acquired independently, and therefore enhance the potential of DMD beyond that what is achievable by an individual dataset.
% respond to user-defined queries, 

A prerequisite to building a user-interactive data platform that supports reusability and reproducibility is sharing protocols. For example, the FAIR (Findability, Accessibility, Interoperability, and Reusability) principles~\cite{wilkinson2016fair} is a concise, domain-independent, and thus generic, guide for data sharing. Exercising the FAIR principles is built upon core elements such as standardized vocabularies, ontologies, and data formats. Despite the foundational role that such general guidelines have played in existing data-sharing platforms, they tend to only specify broad guidelines of data quality assessment at a high level, leaving the needs of discipline-specific standards unaddressed~\cite{boeckhout2018fair}. For DMD, the proper format of (meta)data could differ wildly across different domains (e.g, nanocomposites vs. mechanical/photonic metamaterials). Even within a given domain, it could be difficult to define a set of commonly structured vocabularies or knowledge representations that accommodates all datasets submitted by users. In this regard, we advocate for an extensible, dynamic platform, which starts with initial vocabularies and schema defined by humans, as showcased by \textit{Nanomine} for polymer nanocomposites~\cite{brinson2020polymer}, and which is then allowed to evolve without supervision as more data is ingested.

%%%%%%%%%%%%%%%

\subsubsection{Other Tasks Involving Data}
\label{Other Tasks on Data}
So far we have primarily covered data acquisition. Other possible tasks include data augmentation, data consolidation, bias reduction, problem/domain adaptation, and exploratory data analysis. Each task plays a unique role that cannot be fully addressed by data acquisition itself. For example, it is highly recommended to perform data augmentation because (1) it increases the amount of training data without further evaluations; (2) it helps a machine learning model to be encoded with operational invariances, such as rotation, scaling, translation; and (3) it tends to mitigate overfitting by serving as a regularizer of model training. Within DMD the efficacy of augmentation has been demonstrated in some works~\cite{Kudyshev2020Machine-learning-assistedOptimization, tanriover2022deep}. We believe that further research on data-related tasks other than data acquisition will underpin the future success of DMD by enhancing the generality, customizability, and reusability of data.

\subsubsection{Public Resources}
\label{3.5.4 Public Resources .}
We share a link to an online webpage of public resources that are associated with data-driven design as follows: \href{https://github.com/ideal-nu/Data-Driven-Design-for-Metamaterials-and-Multiscale-Systems-Status-and-Opportunities/}{https://github.com/ideal-nu/Data-Driven-Design-for-Metamaterials-and-Multiscale-Systems-Status-and-Opportunities/}.

%%%%%%%%%%%%%%%%%%%%%%%%%%%%%%%%%%%%%%%%%%%%%%%
\section{Data-Driven Unit Cell Design of Metamaterials}
\label{4 Learning and Generation: Data-Driven Metamaterials Design}

In Section~\ref{3. Data Acquisition}, we reviewed past works related to metamaterials data acquisition and discussed some challenges at the data acquisition stage. In this section, we introduce how past works used data-driven methods to solve unit-cell-level metamaterials design problems.

%%%%%%%%%%%%%%%%%%%%%%%%%%%%%%%%%%%%%%%%%%%%%%%%%%%%%%%%%%%%%%%%%%%%%%%%%%%%%%%%%%%%%%%%%%%%%
\subsection{Overview}

The advance of machine learning has motivated researchers to seek data-driven solutions to many real-world design challenges. These challenges mainly originate from the following factors:
\begin{enumerate}
\item Analysis using physical experiments or high-fidelity simulations usually requires high cost. For example, numerical nanophotonic simulations can take hours or even days for complex systems~\cite{smajic2009comparison}.
\item Advanced fabrication technology (e.g., additive manufacturing, micro/nanofabrication) enables high degrees of design freedom, but exploring a high-dimensional design space to find optimal solutions is challenging.
\item While methods, like adjoint-based shape optimization and TO, can address high-dimensional design problems by using sensitivities to guide optimization, they are usually not applicable when the physics governing the problem is non-differentiable with respect to design variables.
\end{enumerate}

In this section, we will introduce how past work used data-driven methods to solve these challenges, particularly in the domain of unit-cell-level metamaterials design. Note that data-driven methods have been gaining increasing attention in engineering design, mainly for shape and topological designs, to address the challenge brought by their high degrees of design freedom. For shape optimization, a large body of work looked at data-driven aerodynamic shape optimization, which mainly focused on dimensionality reduction or representation learning (e.g., \cite{poole2015metric,allen2018wing,li2019surrogate,chen2020airfoil,li2020efficient}) and inverse design (e.g., \cite{sekar2019inverse,chen2022inverse,glaws2022invertible}). Compared to only considering shape variation in shape design, metamaterials design can usually accommodate topological changes depending on their functional requirements. These extra degrees of freedom make it more difficult for traditional design methods to solve metamaterials design problems. Many works on machine learning-assisted TO serve various purposes such as reparameterization, objective function prediction, sensitivity prediction, direct prediction of TO solutions, and enhancing diversity for generative design. We refer interested readers to Ref.~\cite{woldseth2022use} for a summary of past contributions of neural network-based TO methods. These TO methods were usually applied to structural design problems, where well-established physics and sensitivity analysis are available. In this section, we will cover metamaterials design under different physics (e.g., mechanical, optical, acoustic/elastic, thermal) scenarios, for some of which sensitivity analysis is either unavailable or too difficult. Due to that reason, traditional gradient-based TO methods might not be applicable to certain types of metamaterials design.

The benefits of data-driven methods highly depend on (1)~the cost of data collection and learning and (2)~the acceleration contributed by the data-driven model. To make such methods cost-effective, the trained model needs to be reusable for different tasks. Problems like structural optimization are usually subject to task-dependent objectives and constraints. In this case, it is difficult to train a single ML model that can be reused in different tasks. In contrast, unit-cell-level metamaterial design problems under the same physics usually share common properties of interest (e.g., elasticity properties in mechanical metamaterials and transmission and phase delay in optical metamaterials), regardless of different functional goals (e.g., design for compliant mechanism, energy absorption, or noise reduction). Thus, a data-driven model trained on the same properties of interest can be reused in different metamaterials design tasks.

\begin{figure}[ht]
\centering
\includegraphics[width=0.9\textwidth]
{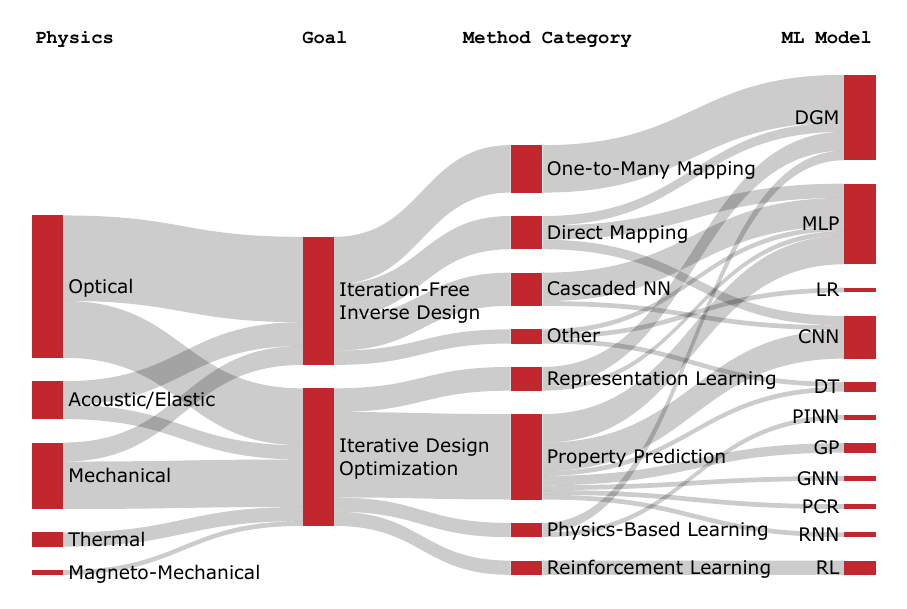}
\caption{Categorization of physics, goals, machine learning models, and their relations based on 56 representative prior works. (LR: logistic regression; RL: reinforcement learning; PCR: principal component regression; DT: decision tree; DGM: deep generative model; GP: Gaussian process; MLP: multi-layer perception; CNN: convolutional neural network; RNN: recurrent neural network; GNN: graph neural network; PINN: physics-informed neural network.)}
\label{fig:metamaterial_design_sankey}
\end{figure}

This section covers five types of metamaterials~\textemdash~optical, acoustic/elastic, mechanical, thermal, and magneto-mechanical. Despite differences in physics, they share similar design scenarios: (1) design allows high-degree of geometric freedom; (2) physical properties are usually the target quantities of interest; (3) physical properties depend on design geometry; (4) generating random design geometries is cheap whereas computing their physics properties is expensive. For this reason, data-driven methods developed for metamaterials are usually applicable under different types of physics. Thus, we focus on the goals of data-driven methods instead of metamaterial types when characterizing past work. Specifically, we categorize the goals into two major types~\textemdash~iterative design optimization (Section~\ref{sec:design_optimization}) and iteration-free inverse design (Section~\ref{sec:inverse_design}). 
% Note that there are other purposes of using data-driven methods, such as representation learning, feature extraction, metamaterials family identification, and data generation. But due to their relatively sparser occurrence in past work, we will only briefly mention them in Section~\ref{sec:other}.
Figure~\ref{fig:metamaterial_design_sankey} illustrates the relationship between the goals of proposed data-driven methods, the physics being considered, and the machine learning models, extracted from 56 representative prior studies. These prior studies were selected from publications from 2018 to 2023, with the aim to use machine learning to assist metamaterials design. We focus on only single-scale metamaterials design in this section, where unit cell designs are arranged periodically in space. In Section~\ref{5 Data-driven Multiscale TO}, we will introduce multi-scale design, where aperiodic unit cell designs are considered to achieve spatially-varying material properties.

%%%%%%%%%%%%%%%%%%%%%%%%%%%%%%%%%%%%%%%%%%%%%%%%%%%%%%%%%%%%%%%%%%%%%%%%%%%%%%%%%%%%%%%%%%%%%
\subsection{Iterative Design Optimization}
\label{sec:design_optimization}

In this section, we review prior works that employed ML methods for metamaterials design optimization. In particular, ML commonly plays roles in accelerating property evaluation (Section~\ref{sec:property_prediction}), learning more efficient design representation (Section~\ref{sec:design_representation}), sequential decision making (Section~\ref{sec:rl}), and physics-informed solution generation (Section~\ref{sec:physics_based}).

\subsubsection{Accelerated Optimization via Data-Driven Property Prediction}
\label{sec:property_prediction}

In unit-cell-level metamaterials design, material properties are usually the design target. For example, absorption spectra and dispersion relations can be properties of interest for optical and acoustic metamaterials design, respectively~\cite{azad2016metasurface,jiang2022dispersion}; for mechanical metamaterials design, properties of interest can be Young's modulus, Poisson's ratio, and volume fraction~\cite{Wang2022IH-GAN:Structures}. We can perform numerical simulations or experiments to evaluate material properties. However, depending on different physics properties, the computational cost can be prohibitive, especially when iterative design optimization is required. Data-driven models can learn complex structure-property relations and hence surrogate time-consuming simulations or experiments, allowing high-throughput property evaluation. This can accelerate the design process when combined with downstream design space exploration methods (e.g., sampling, screening, and optimization). 

Based on the reviewed literature (Figure~\ref{fig:metamaterial_design_sankey}), most ML-accelerated design optimization works were based on this approach, among which the most commonly used ML models were convolutional neural networks (CNN)~\cite{gu2018bioinspired,sajedian2019finding,wiecha2019deep,an2020deep,donda2021ultrathin,garland2021pragmatic,ma2022deep} and multilayer perceptron (MLP)~\cite{inampudi2018neural,zhang2021genetic,zhelyeznyakov2021deep,ji2022design,pahlavani2022deep,lee2022generative}. Typically, CNNs were used for pixelated design representations with high geometric freedom or design dimensionality (allowing more complex designs), while MLPs were used for parametric or shape designs with lower design dimensionality. These two models were mostly used when the training data size is larger than 700. One special case is Ref.~\cite{zhelyeznyakov2021deep}, where the local design variables of only 10 metasurfaces along with the local patches of their electromagnetic (EM) fields response were used as training data of a property-prediction MLP, resulting in 123,210 actual training samples.

While the training of most neural network-based models normally requires large datasets, Gaussian process (GP) was employed when there was a relatively small amount of training data~\cite{wang2021data,wang2022design}. GP's ability to estimate uncertainty makes it well-suited for adaptive sampling and Bayesian optimization. These are useful design techniques especially when the computation of responses is time-consuming. However, one limitation of standard GP models is the difficulty in handling large datasets due to its $\mathcal{O}(N^3)$ time complexity and $\mathcal{O}(N^2)$ memory complexity, where $N$ is the training data size. This also limits the dimensionality of design problems to be considered, because the required amount of data scales exponentially with the dimension due to the curse of dimensionality. Past work proposed ways to make GP more scalable with larger datasets~\cite{hensman2013gaussian,evans2018scalable,bostanabad2019globally,wang2019exact,liu2020gaussian,Wang2022ScalableFactors}, which can potentially expand the use cases of GP to larger datasets and higher problem dimensions in metamaterials design. 

Dimensionality reduction (DR) was also applied to reduce the dimensionality of the original design or property space before applying regression models for property prediction. Wang et al.~\cite{wang2022design} used a Gaussian mixture beta variational autoencoder (GM-$\beta$VAE) to reduce the dimension of the pixelated metamaterials design representation. Chen et al.~\cite{Chen2018ComputationalFamilies} employed principal component regression (PCR) to reduce the 3D metamaterial design parameters, where the principal component analysis (PCA) was followed by linear regression to predict the elastic material properties of mechanical metamaterials. Zhelyeznyakov et al.~\cite{zhelyeznyakov2021deep} reduced the dimension of the near-field response of the metasurface using singular value decomposition (SVD) before applying property prediction models.

In some works, decision trees (DTs) were used as the property prediction model due to their interpretability and flexibility in learning non-linear structure-property relations. Elzouka et al.~\cite{elzouka2020interpretable} predicted the spectral emissivity of dielectric and metallic particles. Chen et al.~\cite{chen2022see} used Generalized and Scalable Optimal Sparse Decision Trees (GOSDT)~\cite{lin2020generalized} to predict band gaps in different frequency ranges based on shape-frequency features extracted from acoustic metamaterial geometries. Particularly, the interpretability of DTs allows us to extract explicit design rules that can guide inverse design, which we will elaborate on in Section~\ref{sec:iteration_free_other}.

In most prior work, metamaterials design variables were either represented as a vector of design parameters or a tensor representing pixelated designs, while the properties of interest normally consisted of single or multiple scalars. Beyond these common representations, Yang et al.~\cite{yang2022high} studied 3D graphene metamaterials with a graph representation and used a graph neural network (GNN) to predict the local atomic stress distributions. Sajedian et al.~\cite{sajedian2019finding} aimed at predicting the absorption curve of plasmonic structures and solved this problem with a recurrent neural network (RNN). 

To obtain the metamaterial design solution, a common approach is to use the property prediction model as a surrogate design evaluation model and incorporate it into any iterative optimization loop~\cite{inampudi2018neural,garland2021pragmatic,zhelyeznyakov2021deep,zhang2021genetic,ma2022deep,wang2022design,lee2022generative}. Besides iterative optimization, past works also employed virtual screening~\cite{yang2022high} and sampling~\cite{gu2018bioinspired,chen2022see,ji2022design} to select design solutions, which also took advantage of the fast design evaluation capability of property prediction models.
% After training a property prediction model, metamaterial designs are often obtained by combining the property prediction model with virtual screening~\cite{yang2022high}, sampling~\cite{chen2022see,ji2022design}, optimization~\cite{inampudi2018neural,garland2021pragmatic,zhelyeznyakov2021deep,zhang2021genetic,ma2022deep,wang2022design}, or learned property-structure mapping~\cite{ma2018deep,Malkiel2018PlasmonicLearning,liu2018training,so2021demand,Yeung2021MultiplexedNetworks,zhen2021realizing}.

% \begin{table}
% \label{tab:property_prediction}
% \caption{Distribution of data-driven property prediction models in past work.}
% \begin{center}
% \begin{tabular}{ cc } 
%  \hline
%  Property Prediction Model & Past Work \\ 
%  \hline
%  GP &  \\
%  DT &  \\
%  MLP &  \\ 
%  CNN &  \\ 
%  RNN &  \\ 
%  GNN &  \\ 
%  \hline
% \end{tabular}
% \end{center}
% \end{table}

\subsubsection{Accelerated Optimization via More Efficient Design Representation}
\label{sec:design_representation}

Another way to accelerate design optimization is through learning an efficient design representation (i.e., latent representation) that is more compact than the original representation but still as expressive, thereby covering the same design space with fewer design variables. This benefits in three ways: (1)~it mitigates the issue caused by the curse of dimensionality when training property prediction models, thus lowering the requirement for training data size and model complexity; (2)~it enables more efficient optimization, since searching for global optimal solutions in a lower-dimensional latent space is easier and faster; (3)~it allows easier downstream analysis such as data visualization, clustering, and arithmetic operations in the latent space. Since we introduced past work on using dimensionality reduction for property prediction in Section~\ref{sec:property_prediction}, this section will focus on the other two benefits.

As shown in Figure~\ref{fig:metamaterial_design_sankey}, most prior works under the category of ``representation learning" used deep generative models (DGMs). Liu et al.~\cite{liu2020hybrid} employed a variational autoencoder (VAE) to learn a lower-dimensional latent space of pixelated optical metasurface designs and leveraged an evolutionary algorithm to optimize designs over the latent space. Wang et al.~\cite{Wang2020DeepSystems} constructed an end-to-end model combining a VAE with a mechanical property regressor, and learned a lower-dimensional, structured latent space organized by mechanical properties of pixelated metamaterials designs. Semantic operations (e.g., moving from ``low stiffness" to ``high stiffness") can be achieved by simply moving in certain directions of the resulting latent space. Chen et al.~\cite{chen2022gan} proposed a generative adversarial network (GAN) with hierarchical latent spaces to simultaneously represent the ``as-designed" and ``as-fabricated" optical metasurfaces. The model not only learned a compact latent representation for ``as-designed" metasurfaces, but also captured the geometric uncertainty of ``as-fabricated" designs, which enables efficient and robust design optimization under fabrication uncertainty. Shen and Buehler~\cite{shen2022nature} used StyleGAN~\cite{karras2019style} to learn disentangled latent spaces that capture attributes and variations at different levels. Design optimization and geometric manipulations (i.e., projection, encoding, and mixing) can be achieved by optimizing the latent vectors.

Autoencoders (AEs) were also used for compressing the dimension of the metamaterials design space~\cite{liu2021thermal,sui2021deep}. Compared to deep generative models like VAEs and GANs, AEs only minimize the reconstruction errors of training samples without considering the continuity of the latent space, which may reduce the efficiency of design optimization and latent space analysis.

\subsubsection{Design as Sequential Decision Making via Reinforcement Learning}
\label{sec:rl}

\begin{figure}[ht]
\centering
\includegraphics[width=1\textwidth]{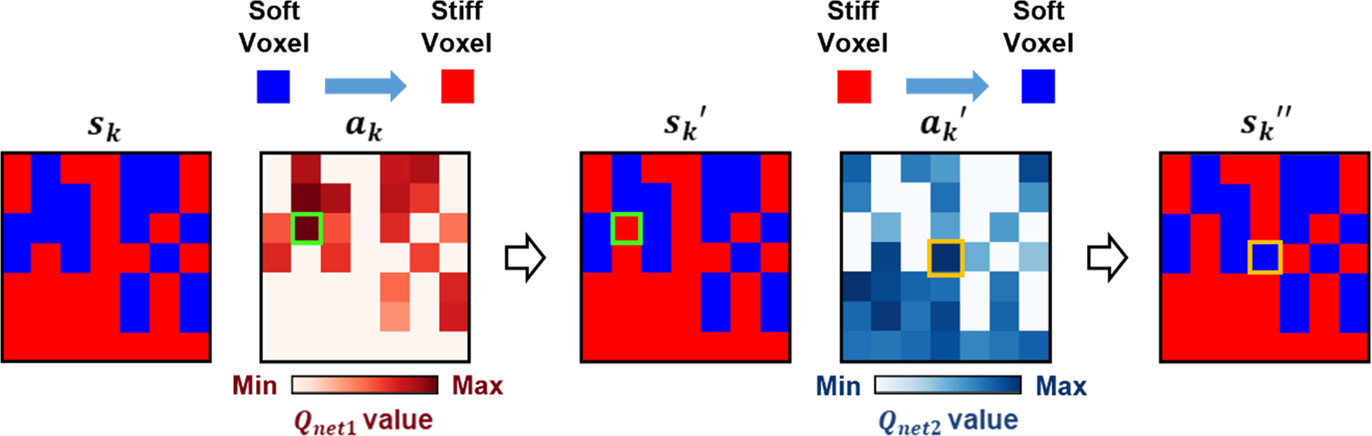}
\caption{ ``Flipping process" of the collaborative DQN method in Ref.~\cite{sui2021deep}: one agent selects a soft voxel and turns it into a stiff voxel, then the other agent does the opposite. Reproduced from Ref.~\cite{sui2021deep} with permission from American Chemical Society.}
\label{fig:sui_rl_flipping}
\end{figure}

The design problem can also be modeled as a sequential decision-making process, which can be solved by reinforcement learning (RL). As introduced in Section~\ref{sec:pre_rl}, the main components to be defined in any RL tasks are state, action, and reward. Defining these three concepts~\textemdash~especially the action~\textemdash~in a design problem needs extra consideration, unlike canonical reinforcement learning problems such as gameplay and adaptive control, since it is more intuitive to treat metamaterials design as a standard optimization problem rather than a dynamic problem that requires sequential decision-making.
Nonetheless, Sajedian~\cite{sajedian2019double} employed the Double Deep Q-Network (DDQN)~\cite{van2016deep} to design both the material type and the geometry of the optical metasurface that maximizes hologram efficiency. The actions were defined as increasing or decreasing each design parameter by a fixed amount; the state is design parameters; the reward considers both the phase-generating capability and the efficiency. 
Similarly, Liu et al.~\cite{liu2021reinforcement} used the DDQN to design a periodic lattice system that achieves thermal transparency. The action was defined similarly to the one in Ref.~\cite{sajedian2019double}; the state is the combination of design parameters and the response of heat fluxes; the reward is represented by how far the simulated system is away from achieving perfect thermal transparency.
Sui et al.~\cite{sui2021deep} proposed a collaborative deep Q network (DQN) that designs mechanical metamaterials (voxelated designs representing the arrangement of soft and stiff materials). The action is the ``flipping process” by two agents: one agent selects a soft voxel and turns it into a stiff voxel, while the other agent does the opposite (Figure~\ref{fig:sui_rl_flipping}); the state is represented by the voxelated design; the reward is the change of the averaged equivalent modulus.

Sui et al.~\cite{sui2021deep} compared deep RL with the genetic algorithm (GA) under different design dimensions. The results show that RL does not have an advantage over GA under small action spaces, but will outperform GA when the action space becomes larger, due to the generalization ability of deep neural networks on large action spaces. On the other hand, with a larger action space, more design evaluations will be required to sufficiently explore the action space, which can quickly lead to a prohibitive computational burden for RL. This was reflected by the design problem dimensions addressed by the reviewed past works, none of which exceeded 50 dimensions. 

Based on these observations, we conclude that it is important to find the ``sweet spot" of action space dimensions where RL can outperform classic optimization methods while not requiring prohibitive computational costs. When proposing RL methods for metamaterials design, one needs to compare it to optimization (either classic or machine learning-accelerated optimization as introduced in Secs.~\ref{sec:property_prediction} and \ref{sec:design_representation}) and justify the necessity of using RL instead of optimization methods which are usually more intuitive and simpler to formulate. Despite the caveats, RL can still be a promising technique in metamaterials design because (1)~compared to data-driven design optimization, RL does not require prior data and hence is not limited by the boundary of existing designs, and (2)~it is easier to formulate the design problem with RL when the design has an unstructured representation (e.g., irregular truss structure represented as graphs~\cite{raina2022learning}).

\subsubsection{Design via Physics-Based Learning}
\label{sec:physics_based}

\begin{figure}[ht]
\centering
\includegraphics[width=1\textwidth]{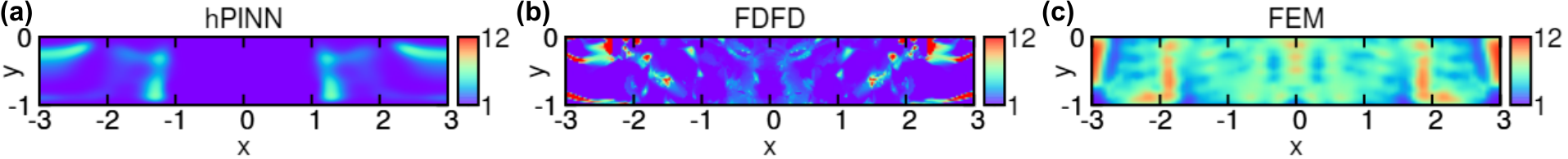}
\caption{Designs of permittivity obtained by a)~the hPINNs method, b)~PDE-constrained adjoint-based optimization with the finite-difference frequency-domain (FDFD) method as the numerical PDE solvers, and c)~PDE-constrained adjoint-based optimization with the finite element method (FEM) as the numerical PDE solvers. The hPINNs method achieved a simpler and smoother solution~\cite{lu2021physics}. Reproduced from Ref.~\cite{lu2021physics} with permission from Society for Industrial and Applied Mathematics.}
\label{fig:hpinn}
\end{figure}

While physics-informed machine learning has drawn huge attention in recent years, its application in metamaterials design is relatively limited. Physics-based learning in design tasks usually uses governing equations to guide the training of machine learning models which produce optimized design solutions. 
Jiang and Fan~\cite{jiang2019global,jiang2020simulator} proposed a generative neural network to generate high-performance dielectric metasurfaces, where the generator training was guided by the gradients from the adjoint electromagnetic simulations of generated designs.
Lu~\cite{lu2021physics} proposed physics-informed neural networks with hard constraints (hPINNs) to solve the topology optimization problem in metamaterials design. The method builds on physics-informed neural networks (PINNs)~\cite{raissi2019physics} but further allows optimizing a design objective function as well as the governing partial differential equations (PDEs) being modeled as hard constraints. The hPINNs method was demonstrated on design problems in optics and fluids. Compared to PDE-constrained adjoint-based optimization methods, the hPINNs method achieved the same objective value, but obtained a simpler and smoother solution with faster convergence (Figure~\ref{fig:hpinn}).

Unlike classic machine learning or data-driven methods, these physics-based learning methods require no training data and are less susceptible to the curse of dimensionality. The optimization can be guided by gradients and hence there is no need to explore the entire solution space. Therefore, design via physics-based learning can be a promising future research direction.

%%%%%%%%%%%%%%%%%%%%%%%%%%%%%%%%%%%%%%%%%%%%%%%%%%%%%%%%%%%%%%%%%%%%%%%%%%%%%%%%%%%%%%%%%%%%%
\subsection{Iteration-Free Inverse Design}
\label{sec:inverse_design}

While the reviewed works in Section~\ref{sec:design_optimization} used ML methods to accelerate design optimization, they still need an iterative optimization process to obtain the final solution. In this section, we review past works that used ML to achieve iteration-free inverse design. There are mainly two application scenarios: (1)~obtaining designs that meet target properties or responses; (2)~obtaining near-optimal solutions under certain constraints or operating conditions. Most prior works addressed the first scenario. Existing ML methods for iteration-free inverse design primarily belong to three categories: one-to-one mapping from target to design (Section~\ref{sec:direct_mapping}), cascaded neural networks (Section~\ref{sec:cascaded_nn}), and conditional generative models (Section~\ref{sec:conditional_generative}). We will also introduce a few other works that do not fall into these three categories (Section~\ref{sec:iteration_free_other}).

\subsubsection{One-to-One Direct Mapping from Target to Design}
\label{sec:direct_mapping}

\begin{figure}[ht]
\centering
\includegraphics[width=1\textwidth]{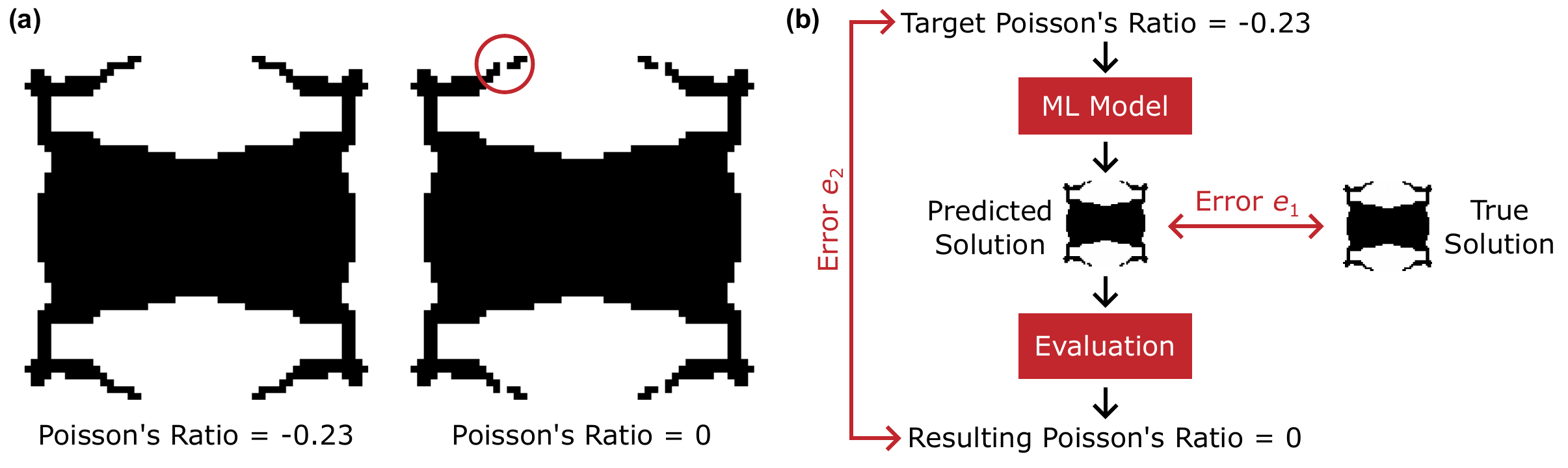}
\caption{The issue of measuring the error in design solutions: a)~Structurally similar design solutions with very different Poisson's ratios. There is 0.16\% difference in the pixelated design but 100\% relative difference between their homogenized Poisson's ratios. b)~The small error $e_1$ between the predicted design and the true solution compared to the large error $e_2$ between the resulting Poisson's ratio and the target Poisson's ratio.}
\label{fig:inverse_1to1}
\end{figure}

Owing to neural networks' capability of approximating any continuous functions, it is possible and straightforward to learn a direct mapping from target quantities of interest (e.g., properties or responses) to design solutions using neural networks. Past work has used MLPs and CNNs to learn such mappings~\textemdash~for example, mapping topology optimization settings (i.e., filter radius, volume fraction, and the type of design objective) to corresponding 2D density maps of mechanical metamaterials~\cite{kollmann2020deep}, target local sound fields to 1D acoustic metasurface designs~\cite{zhao2021machine}, sets of target scattering parameters to optical metasurface patterns~\cite{qiu2019deep}, transmission spectrum to photonic nanostructure geometry~\cite{Malkiel2018PlasmonicLearning}, and target band gaps to phononic crystal designs~\cite{li2020designing}. To train the neural network models, most of these studies used the difference between the predicted and the ``ground-truth" design solutions as the training loss, quantified by metrics such as the mean squared error (MSE), mean absolute error (MAE), or binary cross entropy. This poses a problem for the faithfulness of the predicted solutions in terms of meeting the target, because even structurally similar solutions can result in very different quantities of interest (Figure~\ref{fig:inverse_1to1}(a)), especially when the structures are in pixelated or voxelated representations, as also illustrated in~\cite{woldseth2022use}.

% Kollmann et al.~\cite{kollmann2020deep} trained a CNN to map topology optimization settings (i.e., filter radius, volume fraction, and the type of design objective) to corresponding 2D density maps of mechanical metamaterials. MSE as loss.

% Zhao et al.~\cite{zhao2021machine} trained a CNN to obtain the 1D metasurface according to the local sound field. Combining with another CNN, it can also achieve intensification and weakening of local sound field. Error as loss.

% Qiu et al~\cite{qiu2019deep} used an MLP to map a set of target scattering parameters to the corresponding optical metasurface pattern. MSE as loss.

% Li et al.~\cite{li2020designing} used an MLP to map the target band gap to corresponding reduced design representation. MSE and MAE as training loss.

There are works that may avoid this issue by comparing the target quantities rather than the design solutions (i.e., measuring $e_2$ instead of $e_1$ in Figure~\ref{fig:inverse_1to1}(b)). 
Malkiel et al.~\cite{Malkiel2018PlasmonicLearning} first trained an inverse design network to predict the photonic nanostructure geometry based on the transmission spectrum, and then fine-tuned the inverse network by combining it with a forward response prediction network.
Liu et al.~\cite{liu2021intelligent} proposed a conditional GAN-based model to map a target holographic image to a corresponding optical metasurface design. The physical operation mechanism between the electric-field distribution and the metasurface was used to reconstruct the target image. Both the MSE and an adversarial loss between the original and the reconstructed target images were employed during training.
Jiang et al.~\cite{jiang2022dispersion} used a conditional GAN to map the target dispersion curves to the structural design of elastic metamaterials. A CNN-based dispersion relation prediction model was used for the fast screening of generated designs based on the predicted dispersion curves.
Instead of measuring ``how well the predicted design meets the ground-truth solution", these works focused on ``how well the predicted design meets the target quantities", thus can lead to the prediction of designs that better match the target quantities.

\subsubsection{Avoiding Nonuniqueness Issue via Cascaded Neural Networks}
\label{sec:cascaded_nn}

\begin{figure}[ht]
\centering
\includegraphics[width=1\textwidth]{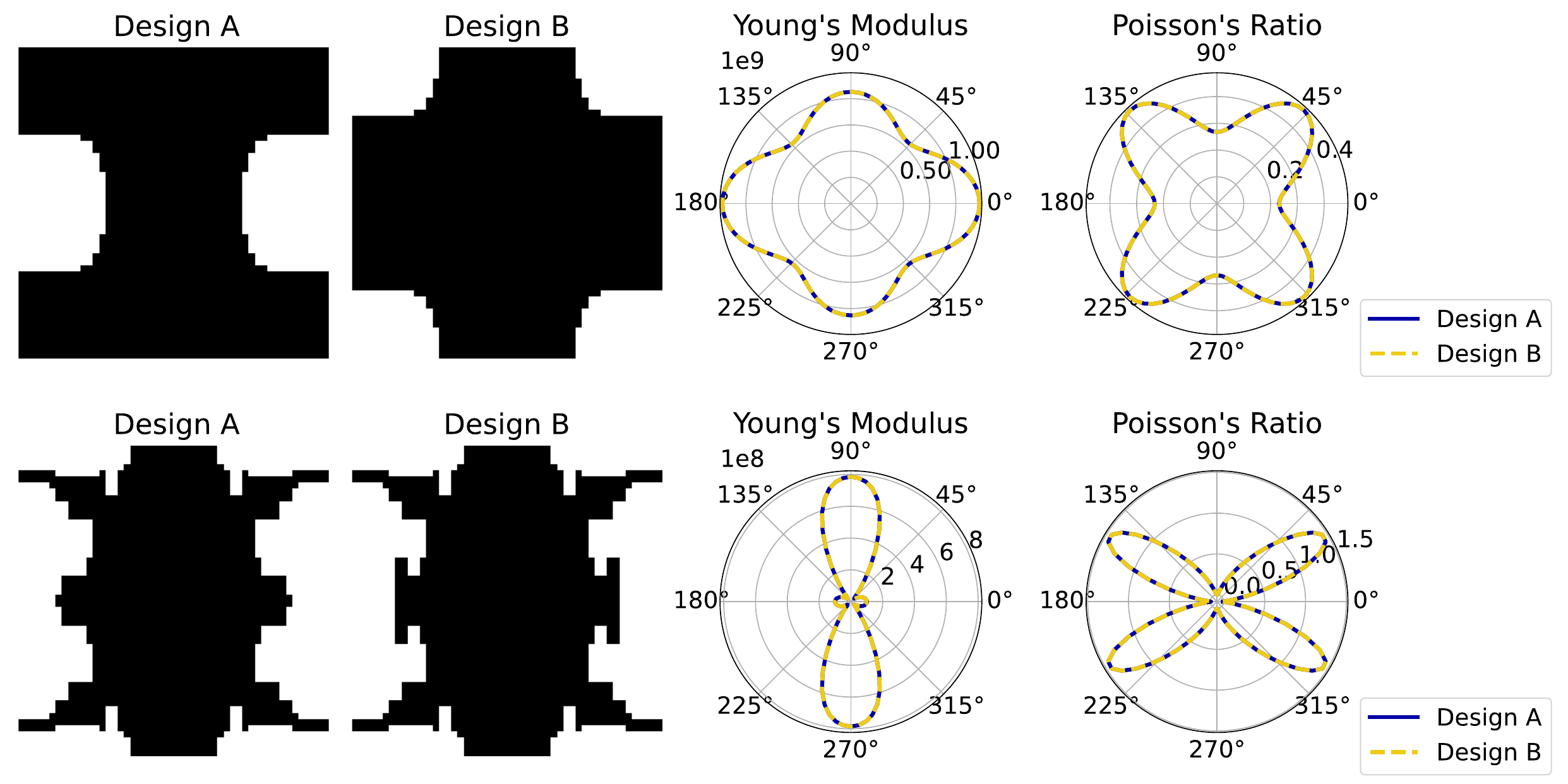}
\caption{Multiple mechanical metamaterials designs correspond to the same properties. This nonuniqueness is either owing to the fact that multiple equivalent structures exist under the periodic boundary condition (top), or because there are parts of the structure that do not contribute to the properties (bottom).}
\label{fig:nonuniqueness}
\end{figure} 

\begin{figure}[ht]
\centering
\includegraphics[width=0.75\textwidth]{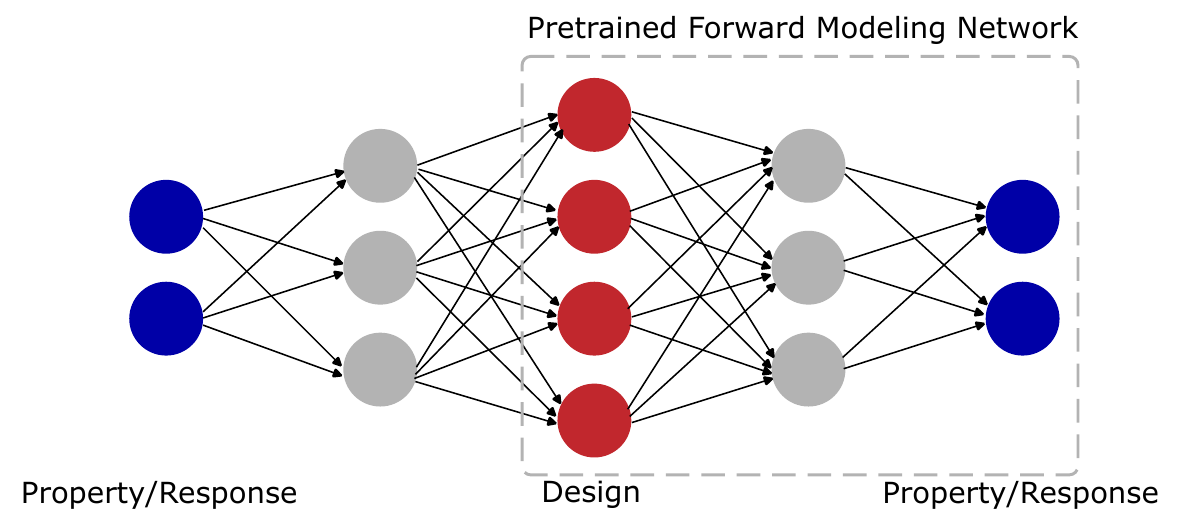}
\caption{Tandem neural network~\cite{liu2018training,an2019deep,kumar2020inverse,so2021demand,Yeung2021MultiplexedNetworks,zhen2021realizing}. The model training is separated into two phases: (1)~training of the forward-modeling network, where each design corresponds to a unique property or response, and (2)~training of the cascaded network by fixing the pretrained forward-modeling network, where the design produced at the intermediate layer does not necessarily belong to training data so that the model is not trained with conflicting designs.}
\label{fig:tnn}
\end{figure} 

Despite the simplicity of learning direct target-to-design mapping for inverse design, the underlying assumption of one-to-one mapping from target to design can be problematic and may lead to convergence issues during ML model training. Because the nonunique solutions will produce conflicting training instances where the same input is associated with different outputs~\cite{liu2018training}. Taking mechanical metamaterials as an example, this nonuniqueness is either owing to the fact that multiple equivalent structures exist under the periodic boundary condition (top of Figure~\ref{fig:nonuniqueness}), or because there are parts of the structure that do not contribute to the properties (bottom of Figure~\ref{fig:nonuniqueness}). Liu et al.~\cite{liu2018training} and An et al.~\cite{An2021MultifunctionalNetwork} also showed a similar phenomenon for 1D nanophotonic structures and 2D optical metasurfaces, respectively. To overcome this nonuniqueness issue, past work proposed the tandem neural network (T-NN) that cascades an inverse-design network with a forward-modeling network (Figure~\ref{fig:tnn})~\cite{liu2018training,an2019deep,kumar2020inverse,so2021demand,Yeung2021MultiplexedNetworks,zhen2021realizing}. The model training is separated into two phases: (1)~training of the forward-modeling network, where each design corresponds to a unique property or response, and (2)~training of the cascaded network by fixing the pretrained forward-modeling network, where the design produced at the intermediate layer does not necessarily belong to training data so that the model is not trained with conflicting designs.

Besides the work using T-NN, there were other model variants with a similar idea of using cascaded neural networks to solve the nonunique mapping problem. For instance, Ma et al.~\cite{ma2018deep} combined two bidirectional neural networks (each of which resembles a T-NN) to learn the relation between optical metamaterial design parameters, reflection spectra, and circular dichroism (CD) spectra, aiming for on-demand inverse design of chiral metamaterials given either the full reflection spectra or the CD spectra.

All the surveyed studies using the aforementioned cascade neural networks were applied to optical metamaterial design with dimensions of design variables not higher than 25. With higher design dimensions, it is more likely that the designs produced in the intermediate layer easily fall out of the training data distribution. When this happens, the forward modeling network is not reliable anymore since it has not seen such out-of-distribution designs. Thus, the cascaded network can still have low error while producing designs that are irrelevant to the target. Besides, given one target, cascade neural networks can only predict one design solution, although there are multiple potential solutions. These limitations motivate the use of generative models (Section~\ref{sec:conditional_generative}).
% For this reason, we need to constrain the design space so that predicted designs are within the data distribution, which motivates the use of generative models (Section~\ref{sec:conditional_generative}).

\subsubsection{One-to-Many Mapping via Conditional Generative Models}
\label{sec:conditional_generative}

Conditional generative models' ability to learn a distribution of designs conditioned on any target quantities of interest makes them the perfect candidates for learning one-to-many mappings in inverse design applications. The conditional generative models also explicitly model the relationship between the target and the designs and thus will not produce designs irrelevant to the target. Most prior works in this direction used conditional generative adversarial networks (cGANs)~\cite{mirza2014conditional} and conditional variational autoencoders (cVAEs)~\cite{sohn2015learning}.

Conditional GANs are the primary model for achieving one-to-many mapping in past inverse metamaterial design studies. The original cGAN relies on the adversarial loss to ensure the generated designs possess properties or produce responses that match the given target, or show optimality under the given condition~\cite{jiang2019free,so2019designing}. However, with the purpose of reducing the distance between two distributions, the adversarial loss alone cannot promote high-accuracy matching between an individually generated design and the corresponding target or condition. To overcome this issue, prior works mainly take three measures: (1)~using a separate prediction network for fast screening of unqualified generated metamaterials design~\cite{An2021MultifunctionalNetwork}, (2)~adding a prediction loss to implicitly maximize the property/response accuracy of generated designs~\cite{liu2018generative,wang2020deeplearning,Wang2022IH-GAN:Structures}, and (3)~progressively updating the training data by adding high-performance generated metamaterial designs and removing low-performance designs~\cite{wen2020robust}.

Conditional VAEs were also employed to achieve the same purpose~\cite{ma2019probabilistic,ma2020data}. While GANs have shown superior performance on approximating high-dimensional and complicated data distributions, VAEs have the advantage of stable training and are able to extract an interpretable latent space from data. Ma et al.~\cite{ma2019probabilistic,ma2020data} showed that the latent space from cVAE automatically learns to distinguish metamaterial geometries from different classes.

Note that although some works mentioned in Section~\ref{sec:direct_mapping} used conditional GANs to generate designs based on target properties, the mapping between target properties and designs is still one-to-one~\cite{liu2021intelligent,jiang2022dispersion}. Because these works used a generator without random noise as its input, the generator can only produce a unique design given a fixed target.

\subsubsection{Other Approaches}
\label{sec:iteration_free_other}

\begin{figure}[ht]
\centering
\includegraphics[width=1\textwidth]{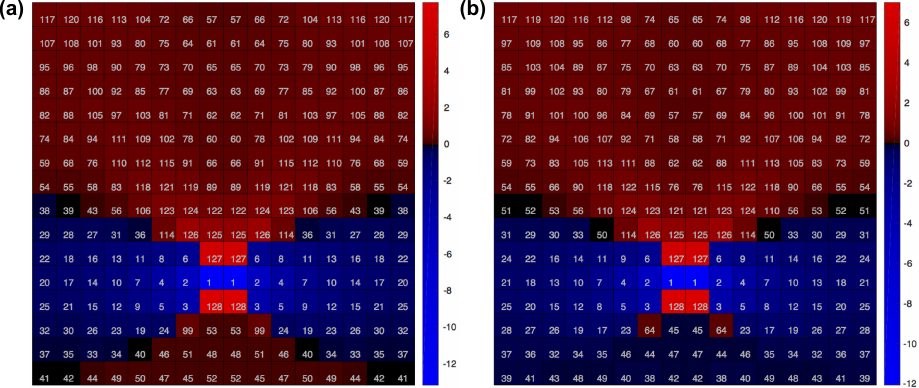}
\caption{Weights outputted from the linear model show how much each element contributes to a) toughness and b) strength~\cite{gu2018novo}. Colors represent the weight of each element: blue represents negative weights and red represents positive weights. Numbers on the elements represent the ranks in terms of weight. Reproduced from Ref.~\cite{gu2018novo} with permission from Elsevier Ltd.}
\label{fig:gu_linear_weights}
\end{figure}

In addition to the three main approaches mentioned above, there are other prior works aiming to achieve iteration-free inverse design. Luo et al.~\cite{luo2020probability} proposed a probability-density-based neural network that predicts the acoustic metastructure design in the form of Gaussian mixture model parameters given the target transmission spectrum. By sampling from the Gaussian mixture, this approach can generate one-to-many mappings between target responses and design solutions. However, it might not work on high-dimensional design problems (e.g., topological design problems) due to the need to model more complex distributions.
Elzouka et al.~\cite{elzouka2020interpretable} proposed to use the decision tree to solve both the forward prediction and inverse design problem. After training a decision tree for forward prediction, one can trace up the tree branches from the target response (at leaf nodes) through all branch-splitting criteria. These criteria can be used as design rules to select designs satisfying the given target. This approach naturally captures the one-to-many mapping behavior of inverse design problems, and the extracted rules are interpretable. However, this approach does not suit high-dimensional design problems either, due to the high computational cost of training decision trees with a large depth. Gu et al.~\cite{gu2018novo} trained a linear model to classify metamaterial geometries into ``good" and ``bad" designs based on their toughness and strength. The weights of the linear model indicate how each element in the design contributes to the performances (Figure~\ref{fig:gu_linear_weights}), based on which new high-performing designs can be sampled. Since the continuous property prediction task was simplified to binary classification, a linear model is sufficient to achieve high
predictive accuracy while having the explainability to guide the generation of new designs.

% %%%%%%%%%%%%%%%%%%%%%%%%%%%%%%%%%%%%%%%%%%%%%%%%%%%%%%%%%%%%%%%%%%%%%%%%%%%%%%%%%%%%%%%%%%%%%
% \subsection{Design by Sampling}

% ~\cite{gu2018novo} trained a linear model to classify metamaterial geometries into ``good" and ``bad" designs based on their toughness and strength. The weights of the linear model indicate how each element in the design contributes to the performances, based on which new high-performing designs can be sampled.

% ~\cite{gu2018bioinspired} trained a CNN to predict the toughness and strength of metamaterial designs, and used it in a self-learning sampling strategy. In each sampling loop, designs were generated both randomly and based on top-ranked designs from the previous loop. High-performing designs can be generated after multiple sampling loops.

% ~\cite{lee2022generative} combined MLPs with the genetic algorithm to generate lattice structures with optimized weight-to-performance ratios. The MLP was updated in each generation.

% \subsection{Meta-NN}

% Design metamaterials/metasurfaces to act as analog neural networks.

%%%%%%%%%%%%%%%%%%%%%%%%%%%%%%%%%%%%%%%%%%%%%%%%%%%%%%%%%%%%%%%%%%%%%%%%%%%%%%%%%%%%%%%%%%%%%
\subsection{Discussion and Future Opportunities}

Based on Figure~\ref{fig:metamaterial_design_sankey}, the most frequently used machine learning models for metamaterials design are DGMs, MLPs, and CNNs, all of which are based on neural networks. Also, almost all ML models for iteration-free inverse design are based on neural networks. The flexibility and scalability of neural networks give them the versatility to address various types of problems with high complexity and ill-posedness (e.g., inverse design). On the other hand, the requirement of large datasets and low interpretability have limited the performance and applications of neural networks. To illustrate and summarize the advantages and disadvantages of different methods, in this section, we compare existing works in terms of their cost-benefit and trustworthiness.

\subsubsection{Cost-Benefit}

\begin{figure}[ht]
\centering
\includegraphics[width=1\textwidth]{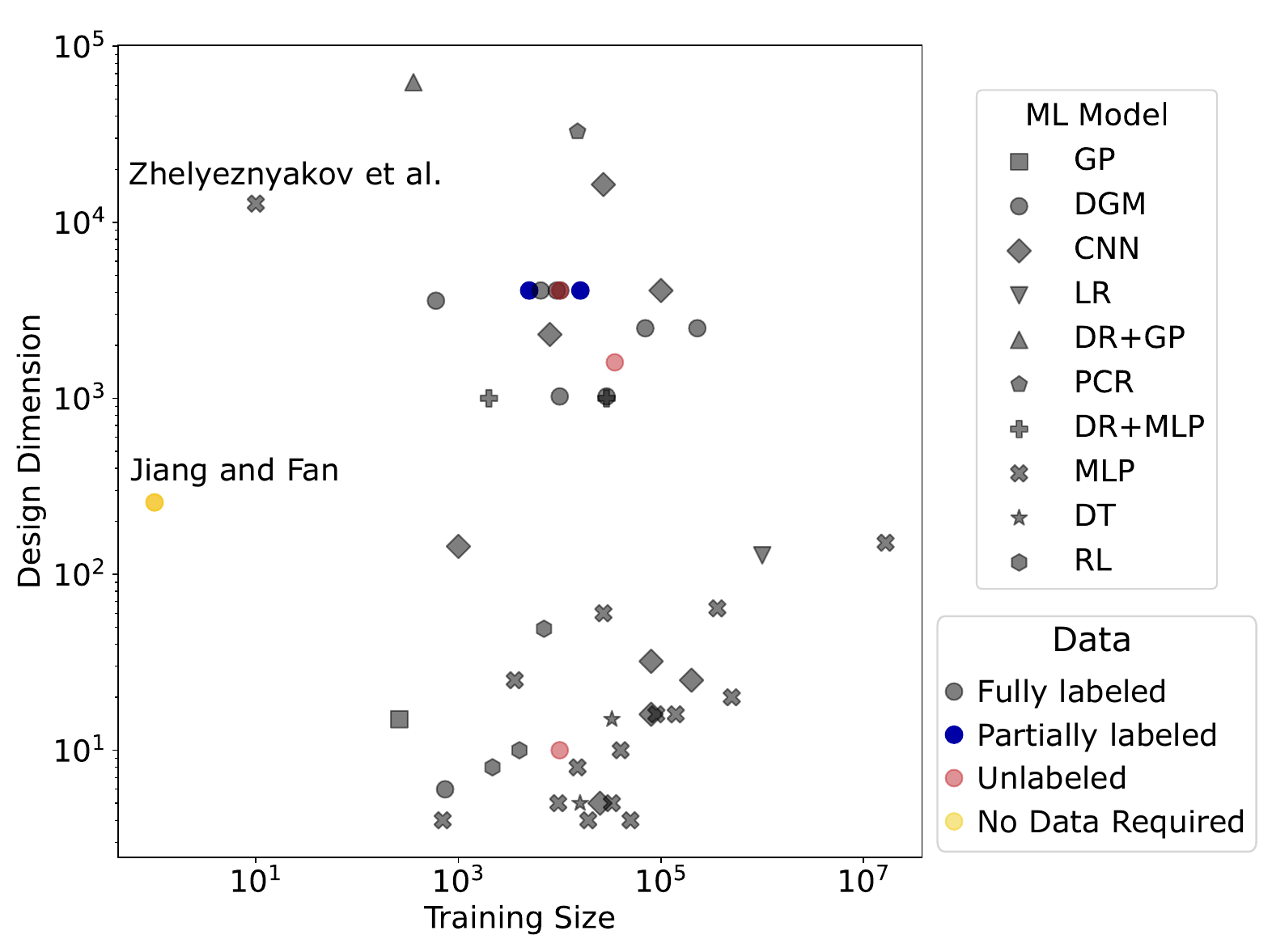}
\caption{Methods proposed in prior work, in relation to the training data size and the design dimension used for demonstration. Note that the training size of RL represents the number of design evaluations performed during training. (LR: logistic regression; DR: dimensionality reduction; RL: reinforcement learning; PCR: principal component regression; DT: decision tree; DGM: deep generative model; GP: Gaussian process; CNN: convolutional neural network; MLP: multi-layer perceptron.)}
\label{fig:metamaterial_design_scatter}
\end{figure} 

The cost of using ML models mainly exists in three stages: data collection, training, and inference. The inference stage is normally much cheaper than the other two stages and hence its cost can be negligible. In most data-driven metamaterials design works, data collection includes expensive physics-based simulations. Thus, the data collection cost usually contributes to most of the total cost and highly depends on the size of the training dataset (with the exception of semi-supervised or unsupervised learning). The training cost also depends on training size, among other factors such as model complexity and the number of training epochs. Therefore, we focus on training size when analyzing the cost. 

The usefulness of ML models also depends on the benefits they can provide. Since ML usually aims at accelerating the design process, we can use the reduction of computational cost as a criterion to evaluate benefits (i.e., the computational cost difference between conventional methods and data-driven methods at inference). However, many prior studies did not include such information. Another important factor that reflects the benefits of ML models is the complexity of the design problem they can address. The complexity highly depends on the dimensionality of the design space. Therefore, we consider the design dimension as an important factor when evaluating benefits. 
% The design dimension can also correlate with the computational cost reduction brought by ML models, since the cost of conventional design optimization methods often increases quickly with the design dimension, whereas it is not necessarily the case for ML-based methods.

Figure~\ref{fig:metamaterial_design_scatter} shows the ML models employed in each prior work, in relation to the training data size and the design dimension used for demonstration. Note that the training size and design dimension are extracted from the experimental settings in prior works and do not necessarily indicate any strict model requirements. This figure shows that when the design problem has a dimension of less than 200, MLPs are the most commonly used machine learning model, usually with the existence of relatively large training datasets. An exception is Zhelyeznyakov et al.~\cite{zhelyeznyakov2021deep}, where the accelerated design of high-dimensional dielectric metasurfaces was achieved by using an MLP and the data of only 10 designs. In that work, instead of treating each entire design as a training sample, local patches of the design geometry and the electromagnetic (EM) field were used. The trained MLP can then be applied to predict the EM field of the entire metasurface. Other models including GP, RL, and DT were also applied in low-dimensional cases. By combining with dimensionality reduction methods such as AEs~\cite{qiu2019deep,li2020designing} and VAEs~\cite{wang2022design}, both MLP and GP show the capability of addressing problems in much higher dimensions. Prior works using CNNs were applied to solving problems with a wide spectrum of dimensions, with training sizes ranging from 1,000 to 200,000. 

For problems with over 1,000 dimensions, DGMs were the most commonly used models. With the capability of representation learning and modeling one-to-many mapping, DGMs are applicable to both iterative design optimization and iteration-free inverse design (Figure~\ref{fig:metamaterial_design_sankey}). When used for representation learning, DGMs can be trained with unlabeled data, such that a more compact design representation is learned from only geometric data~\cite{liu2020hybrid,chen2022gan,shen2022nature}, which avoids time-consuming simulations when preparing training data and produces reusable design representations for problems with different properties or responses of interest.
% DGMs for representation learning can be both unsupervised~\cite{liu2020hybrid,chen2022gan,shen2022nature} and supervised~\cite{Wang2020DeepSystems,liu2021thermal}, with the purpose of learning a more compact representation for either the metamaterial design or the response.
When using DGMs for iteration-free inverse design, labeled data are normally required for the DGMs to learn the mapping from properties or responses to metamaterial designs. But there are exceptions where partially labeled or even unlabeled data were considered. Ma et al.~\cite{ma2019probabilistic,ma2020data} proposed a framework that can use both labeled and unlabeled data for data-driven inverse design of optical metasurfaces, where adding unlabeled data was shown to improve model performance. For the same purpose of optical metasurface inverse design, Liu et al.~\cite{liu2021intelligent} incorporated the physics-based operation between the electric-field distribution and the metasurface design into the decoder of the conditional generative model, which eliminates the need for providing labeled training data. 

By infusing physics into ML models, we can even eliminate the requirement of training data. As discussed in Section~\ref{sec:physics_based}, Jiang and Fan~\cite{jiang2019global,jiang2020simulator} proposed a generative model whose training was guided by the gradient from the adjoint simulation, so that high-performance dielectric metasurface designs can be generated without using training data. Lu et al.\cite{lu2021physics} proposed hPINNs that can optimize a design objective under constraints of governing PDEs\footnote{This work does not appear on Figure~\ref{fig:metamaterial_design_scatter} since the design is not limited to fixed dimensionality.}.

Overall, compared to iterative design optimization, iteration-free inverse design trades off accuracy (i.e., how well the solution matches the true target) for time. Nonetheless, to improve accuracy while still keeping a low computation time, we can use inverse design methods to generate near-optimal solutions as warm starts and further refine the solutions by using optimization with a relatively small number of iterations~\cite{jiang2019free}.

Besides design dimensionality, we need to consider another important factor in evaluating the benefits of machine-learning methods~\textemdash~whether the trained model is applicable to sufficiently many scenarios. 
To quantify this generality, Woldseth et al.~\cite{woldseth2022use} proposed a generality score for neural network-based methods used in topology optimization, which accounted for the required similarity of test and training problems (i.e., higher similarity indicates lower generality), in addition to other topology optimization-related criteria. 
% which considers the level of reusability of trained models under different boundary conditions, mesh dimensions, and loading conditions, as well as how different the test and training problems can be. In the case of metamaterials design, however, it is difficult to quantify the generality due to the diversity of design representation and problem definition. 
This test-training similarity is also transferable to measuring generality in ML-based metamaterial design. Among the reviewed past works, the majority require the training and test problems to be similar (i.e., having the same design representation and the same properties or responses of interest). One exception is the works on representation learning, where the learned representation can be applied to design problems with different properties or responses of interest. Another exception is Zhelyeznyakov et al.~\cite{zhelyeznyakov2021deep}, where the design dimension of test problems can vary since the ML model only cares about the local patches of the design geometry.

Note that although the aforementioned physics-based models have the benefit of requiring no training data, the fact that they need fully specified problem settings (e.g., design constraints, boundary conditions, and operating conditions) for training makes the trained models difficult to generalize beyond the problem specification considered during training, i.e., we need to retrain the model for different problem settings. 

% ablation study

% Dimensionality: low, high

% Simulation/Experiment cost: low, high

% What are suitable methods for each scenario?

% What if one requires real-time control?

% What if output dimension is high (e.g., acoustic/optical metamaterials)?

\subsubsection{Trustworthiness}
\label{trustworthiness}
The trustworthiness of ML-based design methods is important in many engineering problems, especially safety-critical and risk-sensitive ones such as metamaterials for blood vessel stents~\cite{barri2021multifunctional} and medical imaging~\cite{li2020designing}. One important aspect of trustworthiness is quantifying the uncertainty of metamaterial designs to obtain robust or reliable solutions. This uncertainty quantification, however, is understudied in past literature. Uncertainty comes from sources including operating conditions and the fabrication process. Machine learning models can also have uncertainty due to insufficient data. Data-driven design methods considering these uncertainties can make more informed decisions and generate more trustworthy solutions. 

\begin{figure}[ht]
\centering
\includegraphics[width=1\textwidth]{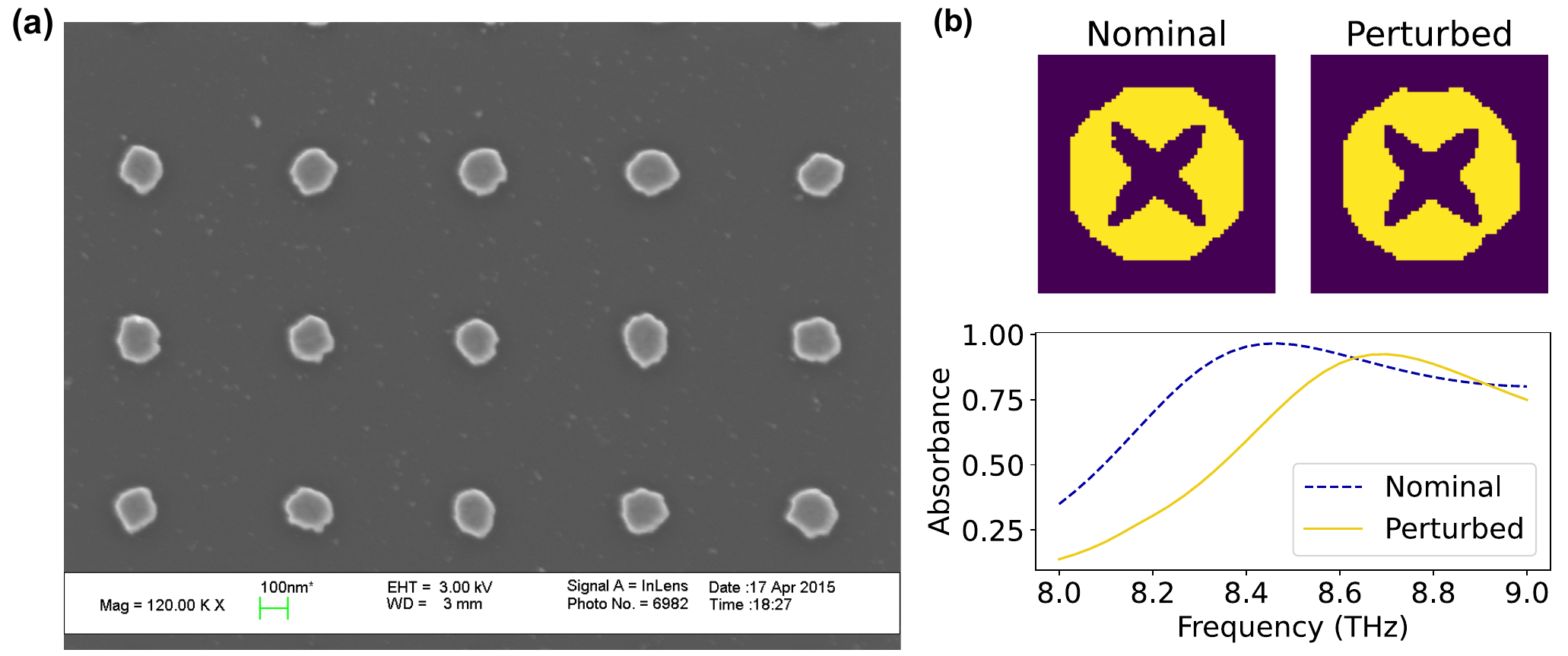}
\caption{Fabrication uncertainty and its effects on design performance. a)~Examples of metasurface patterns fabricated through the electron-beam lithography~\cite{roxworthy2014reconfigurable}, where the nominal design is a nanocylinder with the circular cross-section (source: Balogun Research Group at Northwestern University). b)~Effects of geometric perturbation on the absorbance profile of a metasurface, where the nominal design is the optimal solution of the deterministic optimization from Chen et al.~\cite{chen2022gan}.}
\label{fig:metasurface_uncertainty}
\end{figure}

Particularly, due to fabrication uncertainties, the properties or responses of as-fabricated metamaterials can largely deviate from the as-designed ones. Figure~\ref{fig:metasurface_uncertainty}(a) shows an example of geometric deviation of fabricated metasurface patterns. Figure~\ref{fig:metasurface_uncertainty}(b) shows how geometric deviation can lead to response changes. The nominal design is represented as $64 \times 64$ binary pixelated images where 1 (yellow) represents material and 0 (dark blue) represents void. The perturbed design is obtained by slightly distorting the nominal design, which mimics the fabrication error. The figure shows that the absorbance spectrum changes significantly due to this small perturbation, indicating the necessity of quantifying fabrication uncertainty. Due to the high dimensionality of variables to be considered in geometric uncertainty quantification, previous metamaterials design work assumes uniform boundary variation where the boundary of the geometry is uniformly ``eroded" (e.g., over-etched) or ``dilated" (e.g., under-etched)~\cite{wang2019robust}. To avoid making simplifying assumptions on the form of uncertainty and preserve the high degrees of freedom of geometric uncertainty, Chen et al.~\cite{chen2022gan} proposed a deep generative model with hierarchical latent spaces to simultaneously model the geometric variation of nominal designs and the freeform uncertainties of fabricated designs. Chen et al. incorporated this generative model in robust design optimization and demonstrated notable improvement in as-fabricated design performances compared to only considering uniform uncertainty.

% \begin{figure}[ht]
% \centering
% \includegraphics[width=0.7\textwidth]{figure/fabricated_metasurface.pdf}
% \caption{Examples of fabricated metasurface patterns with the nominal design being a circular pattern.}
% \label{fig:fabricated_metasurface}
% \end{figure} 

Another key aspect of trustworthiness is interpretability, where humans can understand the reasoning behind an ML model's prediction or decision. Past metamaterial design works have either used inherently interpretable models like linear models~\cite{gu2018novo}, decision trees~\cite{elzouka2020interpretable,chen2022see}, or physics-assisted models that infuse governing PDEs~\cite{lu2021physics} and physics-based solvers~\cite{jiang2019global,jiang2020simulator} into neural networks. These methods improve interpretability in different ways: inherently interpretable models such as decision trees and generalized linear models allow us to extract design rules or investigate the importance of design variables; while physics-assisted models use physics rules to constrain the search of model parameters and solutions during model training. Existing studies using inherently interpretable models (i.e., decision trees) only considered very simple (low-dimensional) designs due to the prohibitive computational cost when using these models for high-dimensional inputs. The application and development of inherently interpretable models~(e.g., neural network-based models~\cite{chen2019looks,yang2021gami}) for complex, high-dimensional metamaterial designs were under-explored.

\subsubsection{Novelty}

Another under-studied challenge in data-driven metamaterials design is how to create novel designs beyond just interpolating existing ones to achieve unprecedented properties or responses. Reinforcement learning can discover novel solutions, but past works only used RL to address low-dimensional design problems, due to the computational cost issue (Section~\ref{sec:rl}). Data-driven methods using classic ML models are built on the \textit{i.i.d.} assumption that training and testing data are independent and identically distributed. To extrapolate outside existing designs, the training data needs to be updated with desirable generated designs to shift the training data distribution, and the model needs to be retrained on the updated training data. One usually needs to repeat this process many times to obtain a significant improvement over originally existing designs~\cite{gu2018bioinspired}. Without retraining, data-driven models cannot learn useful information that generalizes to scenarios outside existing training data distribution, and therefore cannot lead to novel solutions beyond data distribution. More sophisticated methods are needed to distill generalizable information and create novel designs. Chen et al.~\cite{Chen2018ComputationalFamilies} extracted parameterized templates of five 3D auxetic metamaterial families from data. The templates can then be used to generate new designs beyond training data, although having the same topologies as the five families. Future research may explore new methods to generate designs that break more limitations prescribed by existing data.

%%%%%%%%%%%%%%%%%%%%%%%%%%%%%%%%%%%%%%%%%%%%%%%
\section{Data-driven Multiscale Metamaterial System Design}
\label{5 Data-driven Multiscale TO}
    \subsection{Overview}
    \label{5.1 Overview of Data-driven Multiscale TO}
\begin{figure}[!htb]
\centering
\includegraphics[width=1\textwidth]{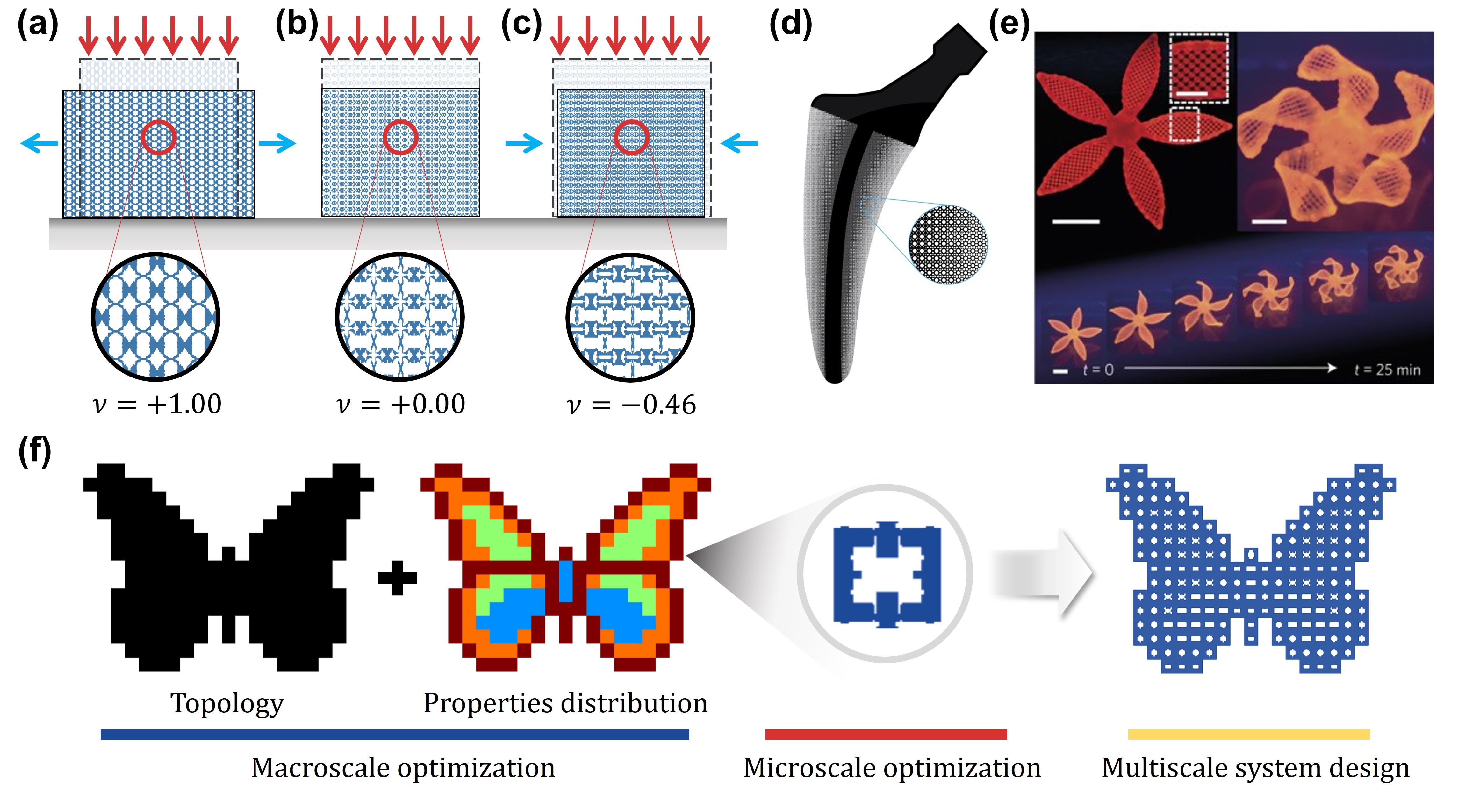}
\caption{Metamaterials and multiscale systems. a)-c) Materials exhibit different Poisson’s ratios \emph{v} derived from different microstructures, i.e., different transverse displacement given the same pressing loading on the top, with the transparent boxes showing the original shapes before distortion. d) Multiscale orthopedic implant design. Reproduced from Ref.~\cite{garner2019compatibility} with permission from Elsevier B.V. e) Multiscale design for shape morphing under thermal excitations. Reproduced from Ref.~\cite{sydney2016biomimetic} with permission from Macmillan Publishers. f) An illustration of the multiscale metamaterial system design process.}
\label{fig:multiscale system}
\end{figure}

%     \liwei{
% \begin{itemize}
%   \item {What's multi-scale TO, why this topic is so important that we need to give it a separate section? \underline{Figure to show hierarchical system} Mention scale separation here}
%   \item {What are the challenges and uniqueness associated with multiscale TO?
%   \\two scales, nested design space, information translating between the two scales, compatibility issues}
%   \item {Two design frameworks, 
%   \\bottom-up \& top-down \underline{Figure to illustrate the two frameworks} }
%     \item {A figure or table to show the classification criteria for Section 5.1 and 5.2 }
% \end{itemize}
% }
 Traditional structural design methods and their data-driven counterparts often focus on homogeneous material distributions. In contrast, some functional engineering structures require heterogeneous property distributions to meet spatially varying requirements, which are critical in achieving better performance and more complex functions. For instance, an invisibility cloak requires heterogeneous properties around an object to prevent it from detection with external physical fields \cite{leonhardt2006optical,pendry2006controlling,milton2006cloaking}. Requirements on heterogeneous properties  can also be found in soft robots, where the goal is to achieve local or global target postures \cite{skylar2019voxelated,wehner2016integrated,zhang2020seamless,kim2019ferromagnetic}. Recently, structural design methods have evolved to optimize both the structure and the distribution of multiple materials for heterogeneous property requirements. Topology optimization (TO) is the most flexible among these methods, enabling freeform changes to the structure and providing greater design freedom than traditional parameter or shape design methods \cite{zuo2017multi,li2022digital}. Despite its promise, it is still challenging to fabricate these multi-material structures in achieving as-designed functions. This issue is caused by the narrow selection of available materials and the constraints of manufacturing processes. In contrast, complex geometries can be more easily manufactured at fine resolutions with additive manufacturing \cite{gu2021material}. This technical revolution has opened up new avenues to realize unprecedented and tailorable material properties by changing the geometry of microstructures rather than constituent materials ~\cite{yu2018mechanical,zheludev2010road,kadic20193d,lumpe2021exploring}, as shown in Figure~\ref{fig:multiscale system}(a)-(c). Therefore, heterogeneous properties can be obtained by spatially varying the microstructures, instead of constituent materials, to assemble a multiscale metamaterials system for intricate structural behaviors (Figure~\ref{fig:multiscale system}(d)-(e)) \cite{garner2019compatibility,sydney2016biomimetic}. In this section, we will discuss data-driven methods for designing multiscale metamaterial systems that determine architectures at both micro and macro scales to achieve the desired metamaterial behaviors. Our focus is on the heterogeneous distribution of (effective) mechanical or thermal properties that originate from the lower material scale, as these physics are involved in most existing multiscale metamaterial designs. For designs with other underlying physics, we refer readers to Refs.~\cite{so2022revisiting, So2020DeepNanophotonics, liu2021tackling, li2022empowering, jin2022intelligent, liu2023deep}.

Multiscale structures with carefully tuned microstructures have been shown to have the edge over single-scale design (macroscale only with homogeneous materials or periodic unit cells discussed in Section~\ref{4 Learning and Generation: Data-Driven Metamaterials Design}) for engineering applications involving multi-physics or spatially varying requirements. Typical applications include strain cloaking \cite{wang2022mechanical,buckmann2014elasto}, target deformation design \cite{jin2020guided,boley2019shape,Panetta2015ElasticFabrication,schumacher2015microstructures}, thermal-elastic property optimization \cite{wu2018multiscale,deng2013multi,yan2016multi}, dynamic behaviors design \cite{wang2022generalizedOptimization,wang2020valley,kadic2020elastodynamic,zhang2020multiscale,zhao2019efficient}, buckling resistance \cite{thomsen2018buckling,wang20213d} and energy absorption \cite{pham2019damage,yin2021strong,jiang2021tailoring}. However, the design of multiscale structures is a complex two-scale problem, as shown in Figure~\ref{fig:multiscale system}(f). At the macroscale, the topology of the structure and its mechanical properties distribution are optimized to meet performance targets, while at the microscale, unit cells need to be designed at different locations to achieve the corresponding properties. Ideally, the design process for a multiscale metamaterial system should be carried out in a way that the two scales are coupled and designed concurrently. This means that the design of the microscale and macroscale architectures should be done simultaneously and interactively, rather than separately and independently. This hierarchical nature of the system poses unique challenges in the design process compared to the singlescale topology optimization methods illustrated in Section~\ref{4 Learning and Generation: Data-Driven Metamaterials Design}. 

Firstly, at the macroscale, property parameters required to fully describe the physical response of materials are generally high-dimensional without strict bounds for the achievable properties. This leads to an ill-defined property design space and a complicated optimization process. As a result, most existing methods are subject to over-simplistic constraints on the design space of properties. Secondly, at the microscale, the design of microstructures is an inverse problem without much \textit{a priori} design knowledge (see Section~\ref{4 Learning and Generation: Data-Driven Metamaterials Design}). It is characterized by its infinite-dimensional geometrical design space and one-to-many mapping from properties to structures. This creates an irregular landscape for the design objective (macroscale properties or performance) with many local optima, making the design sensitive to the initial guess and constraints. Finally, the synthesis of micro-and macro-designs suffers from the ``curse of dimensionality" induced by the hierarchical multiscale design space, complex combinatorial search associated with the unit-cell selection, and adjacent microstructures whose shapes are incompatible at their interfaces (geometric frustration). Due to these issues, traditional multiscale structural methods are either overwhelmingly time-consuming or rather restrictive in design flexibility.

\begin{figure}[!htb]
\centering
\includegraphics[width=1\textwidth]{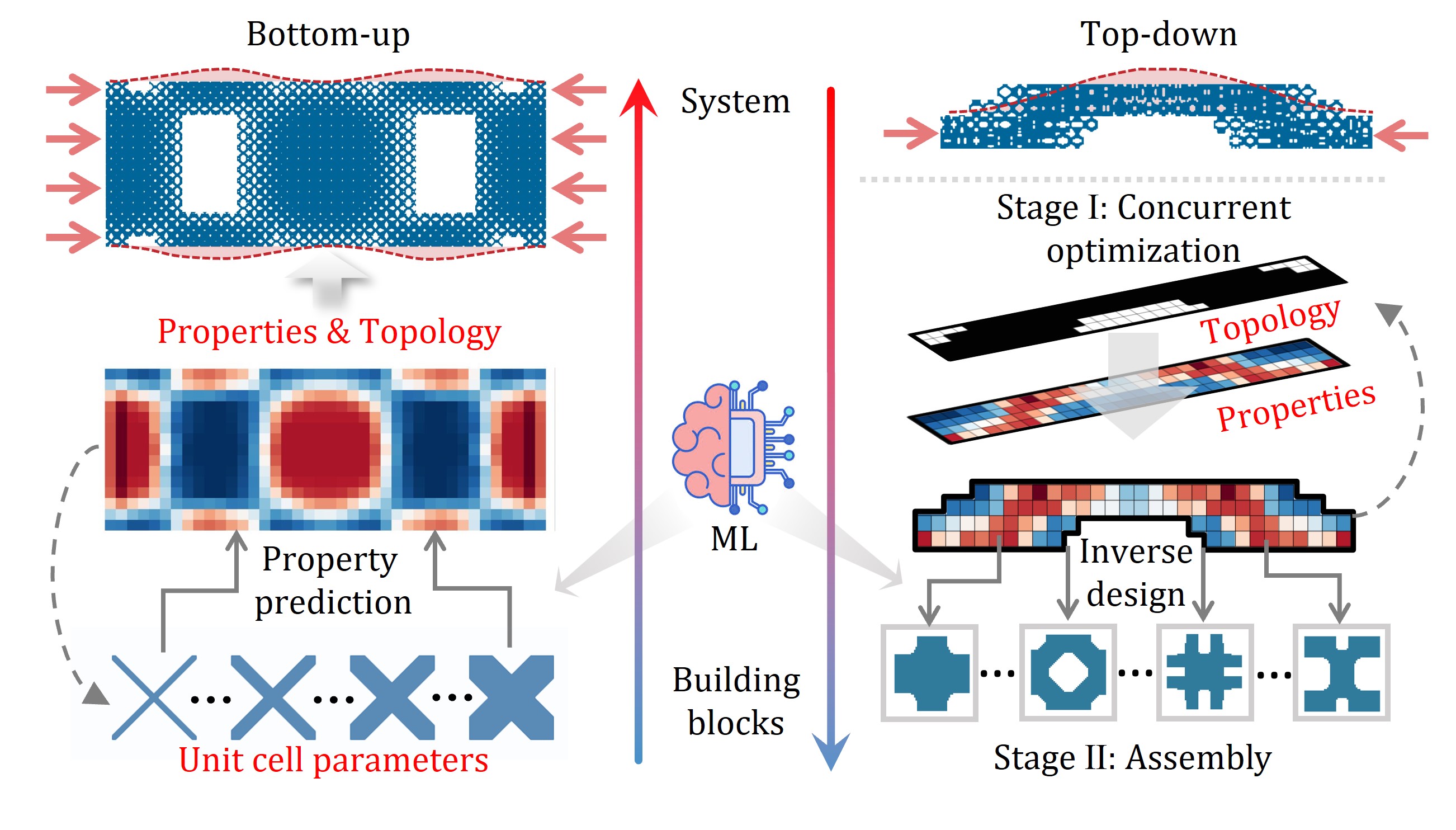}
\caption{Illustration of bottom-up and top-down frameworks. The target displacement profile (red dashed lines) is used as an example of the system-level design objective.}
\label{fig:two frameworks}
\end{figure}

Capitalizing on the growth of data resources and computational capability, data-driven design based on machine learning models is recognized as a promising tool to address the aforementioned challenges for multiscale systems. As depicted in Figure~\ref{fig:two frameworks}, we propose to divide existing data-driven multi-scale design frameworks into two main categories, i.e., bottom-up and top-down frameworks, based on the relations between design variables at the two scales.  The bottom-up framework directly uses the parameters at the microscale level, e.g., volume fraction and unit cell type, as design variables. Costly nested calculation of effective properties of the unit cells is replaced by a surrogate model of the structure-property mapping. In contrast, with the top-down framework, the macroscale topology and spatial distribution of homogenized material properties are concurrently optimized first. Then, to assemble a full multiscale structure, the optimized properties will serve as targets to retrieve the corresponding building blocks. To accelerate this assembly process, ML models trained at the microscale level (Section~\ref{4 Learning and Generation: Data-Driven Metamaterials Design}) can be used to compactly represent and/or efficiently generate unit cells. In the remainder of this section, we will illustrate and review the state-of-the-art of these two frameworks.

    \subsection{Bottom-Up Framework}
    \label{5.2 Bottom-Up Framework}
    % \liwei{
    %     \begin{itemize}
    %         \item {Single-class graded unit cells:
    %         \\ - linear case 
    %         \\ homogenization-based, FE2 method, reduced-order model(nodal condensation)
    %         \\ - nonlinear case}
    %         \item {Variant 1: consider unit-cell orientation
    %         \\ - conformal mapping 
    %         \\ - de-homogenization 
    %         \\ original method, generalized methods to consider complex unit cells, combined with NN
    %         }
    %         \item {Variant 2: Increase the diversity of unit-cell prototypes
    %         \\ - multi-class
    %         \\  discrete-material-based, clustering-based,LVGP-based
    %         \\ - remixed class
    %         \\ Interpolated(kriging), Yu-Chin's method
    %         \\ - Special unit-cell designs
    %         \\ Gaussian random field (spindoid), Voronoi foam, Latent-variable + generated model (Wayne's GAN， focus on design, not population)
    %         }
    %     \end{itemize}
    % }
        \subsubsection{Data-Driven Design with Single-Class Graded Unit Cells}
        \label{5.2.1 single-class}

\begin{figure}[!htb]
\centering
\includegraphics[width=1\textwidth]{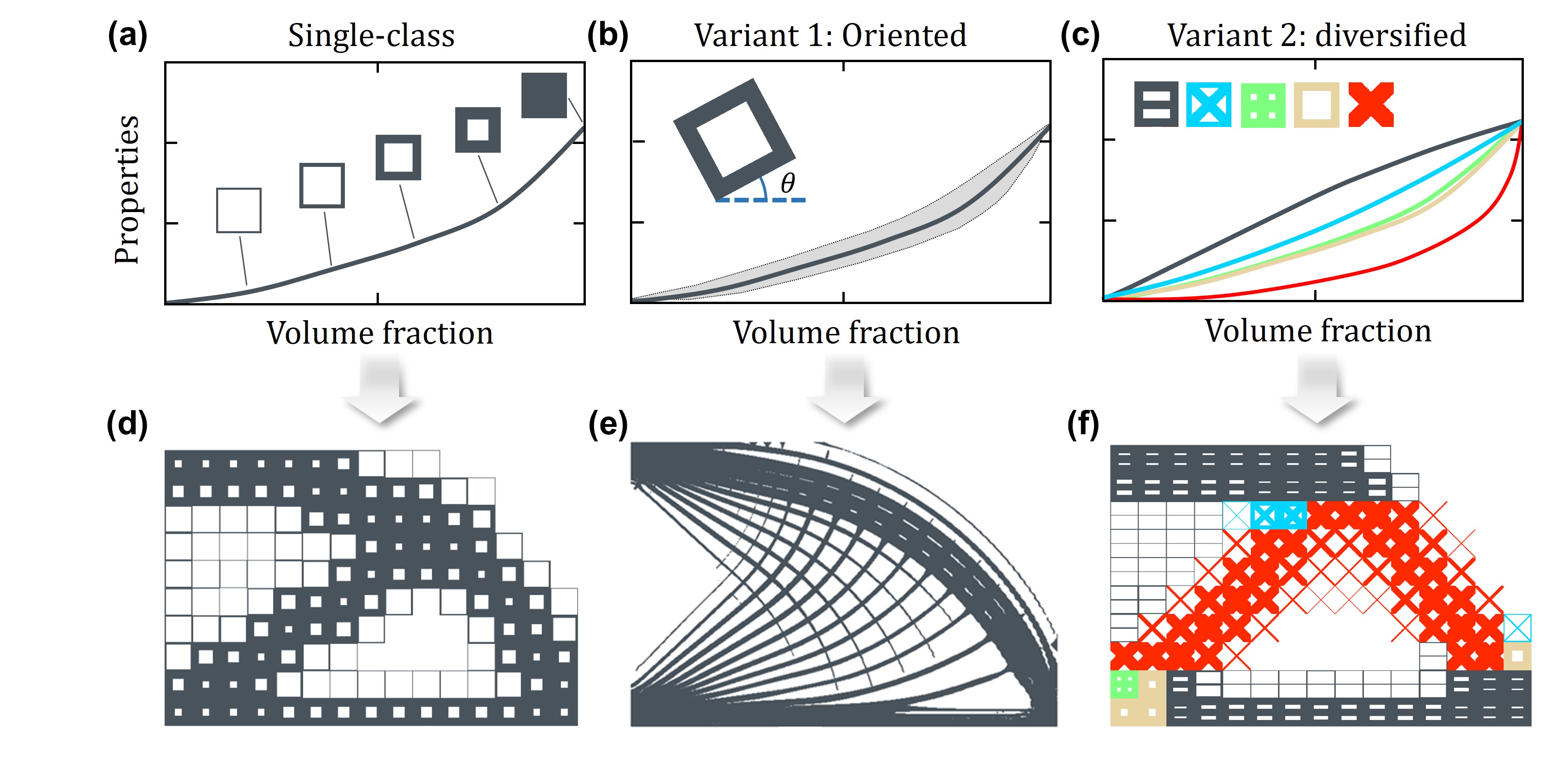}
\caption{Three bottom-up frameworks. a) Property space of single-class unit cell. b) Property space of oriented unit cell. The shaded regions are for unit cells with non-zero oriented angles. c) Property space of a diversified set of unit cells. The property curves are colored in accordance with the colors of unit cell classes. Corresponding multiscale designs with d) single-class unit cells, e) oriented unit cells, f) diversified unit cells.}
\label{fig:three bottom-up frameworks}
\end{figure}

Most bottom-up data-driven methods assumed the same topological concept (single-class) for all the microstructures and only spatially vary, i.e., grade, the geometrical parameters as shown in  Figure~\ref{fig:three bottom-up frameworks}(a) and  Figure~\ref{fig:three bottom-up frameworks}(d). Lattice-based
 \cite{cheng2019functionally} and surface-based microstructures were commonly used due to their simplicity and good manufacturability. One could change the thickness of rods or surfaces \cite{liu2022kriging} to obtain different geometries, with each corresponding to a specific volume fraction. The thickness or the volume fraction can be used as design variables in the optimization process. An advantage of this graded single-class assumption is that one could leave out the microscale details during the optimization and directly optimize the spatial distribution of geometrical parameters at the macroscale instead. This is also called homogenization-based design. The data-driven aspect of this framework is that the time-consuming homogenized properties evaluation in each iteration was replaced by a surrogate model, capturing the relation between geometrical parameters and the precomputed properties. Some examples of such models are exponential function \cite{li2019design}, polynomial \cite{wang2017multiscale,cheng2019functionally}, Kriging \cite{liu2022kriging}, diffuse approximation \cite{wu2019topology} and neural network \cite{white2019multiscale}. After optimization, the corresponding multiscale structure can be obtained by filling the macroscale design with the microstructures specified by the optimal parameters, a process known as de-homogenization \cite{groen2018multi}. 

Despite its simplicity and high efficiency, single-class data-driven graded design usually leads to sub-optimal solutions since the microstructrues belong to the same topological class with fixed unit cell orientations. For example,  to optimize compliance for both single-loading and multi-loading cases, microstructures must consist of oriented two and three alternating layers of orthotropic materials, known as rank-2 and rank-3 materials, respectively~\cite{avellaneda1987optimal}. These microstructures are referred to as rank-2 and rank-3 materials, respectively. Achieving these designs requires spatial changes to the unit cell topology and orientation. Although most existing designs used compliance minimization as the demonstrative case, the single-class graded microstructures did not meet the optimum requirement \cite{WuZijun2020Tsto}. As a result, the performance of the multi-scale design was even worse than the single-scale design. Similar observations were also reported in various applications, such as frequency response control \cite{wang2022generalizedOptimization} and natural frequencies maximization \cite{wang2022data}. To increase design flexibility, many studies were dedicated to 1) increasing the diversity of microstructures and/or 2) considering unit cell orientation designs. As shown in Figure~\ref{fig:three bottom-up frameworks}(c)-(d),(f)-(g), these two variants of bottom-up methods can both expand the property space of the database. The rest of Section~\ref{5.2 Bottom-Up Framework} reviews these approaches.

        \subsubsection{Variant 1: Considering Unit-Cell Orientation}
        \label{5.2.2 unit-cell orientation}
\begin{figure}[!htb]
\centering
\includegraphics[width=1\textwidth]{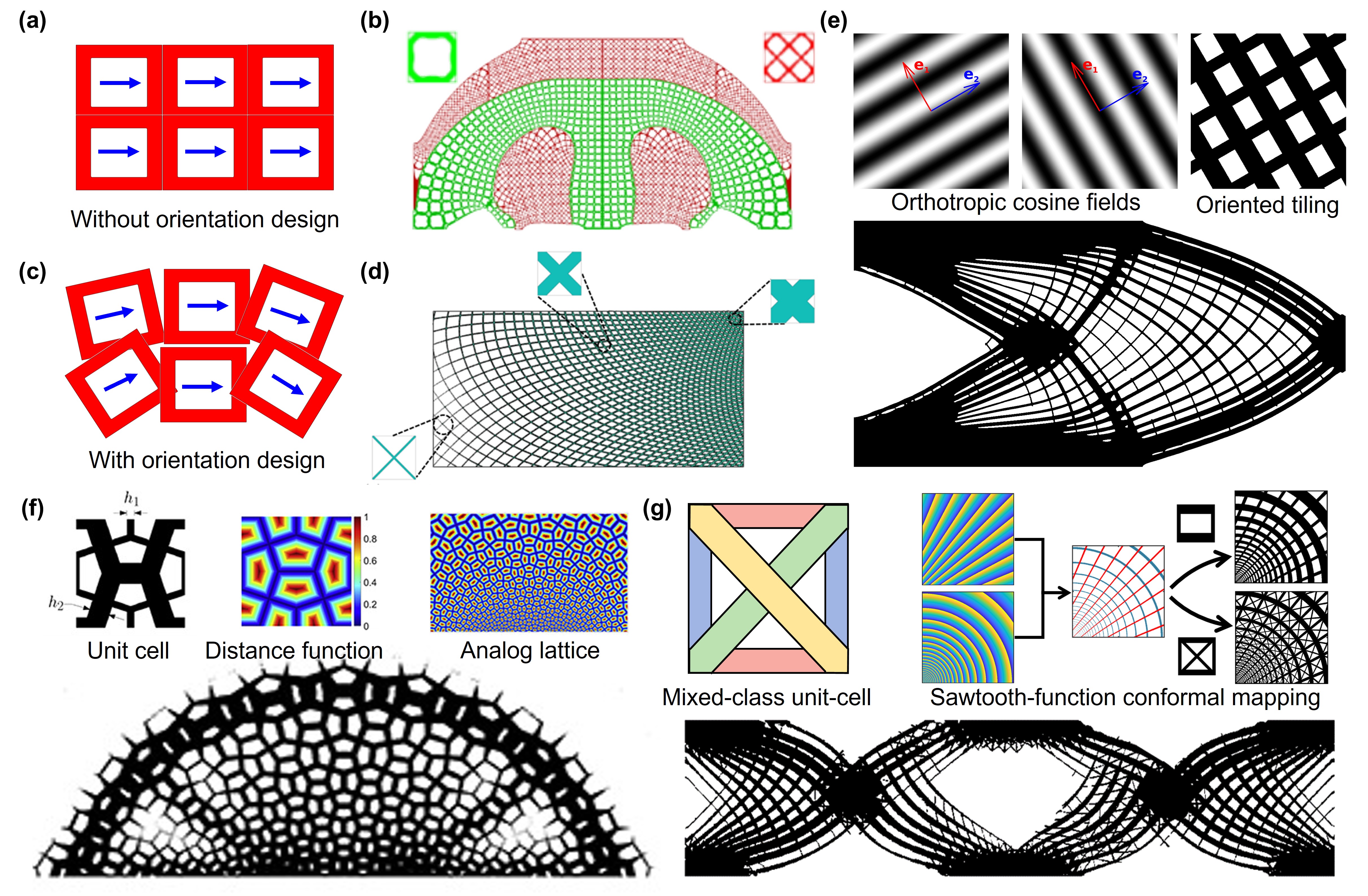}
\caption{Examples of oriented two-scale designs. a) Compatible tiling without unit cell orientation design. b) Incompatible tilling with unit cell orientation design. c) Oriented design using conformal mapping. Reproduced from Ref.~\cite{jiang2021generative} with permission from ASME. d) Oriented design by optimizing parameterized mapping functions. Reproduced from Ref.~\cite{ma2022compliance} with permission from Elsevier B.V. e) De-homogenized design via cosine-based mapping. Reproduced from Ref.~\cite{groen2018homogenization} with permission from John Wiley \& Sons, Ltd. f) De-homogenized design via Fourier-series-based mapping. Reproduced from Ref.~\cite{TamijaniAliY2020Tamd} with permission from Elsevier Ltd. g) De-homogenized design via saw tooth-function-based mapping. Reproduced from Ref.~\cite{wang2022generalizedOptimization} with permission from Elsevier B.V.}
\label{fig:Oriented designs}
\end{figure}

The major obstacle in designing graded structures with oriented microstructures lies in the de-homogenization process. When the unit-cell orientations vary across the macroscale structure, the corresponding microstructures need to be rotated accordingly. However, neighboring microstructures might not connect with each other after rotation (Figure~\ref{fig:Oriented designs}(a) and ~\ref{fig:Oriented designs}(c)). This causes the multiscale structure to fail to attain the designed performance and, furthermore, impossible to manufacture. The key to mitigating this issue is to construct a smooth mapping from the stand-alone regular unit cells (e.g., unit cells shown in the upper corners of Figure~\ref{fig:Oriented designs}(b)) to an assembled tiling (e.g., the multiscale structure shown in the Figure~\ref{fig:Oriented designs}(b)), which can distort the micro-structure to ensure compatibility but at the same time retain their effective properties.

Conformal mapping is considered a powerful tool to realize this mapping since it can preserve the angle of the geometrical features and thus minimize the variation in effective properties \cite{vogiatzis2018computational}. Jiang et al. \cite{jiang2021generative} constructed conforming mapping after homogenization-based design to morph a rectangular tiling of periodic microstructures into corresponding irregular regions in the multi-scale structure (Figure~\ref{fig:Oriented designs}(b)). However, the design of unit-cell orientations was not considered  in this study. Ma et al.\cite{ma2022compliance} used a linear combination of a set of basis functions to parameterize the mapping from a predefined unit cell to a multi-scale oriented tiling (Figure~\ref{fig:Oriented designs}(d)). The coefficients of the basis functions were used as design variables, and used to train an ANN model to predict effective properties from the local Jacobian matrix of the corresponding mapping. While this method allowed the implicit design of unit-cell orientation, the design space was restricted by the form and orders of basis functions, as well as the predefined shapes of unit cells. 

In a major advance in 2008, Pantz and Trabelsi \cite{pantz2008post} focused on square microstructures with rectangular holes and proposed a method to project a homogenized design to a multiscale structure with oriented microstructures on a high-resolution mesh. They achieved this de-homogenization process by constructing implicit mapping functions from a pair of cosine fields to approximate the oriented unit cells in different spatial locations. Later, Groen and Sigmund 
 \cite{groen2018homogenization} simplified this de-homogenization process by introducing the connected component labeling method to obtain a consistent orientation field, and by relaxing the optimization problem for the mapping function (Figure~\ref{fig:Oriented designs}(e)) \cite{groen2018homogenization}. This simplified method was further extended to enable the efficient design of 3D multiscale structures \cite{Groen2020De-homogenizationTopologies,Groen2021Multi-scaleModeling}. The de-homogenization process was accelerated by training a neural network to obtain the mapping function from the optimized unit-cell orientations without any extra optimization process \cite{elingaard2022homogenization}. Although the de-homogenization method is appealing in terms of efficiency and performance, it is still confined to simple static compliance minimization problems \cite{Geoffroy-DondersPerle2020Coom,Geoffroy-DondersPerle20203too}. The reason is that its original version can only handle square cells with rectangular holes, simply making the unit-cell orientation align with the principal strain direction. While these designs are optimal for static compliance minimization under a single loading, they would become sub-optimal for general design cases, such as multi-loading, dynamic response optimization, and multi-physics problems. 

To extend the applicability of de-homogenization, various enhanced methods have been proposed to accommodate more complicated unit cells. Ladegaard et al.\cite{jensen2022homogenization} combined three cosine fields to construct the implicit mapping functions for oriented rank-3 microstructures. It enables the de-homogenization to find the optimal structures for static compliance problems in multi-loading cases. To handle freeform unit cells, Tamijian et al. \cite{TamijaniAliY2020Tamd} represented the complex unit-cell geometries as a Fourier series and optimized the spatial distribution of each Fourier basis to orient unit cells in a compatible way (Figure~\ref{fig:Oriented designs}(f)). Groen \cite{groen2018multi} suggested using cosine functions to construct a similar implicit mapping from regular unit cell regions into oriented patches. Inspired by the texture mapping in computer graphics, Kumar et al. \cite{KumarTej2019AdKa} adopted a finite-element mesh to parameterize this implicit mapping. The mapping was obtained by solving a set of linear equations associated with the discrete mesh. While these extensions allowed the use of more complicated unit-cell geometries, their constructed implicit mapping was not conformal and may deteriorate the performance of the multi-scale designs. In a more recent work, Wang et al. \cite{wang2022generalizedOptimization} proposed to construct a conformal mapping in the de-homogenization process by using Sawtooth function fields (Figure4g). Unit cells with mixed-class topologies were then used to broaden the property space, with neural networks as surrogate models.

Overall, this branch of variants enabled the de-homogeniztaion process of oriented unit cells without much loss in efficiency and has been extended to handle complex unit-cell geometries. However,  the works under this category all focused on static compliance minimization problems with a few exceptions. This is due to the smoothness requirement of the orientation field and the restriction of the unit cell topologies.

        \subsubsection{Variant 2: Increasing the Diversity of Unit Cells}
        \label{5.2.3 diversity of unit cells}
The second branch of the variants aims to increase design flexibility by considering diverse micro-structure topologies in the optimization. As illustrated earlier, the success of data-driven graded design relies on the low-dimensional descriptor of unit cell geometries. Therefore, the key challenge addressed in this branch of variants is to represent broader sets of unit-cell topologies without significantly increasing the dimension of descriptors.  

\begin{figure}[!htb]
\centering
\includegraphics[width=1\textwidth]{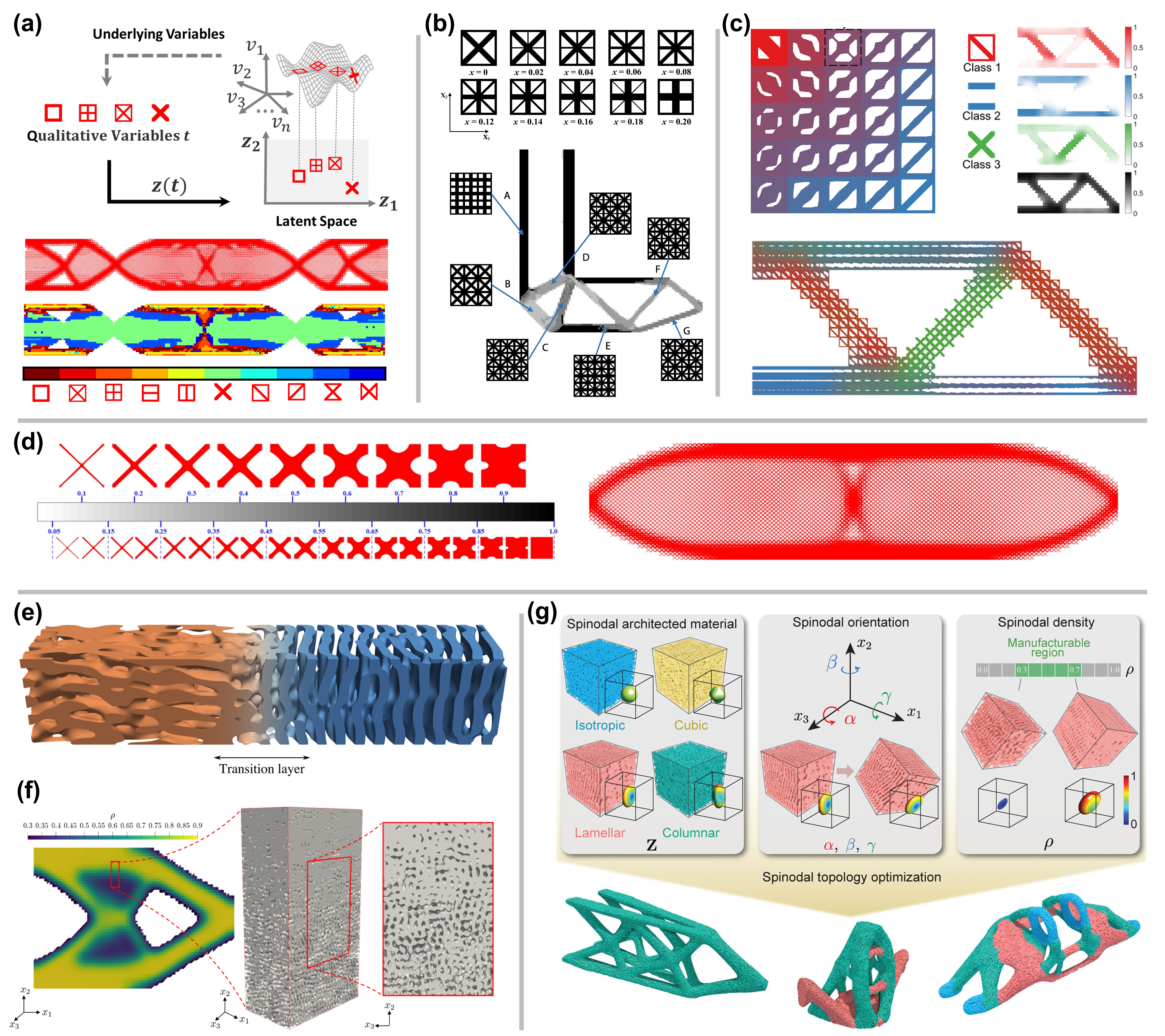}
\caption{Multiscale designs with diversified unit cells. a) Design with multiple unit cell classes. Reproduced from Refs.~\cite{Wang2022ScalableFactors, wang2022data} with permission from (top) ASME and (bottom) Elsevier Ltd. b) Multiscale design with unit cells that allow a smooth transition between two classes. Reproduced from Ref.~\cite{wang2018concurrent} with permission from Springer-Verlag GmbH Germany. c) Multiscale design with blended unit cells. Reproduced from Ref.~\cite{Chan2021RemixingBlending} with permission from Springer-Verlag GmbH Germany. d) Multiscale design with concurrently optimized unit cell prototypes. Reproduced from Ref.~\cite{zhang2020multiscale} with permission from Elsevier Ltd. e) Smooth transition of spinodal metamaterials. Reproduced from Ref.~\cite{kumar2020inverse} with permission from Creative Commons CC BY. f) Multiscale design with spatially varying spinodal microstructures. Reproduced from Ref.~\cite{ZhengLi2021Dtoo} with permission from Elsevier B.V. under the CC BY license. g) Multiscale design with selected symmetric types of candidate spinodal unit cells. Reproduced from Ref.~\cite{senhora2022optimally} with permission from Wiley-VCH GmbH.}
\label{fig:Diverse designs}
\end{figure}

As shown in Figure~\ref{fig:three bottom-up frameworks}(f) and Figure~\ref{fig:Diverse designs}, a relatively straightforward idea is to include multiple unit cell classes in the optimization, each with its own low-dimensional parameterization. By considering a unit cell class as a special type of discrete material, this idea naturally fits into the discrete material optimization framework. In this framework, the unit cell class was represented by one-hot encoding and relaxed to be a continuous variable by adding penalization or constraints \cite{liu2020data}. This enabled the automatic selection of the optimal unit-cell classes for different spatial regions in the optimization. Due to its simplicity, it has become the most commonly used framework to accommodate multiple unit-classes in multiscalse system design.

However, the dimension of design variables (one-hot encoding) for each microstructure grows linearly with the number of classes being considered. This will greatly increase the execution time and complexity. Meanwhile, the one-hot encoding only represents the unit-cell classes in a qualitative way without any physical meaning. As a result, the surrogate models and the optimization process cannot explicitly exploit the correlation or similarity between different classes for better performance. To address these issues, Wang et al. \cite{wang2021data,wang2022data,Wang2022ScalableFactors} proposed multi-response latent variable Gaussian process (MR-LVGP) and its enhanced variants to transform the discrete classes into continuous 2D latent variables through statistical inference. The constructed latent space captured the effects of classes on the mechanical properties, which induced an interpretable distance metric that reflects the similarity with respect to properties (Figure~\ref{fig:Diverse designs}(a)). Moreover, with the nonlinear embedding, the dimension of the latent space remained when the number of unit cell classes increased. By integrating the MR-LVGP models with TO, an efficient data-driven optimization process was developed that can concurrently explore multiple classes and/or constituent materials and their associated geometric parameters for better structural performance.

To further increase the diversity in unit-cell geometries, it is desirable to enable the transition or blending between discrete unit-cell classes (Figure~\ref{fig:Diverse designs}(b)-(d)). In this way, new micro-structure prototypes can be created that go beyond the geometries and properties of given classes.  To achieve this, Wang et al. \cite{wang2018concurrent,wang2020concurrent} established a sophisticated parameterization method for selected classes of microstructures formed by multiple groups of rods (Figure~\ref{fig:Diverse designs}(b)).  By changing the relative ratio among the rod thicknesses of  different groups, a smooth transition between multiple unit-cell topologies can be achieved. However, this parameterization technique is difficult to be generalized to other microstructures with freeform topologies. Chan et al. \cite{Chan2022RemixingBlending} proposed a more general shape blending scheme that can accommodate freeform unit cell classes with distinct and even incompatible topologies, generating smoothly graded microstructures (Figure~\ref{fig:Diverse designs}(c)). The interpolation scheme only had a few extra parameters, which can be directly integrated into the data-driven graded design framework with a neural network as the surrogate model for properties. While blending multiple unit classes expands the design space, it requires a set of prespecified unit-cell prototypes. How to obtain an optimal set of prototypes is problem-dependent and generally unknown beforehand. In Chan et al., the sets were selected either using domain knowledge or an autonomous set selection method that maximizes diversity metrics \cite{Chan2022RemixingBlending}. Instead of using prespecified prototypes, Zhang et al. \cite{zhang2019concurrent,zhang2020maximizing,zhang2020multiscale} propose to simultaneously evolve the prototypes during the optimization. The surrogate model to predict the properties of interpolated unit cells, i.e., a GP model, was also updated in each iteration after evolving the prototypes (Figure~\ref{fig:Diverse designs}(d)). By doing so, the initial prototypes can change their shapes to better handle customized design scenarios. 

Besides considering multiple classes and their interpolation, special types of materials, i.e., spinodal materials, are emerging as a promising choice in data-driven multiscale design. These stochastic self-assembled can easily achieve diverse microscale geometries with inherent connectivity when the volume fraction is above a given theoretical threshold \cite{soyarslan20183d}. Zheng et al. \cite{ZhengLi2021Dtoo} used Gaussian random fields to describe micro-structures of spinodal materials and train a fully connected neural network to associate the field parameters with the effective mechanical properties (Figure~\ref{fig:Diverse designs}(e)-(f)). By using the field parameters as design variables, a multiscale structure with spatially varying but smoothly graded microstructures can be achieved \cite{kumar2020inverse}. Senhora et al. \cite{senhora2022optimally} simplified the design of spinodal materials by focusing on selected symmetric types of candidate spinodal architected materials, extending to accommodate various complex 3D designs (Figure~\ref{fig:Diverse designs}(g)).  

    \subsection{Top-Down Framework} 
    \label{5.3 Top-Down Framework}
        % \liwei{
        %     \begin{itemize}
        %     \item {- unit cells with fixed connector \& isotropic properties}
        %     \item{ - unit cells with isotropic properties + multiple families}
        %     \item{ - unit cells with an-isotropic properties + random tiling optimization} 
        %     \item{ - unit cells with an-isotropic properties + GRF tiling optimization} 
        %     \end{itemize}
        %     }
            
\begin{figure}[!htb]
\centering
\includegraphics[width=1\textwidth]{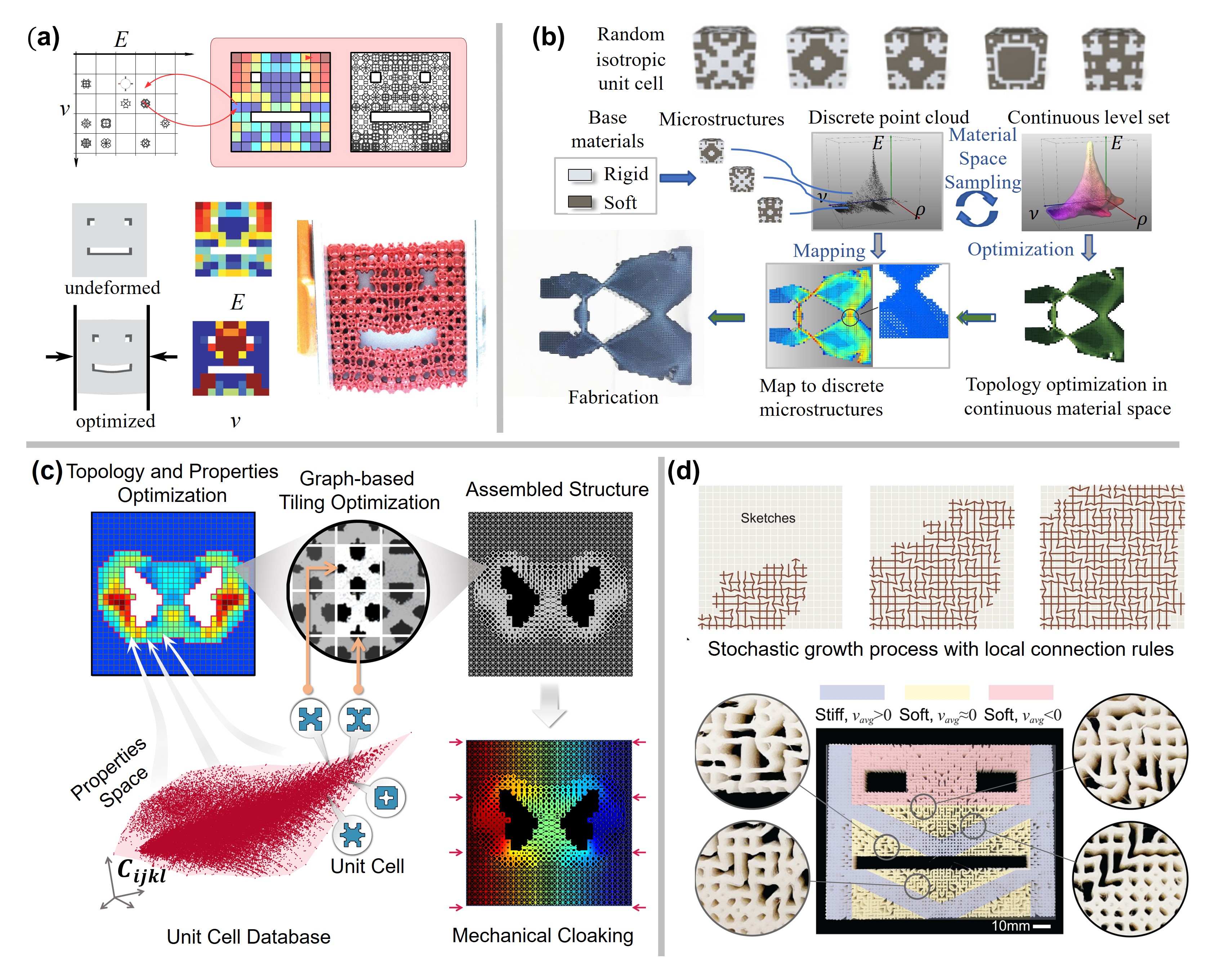}
\caption{Top-down methods. a) Design framework using discrete unit cells with cubic symmetry and predefined compatibility. The top row shows two pairs of compatibly connected isotropic unit cells in the database. Reproduced with permission from the authors~\cite{Panetta2015ElasticFabrication}. b) Design framework using randomly-generated isotropic unit cells with two constituents. ~\cite{Zhu2017} Reproduced from Ref.~\cite{Zhu2017} with permission from Association for Computing Machinery. c) Design framework using randomly-generated anisotropic unit cells. Reproduced from Ref.~\cite{wang2020deep} with permission from Elsevier B.V. d) Multiscale design with stochastic growth rules. Reproduced from Ref.~\cite{liu2022growth} with permission from American Association for the Advancement of Science.}
\label{fig:Top-down methods}
\end{figure}

    The bottom-up framework illustrated in the last section depends on a properly parameterized unit cell model with low-dimensional variables. While this constrained design space greatly expedites the design process, it also shrinks the property space, which will fall short for advanced applications such as soft robots \cite {wehner2016integrated} and mechanical cloaking \cite{buckmann2015mechanical}. The top-down design framework, the second branch of multiscale system design, aims to remove the constraints imposed on the microstructure geometries to allow the use of either pre-specified or freeform unit cells in assembling the full structure (Figure~\ref{fig:Top-down methods}) \cite{Panetta2015ElasticFabrication,BickelBernd2010Dafo,ChuChen2008DfAM,CramerAndrewD2015Mifm,HanYafeng2018ANDM,MironovVladimir2009OpTs}. This framework, in its ideal form, can unleash the highest potential of metamaterials. Specifically, a large database of microstructures is first constructed, containing different geometries and precomputed properties (Figure~\ref{fig:Top-down methods}(a)). Since the complex unit cell geometries do not have inherent low-dimensional representations as in the bottom-up framework, the property space of the database will serve as the design space for the property distribution optimization at the macroscale. After that, the property distribution at the macroscale cascades to the microscale (Figure~\ref{fig:two frameworks}) and, based on this, the corresponding microstructures are generated or fetched from the database to fill each element in the full structure. Therefore, there is no need to do the nested optimization and property evaluation at the microscale during the design process, which significantly improves the efficiency during the design of structures. However, to ensure compatibility between adjacent unit cells, these methods still need to force unit cells to be similar in geometries or compatible on the shared boundaries, which limits the range of achievable properties. 

    To address this issue, Schumacher et al. \cite{schumacher2015microstructures} proposed to construct a database with different metamaterial families. Unit cell geometries are similar within each family but distinctly different across families. They recognized that these families cover different regions but would overlap in the property space. The overlapped regions contain candidates with various geometries for the same properties. The best match can then be selected from those candidates for compatible boundaries. This compatible tiling keeps the design performance at the macroscale matches with the evaluation with homogenization. Later, Zhu et al. \cite{Zhu2017, chen2018computational} discarded the concept of families and generated a larger and richer database by stochastically adding and removing materials from microstructures in the database (Figure~\ref{fig:Top-down methods}(b)). In the assembly stage, a random-search based method was used to select microstructures with compatible boundaries to realize the target mechanical properties at each point in the macroscale structure. 
    
    However, due to the large amount and diverse shapes of microstructures, an immense combinatorial space needs to be explored through stochastic methods to form compatible boundaries between adjacent unit cells. Moreover, while these studies proposed elaborate methods for database construction, they lacked an effective representation and retrieval method for microstructures. As a result, it is challenging to incorporate these large databases into the multiscale design in a scalable way. This might also be the reason that these studies only focused on unit cells with isotropic or cubic symmetry. 
    
    In contrast, Wang et al.\cite{wang2020deep} focused on anisotropic microstructures with high-dimensional stiffness tensor (Figure~\ref{fig:Top-down methods}(c)). To tackle the aforementioned challenges, they simultaneously trained a VAE and an NN-based property predictor to map complex microstructures into a low-dimensional, continuous, and organized latent space. They found that the latent space of VAE provided a distance metric to measure shape similarity, enabled interpolation between microstructures, and encoded meaningful patterns of variation in geometries and properties. These characteristics enabled an effective selection of diverse unit cell candidate sets from the database to increase the chance of compatible assembly. The shape similarity metric was also utilized as a metric for geometrical compatibility between unit cells. By combining both geometrical and mechanical compatibility measures, the assembly process was formulated as an energy-minimization problem on a grid-like graph and solved efficiently by dual decomposition. This method has been successfully applied to achieve target deformation profiles \cite{wang2020deep},  mechanical cloaking \cite{wang2022mechanical}, and fracture resistance \cite{DaDaicong2022Datd}. Following a similar direction, Wang et al. \cite{Wang2022IH-GAN:Structures} used a GAN model to learn the distribution of implicit-surface-based geometries conditioned on given properties. By utilizing the continuity of the GAN-generated structures and a transition layer blending technique, compatible microstructures are inversely generated to achieve the designed properties in the assembled structure. Liu et al. \cite{liu2022growth} adopt a different strategy by devising a stochastic growth rule, similar to cellular automaton \cite{wolfram1984cellular}, to blend different graph-based features into irregular microstructures (Figure~\ref{fig:Top-down methods}(d)). Good compatibility can be ensured by devising special local growth rules. With the relation between the homogenized properties and the growth rule, the full structure can achieve the target property distribution through an automatic growth process.
    \subsection{Discussion}
    \label{5.4 Discussion}
            % \liwei{
            % \begin{itemize}
            % \item {- efficiency vs design flexibility (compare both two frameworks and methods within each framework \\\underline{Figure: x- efficiency, y-design flexibility}}
            % \item{ - Generalizability(nonlinear case, multi-physics case)}
            % \item{ - Manufacturability} 
            % \end{itemize}
            % }
        Both data-driven multiscale frameworks have shown promise in achieving efficient multiscale design in various applications, but they still have limitations and are not yet ideal. In this section, we will compare these frameworks to illustrate their respective strengths and weaknesses, providing insights into their applicability. We will also highlight critical knowledge gaps and challenges in existing research.
         \subsubsection{Comparison of Bottom-Up and Top-Down Frameworks}
         \label{5.4.1 Comparison of two framework}
         \begin{itemize}
         \item {\textbf{Efficiency} The two types of frameworks both ignore the microscale details in multiscale optimization by using the unit cell parameters (bottom-up) or effective properties (top-down) as the design variables. This allows them to bypass nested optimization across different scales and numerous homogenization evaluations, enabling much higher computational efficiency than traditional multiscale designs. Among them, the bottom-up framework is generally more efficient than its top-down counterpart because the compatibility between neighboring unit cells is guaranteed by the parameterized unit cell or easily handled by adding local constraints. In the top-down framework, when diverse microstructures are considered, an extra tiling optimization is usually needed after the homogenization-based optimization to guarantee compatibility, which is relatively time-consuming.}
         \item {\textbf{Design Freedom} The bottom-up framework uses microscale parameters as the design variables and thus requires a parameterized model for unit cells. This restricts the change of unit cell topologies. In contrast, the top-down framework directly optimizes the effective properties, which can be adapted to any unit-cell geometries. Therefore, the top-down framework has higher design freedom than its bottom-up counterpart. The flexibility of the bottom-up framework can be improved by considering multiple classes of microstructures or the unit cell orientation, as previously illustrated. However, this will sacrifice some efficiency and may lead to a complex optimization problem with more local optima. 
         }
         \item {\textbf{Manufacturability} The parameterization of unit cells in the bottom-up framework makes it easier to impose functionally graded constraints or filters on the geometries, which can benefit the manufacturability. The manufacturing restriction can also be considered in selecting the unit cell classes. In the top-down framework, manufacturing constraints of a single unit cell can be added to the construction of the database. However, the manufacturability of the assembled structure cannot be explicitly considered when designing the property distribution. Additional steps of compatibility optimization and post-processing are required in the assembly stage to obtain manufacturable full structures\cite{Zhu2017, chen2018computational, wang2020deep}.}
         \item {\textbf{Generalizability} The top-down method assumes weak mechanical coupling between unit cells and has been confined to material designs in the realm of linear elasticity. For dynamic applications or nonlinear cases, this coupling might not be negligible \cite{florijn2014programmable,ma2019valley,nguyen2011multiscale,wallin2020nonlinear,nakshatrala2013nonlinear}. Since the unit cells in the bottom-up framework are parameterized, it is easier to consider the coupling between unit cells by modeling the interaction as a function of their parameters. The bottom-up framework can also be extended to consider strain-dependent properties in accommodating nonlinear cases \cite{white2019multiscale}.}
         \end{itemize}
        \subsubsection{Assumptions of Homogenization}
        \label{5.4.2 Assumptions of homogenization}
         The first-order homogenization method is the basis of most data-driven multiscale designs to obtain the effective properties of unit cells in the database \cite{li2008introduction}. It assumes that the stress of each point in the macroscale only depends on its local strain value and is not affected by neighboring unit cells \cite{hassani1998review}. This is only valid when the microscopic length scale is much smaller than that of the macrostructure, and when the microstructures are periodically distributed\cite{da2021design}. For example, in a periodic design for compliance minimization, the ratio between the sizes of macro- and micro-structures should be of 5-6 to ensure a relatively accurate evaluation of the performance with homogenization. Most existing data-driven multiscale designs do not meet the first-order homogenization assumption due to the aperiodic tiling and the large unit cell size restricted by manufacturability. However, if neighboring unit cells have a smooth change or compatible connections within a large entire structure, homogenization can still provide relatively satisfying results as reported in some studies \cite{Chan2022RemixingBlending,schumacher2015microstructures}. Nevertheless, it is still advisable to present the performance metrics obtained via full-scale simulation as a validation, which was rarely reported in existing papers.
        
        Meanwhile, reduced-order models combined with ML are becoming promising alternatives to replace homogenization in multiscale deigns which does not rely on first-order homogenization assumption and can thus simulate the response more accurately. For example, Wu and Fu et al. \cite{fu2019topology,wu2019topology} condensed the fine FE model of unit cells into a super element with only boundary nodes. Since no periodicity or scale separation is assumed, the macro-response remains accurate for heterogeneous design even with the size of the unit cell close to the macrostructure. By combing proper orthogonal decomposition and diffusion approximation methods, the relation between volume fraction and the condensed stiffness matrix can be directly obtained for the previous graded structural design framework.
        
        Physics-based machine learning (PBML) model is emerging as another intriguing direction to accelerate the full-scale simulation without resorting to homogenization (see \ref{sec:physics_based}). This can be a potentially useful tool to bypass the first-order homogenization assumption. Currently, PBML mainly focuses on forward analysis instead of inverse optimization. Yao et al. \cite{yao2020fea} considered the finite element model as a special type of convolution layer to construct FEA-Net, predicting the mechanical response of metamaterials under external loading. Following similar ideas, Saha et al. \cite{saha2021hierarchical} used neural networks to replace the interpolation functions in FEA models, which can provide fine-resolution results with lower computational expenses. While these physics-informed methods require fewer data, they are either less efficient or restricted to a single type of microstructure, compared to the aforementioned fully data-driven methods. Although many existing studies claim that PBML is more efficient and easy to use than traditional FEA, most of these models can still be considered as special FE models or PDE solvers. They used a neural network to replace those classical approximation functions, which transform the original linear weighted-form equation into a highly non-linear optimization problem. Stochastic gradient descents and their variances were then used to solve the problem, which does not guarantee convergence to the real solution. The universal applications and higher efficiency were obtained at the cost of rigorous theoretical foundations. Also, most existing studies compare PBML with naive FE models, instead of some more advanced finite element methods, which might not be a fair comparison \cite{efendiev2009multiscale, liu2018narrow,mukherjee2021accelerating}. It is advisable for the researchers to explore both PBML and advanced finite element methods in accelerating the data-driven designs for multiscale systems.
        %%%%
        \subsubsection{Task Specificity of Data Acquisition for Multiscale Design}
        \label{5.4.5 Task Specificity of Data Acquisition for Multiscale Design}
\begin{figure}[t!]
        \centering
        \includegraphics[width=.8\textwidth]{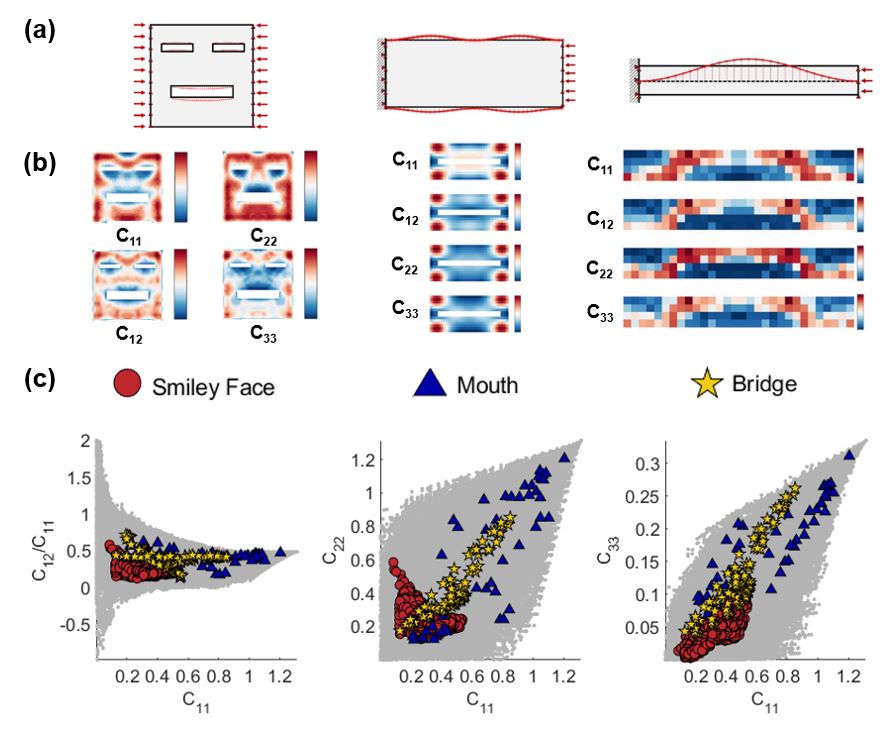}
        \caption{Task specificity of data acquisition and assessment based on the case study in Wang et al.~\cite{wang2020deep}. a) Illustration of three system-level design tasks associated with different target displacements: smiley face (left), mouth (center), and bridge (right). b) On-demand distributions of homogenized elasticity components $\{C_{11}$, $C_{12}$, $C_{22}$, $C_{33}\}$. c) The distributions of required property for all the individual tasks plotted with respect to that of the orthotropic dataset (gray), created through Pixel/Voxel and Perturbation. All the results are Reproduced from Ref. Wang et al.~\cite{wang2020deep} with permission. Copyright 2020, Elsevier B.V.        }
        \label{fig: task specificity}
        \end{figure}

        In Section~\ref{task-related metrics} we have briefly discussed the task-specificity of data assessment within metric-level assessment. Enlarging the scope to the system-level design optimization, we reiterate the task-specificity based on some concrete results reproduced from the literature. Figure~\ref{fig: task specificity} shows a set of target tasks at the system level defined by Wang et al.~\cite{wang2020deep}. Assuming linear elasticity deformation in 2D mechanical metamaterials, three design tasks are prepared to achieve different target displacements in Figure~\ref{fig: task specificity}(a). Required distributions of homogenized elasticity components ($C_{11}$, $C_{12}$, $C_{22}$, $C_{33}$) are pre-computed for each target pattern (Figure~\ref{fig: task specificity}(b)). The 240k-size orthotropic mechanical metamaterial dataset, generated with Pixel/Voxel (Section~\ref{3.2.1 Representation . .}) and Perturbation (Section~\ref{3.2.2 reproduction . .}), is used.

        Figure~\ref{fig: task specificity}(c) manifests that the required data distributions vary significantly contingent upon tasks. Overall, the 240-k dataset can cover all three tasks, even if no information on them was given before the data acquisition. The on-demand property distribution for each task is quite disparate across the tasks. For example, the $50\times 50$ properties to achieve smiley face (red) are relatively clustered. This implies that the wide coverage does not benefit this particular task as much as the other two cases. But the smiley face task demands a large portion of anisotropic samples, i.e., those having large either $C_{11}/C_{22}$ or $C_{22}/C_{11}$, as shown in the $C_{11}-C_{22}$ space. Thus, if known before the data acquisition, such samples can be prioritized with an associated metric during the data acquisition, as shown in Lee et al.~\cite{lee2023t} and Wang et al.~\cite{Wang2020}. Meanwhile, the required $4 \times 20$ properties to produce the bridge-like deformation (blue) tend to widely spread in the associated property space. Among the tasks, the bridge-like deformation design benefits most from wide coverage, which is congruent with a generic goal of data acquisition. Even in this task, however, samples having negative Poisson's ratio are not used; this indicates such uniform coverage of property is not unconditionally favored, but depends on the intended tasks. We remark that herein we intentionally omit considering geometric/mechanical compatibility among building blocks to convey our point with particular focus on data. 

        In summary, for multiscale design purposes, the data acquisition and assessment are intrinsically open to subjectivity, in part due to task-specificity. Not all data holds equal utility for general design tasks. Given a target task(s) at the system level, the data acquisition and assessment should involve the specified task(s) and even intentionally introduce bias, in addition to data uniformity, to meet the task requirement and ensure an efficient data collection. Herein, the uniformity is meaningful only within the domains associated with the target tasks, as opposed to generic scenarios of data acquisition.
        % Built on these observations, we summarize our general view on data acquisition and data assessment of DMD for commonplace scenarios as follows:
        % \begin{itemize}
        %     \item If no target tasks have been specified, the data acquisition can aim for generic use, i.e., uniformity and wide coverage, in both shape and property space of building blocks. The data assessment can also follow the same criteria without any preferences on a certain region.
        %     \item Even without any on-demand properties specifically given \textit{\textit{a priori}}, instance-wise preferences related to shape (e.g., fabrication feasibility), property (e.g., high physical anisotropy), or both (e.g., performance-to-mass ratio) can be enforced during data acquisition to tailor the data distribution as desired with minimal trial-and-error. The data assessment must address both distributional metrics and task-related metrics.
        %     \item In case a target task at the downstream, or a set of target tasks, is given, the data acquisition and assessment can be prepared with regard to the specified task(s), in addition to data uniformity. Under this case, the data uniformity is useful only within the domains associated with the tasks. The assessment is subject to the definition of target tasks.
        % %    \item When a new task comes in addition to existing ones, additional data acquisition could follow to address it.
        % \end{itemize}
        
        %%%%
         \subsubsection{Other Challenges}
         \label{5.4.3 Other Challenges}
         Overall, the data-driven paradigm has shown its promise in multiscale metamaterial design, with a superior capability to excavate the underlying relations between properties and geometries. However, there is a tendency in existing studies to apply off-the-shelf models as black-box tools for multiscale design, claiming that the universal fitting capability and high efficiency can benefit the design process. The underlying assumptions and constraints of the models have been constantly ignored, leaving the applicability questionable. 
         
         Before applying a specific model to multiscale design, it would be helpful to consider how it will influence the optimization solver, and whether its outputs and assumptions agree with the physics. For example, various training losses have been used to evaluate the performance of ML models, which mainly focus on the mean errors. It is not guaranteed that models trained with these loss functions will always produce physically feasible outputs in the whole input domain. In some cases, this could also lead to issues that are detrimental for optimization solvers, e.g., singular and negative-defined stiffness matrix, unit cells with disconnected components, the discontinuity between different cells, non-differentiability and local fluctuations of the predicted responses. Some pre-trained deep NNs devised in computer graphics were transferred to the applications in multi-scale design. However, hierarchical features extracted by the latent layers of these models might not be suitable for the structure designs. An important characteristic in multi-scale meta-material designs is that a given macro-scale property can be achieved by multiple micro-structures \cite{wang2020deep, schumacher2015microstructures}. Nevertheless, in most existing methods, a one-to-one mapping is assumed between properties and geometries, which fails to accommodate the one-to-many nature. Most databases and ML models only consider constant property parameters of a single unit cell or periodically assembled unit cells. They do not take into account state-dependent and history-dependent properties, e.g. strain-dependent stiffness tensor in nonlinear cases, and the interactions between different unit cells. Moreover, existing research mainly focuses on simple regression models to predict the homogenized properties and generative models to reduce the dimension of shape descriptors for the unit-cell design. How to use data-driven methods to improve the optimization procedure itself, e.g., extraction of design rules and underlying physical knowledge, initial designs recommendation, iterative optimization strategy, and combinatorial assembly, generally remains unexplored.

%%%%%%%%%%%%%%%%%%%%%%%%%%%%%%%%%%%%%%%%%%%%%%%
\section{Conclusion}

We presented a comprehensive, critical review of the data-driven design of metamaterials and multiscale systems (DMD). Through our analysis, we categorized previous research endeavors into the distinct modules of a cohesive data-driven design framework, including data acquisition (Section~\ref{3. Data Acquisition}), unit-cell level learning and optimization (Section~\ref{4 Learning and Generation: Data-Driven Metamaterials Design}), and multi-scale system designs (Section~\ref{5 Data-driven Multiscale TO}). In Section~\ref{3. Data Acquisition}, we examined the common practices of data acquisition strategies, with special attention to the methodological components of shape generation and property-aware sampling, and to data assessment. In Section~\ref{4 Learning and Generation: Data-Driven Metamaterials Design}, we covered prior works that utilized datasets and ML to enable acceleration of, or higher design freedom in, unit cell design. In Section~\ref{5 Data-driven Multiscale TO}, we reviewed data-driven multiscale design that employed databases and ML at the unit-cell level to optimize structures across multiple scales, meeting heterogeneous properties requirements in a top-down or bottom-up manner. For these multiscale design efforts, the main focus was to replace homogenization with surrogate modeling, and increase design flexibility by accommodating diverse and oriented unit cells.      

Based on our literature survey, we shared our perspectives on the current practices and suggested new avenues for future research efforts. In Section~\ref{3. Data Acquisition}, we disclosed that the current research trend in DMD is arguably biased towards the final products at the downstream modules and lacks principled methods dedicated to data acquisition, which is critical to the successful and robust deployment of DMD frameworks. To address this, more benchmark datasets need to be made publicly available and standard dataset assessment protocols should be established so that the contributions of future works can be rigorously appreciated. In Section~\ref{4 Learning and Generation: Data-Driven Metamaterials Design}, we discussed the cost-benefit trade-off of the reviewed methods, which is usually neglected in prior works. In addition, we noted that the trustworthiness and creativity of ML models for unit cell designs are also under-studied. There are ample opportunities to develop interpretable or physics-informed ML methods, as well as to enhance their uncertainty quantification and generalization capabilities. In Section~\ref{5 Data-driven Multiscale TO}, we remarked that there are still many unexplored opportunities for data-driven multiscale system design, particularly in areas such as extracting design rules, discovering physical knowledge, providing initial design recommendations, facilitating iterative optimization strategies, and enabling combinatorial assembly.

Overall, ML has shown promise in metamaterials design with its superior capability to excavate the complex relations between properties and geometries.  Despite the potential of the data-driven design approaches we reviewed, it is important to acknowledge that the field is still in its early stages and faces many grand challenges. Addressing these challenges means that the disconnect between the two primary approaches~--~data-driven and physics-based~--~needs to be resolved, which will require careful consideration of the cost-benefit and trustworthiness of ML methods, as well as collaboration between various disciplines. We believe that the tighter integration of physics and data-driven methods can unlock exciting possibilities for metamaterials design. Towards this future, we hope our review contributes to promoting interdisciplinary collaborations and bridging the gap between the two camps.

%%%%%%%%%%%%%%%%%%%%%%%%%%%%%%%%%%%%%%%%%%%%%%%
\section*{Acknowledgements}
This work was supported by the NSF BRITE Fellow program (Grant No. CMMI 2227641), the NSF CSSI program (Grant No.
OAC 1835782), and the Northwestern McCormick Catalyst Award.

\section*{Conflict of Interest}
The authors declare no conflict of interest.

\bibliographystyle{unsrt}

\begin{thebibliography}{100}

\bibitem{skylar2019voxelated}
Mark~A Skylar-Scott, Jochen Mueller, Claas~W Visser, and Jennifer~A Lewis.
\newblock Voxelated soft matter via multimaterial multinozzle 3d printing.
\newblock {\em Nature}, 575(7782):330--335, 2019.

\bibitem{wehner2016integrated}
Michael Wehner, Ryan~L Truby, Daniel~J Fitzgerald, Bobak Mosadegh, George~M
  Whitesides, Jennifer~A Lewis, and Robert~J Wood.
\newblock An integrated design and fabrication strategy for entirely soft,
  autonomous robots.
\newblock {\em nature}, 536(7617):451--455, 2016.

\bibitem{zhang2020seamless}
Yubai Zhang, Zhenhua Wang, Yang Yang, Qiaomei Chen, Xiaojie Qian, Yahe Wu, Huan
  Liang, Yanshuang Xu, Yen Wei, and Yan Ji.
\newblock Seamless multimaterial 3d liquid-crystalline elastomer actuators for
  next-generation entirely soft robots.
\newblock {\em Science advances}, 6(9):eaay8606, 2020.

\bibitem{kim2019ferromagnetic}
Yoonho Kim, German~A Parada, Shengduo Liu, and Xuanhe Zhao.
\newblock Ferromagnetic soft continuum robots.
\newblock {\em Science Robotics}, 4(33):eaax7329, 2019.

\bibitem{yu2018mechanical}
Xianglong Yu, Ji~Zhou, Haiyi Liang, Zhengyi Jiang, and Lingling Wu.
\newblock Mechanical metamaterials associated with stiffness, rigidity and
  compressibility: A brief review.
\newblock {\em Progress in Materials Science}, 94:114--173, 2018.

\bibitem{zheludev2010road}
Nikolay~I Zheludev.
\newblock The road ahead for metamaterials.
\newblock {\em Science}, 328(5978):582--583, 2010.

\bibitem{kadic20193d}
Muamer Kadic, Graeme~W Milton, Martin van Hecke, and Martin Wegener.
\newblock 3d metamaterials.
\newblock {\em Nature Reviews Physics}, 1(3):198--210, 2019.

\bibitem{lumpe2021exploring}
Thomas~S Lumpe and Tino Stankovic.
\newblock Exploring the property space of periodic cellular structures based on
  crystal networks.
\newblock {\em Proceedings of the National Academy of Sciences},
  118(7):e2003504118, 2021.

\bibitem{leonhardt2006optical}
Ulf Leonhardt.
\newblock Optical conformal mapping.
\newblock {\em science}, 312(5781):1777--1780, 2006.

\bibitem{leonhardt2009broadband}
Ulf Leonhardt and Tom{\'a}s Tyc.
\newblock Broadband invisibility by non-euclidean cloaking.
\newblock {\em Science}, 323(5910):110--112, 2009.

\bibitem{liu2008three}
Na~Liu, Hongcang Guo, Liwei Fu, Stefan Kaiser, Heinz Schweizer, and Harald
  Giessen.
\newblock Three-dimensional photonic metamaterials at optical frequencies.
\newblock {\em Nature materials}, 7(1):31--37, 2008.

\bibitem{ma2013first}
Yungui Ma, Yichao Liu, Lu~Lan, Tiantian Wu, Wei Jiang, CK~Ong, and Sailing He.
\newblock First experimental demonstration of an isotropic electromagnetic
  cloak with strict conformal mapping.
\newblock {\em Scientific Reports}, 3(1):1--5, 2013.

\bibitem{pendry2006controlling}
John~B Pendry, David Schurig, and David~R Smith.
\newblock Controlling electromagnetic fields.
\newblock {\em science}, 312(5781):1780--1782, 2006.

\bibitem{fan2008shaped}
CZ~Fan, Y~Gao, and JP~Huang.
\newblock Shaped graded materials with an apparent negative thermal
  conductivity.
\newblock {\em Applied Physics Letters}, 92(25):251907, 2008.

\bibitem{schittny2013experiments}
Robert Schittny, Muamer Kadic, Sebastien Guenneau, and Martin Wegener.
\newblock Experiments on transformation thermodynamics: molding the flow of
  heat.
\newblock {\em Physical review letters}, 110(19):195901, 2013.

\bibitem{han2013homogeneous}
Tiancheng Han, Tao Yuan, Baowen Li, and Cheng-Wei Qiu.
\newblock Homogeneous thermal cloak with constant conductivity and tunable heat
  localization.
\newblock {\em Scientific reports}, 3(1):1--5, 2013.

\bibitem{hu2020machine}
Run Hu, Sotaro Iwamoto, Lei Feng, Shenghong Ju, Shiqian Hu, Masato Ohnishi,
  Naomi Nagai, Kazuhiko Hirakawa, and Junichiro Shiomi.
\newblock Machine-learning-optimized aperiodic superlattice minimizes coherent
  phonon heat conduction.
\newblock {\em Physical Review. X}, 10(2), 2020.

\bibitem{chen2014periodic}
Yanyu Chen and Lifeng Wang.
\newblock Periodic co-continuous acoustic metamaterials with overlapping
  locally resonant and bragg band gaps.
\newblock {\em Applied Physics Letters}, 105(19):191907, 2014.

\bibitem{deymier2013acoustic}
Pierre~A Deymier.
\newblock {\em Acoustic metamaterials and phononic crystals}, volume 173.
\newblock Springer Science \& Business Media, 2013.

\bibitem{phani2017dynamics}
A~Srikantha Phani and Mahmoud~I Hussein.
\newblock {\em Dynamics of lattice materials}.
\newblock John Wiley \& Sons, 2017.

\bibitem{ma2019valley}
Jihong Ma, Kai Sun, and Stefano Gonella.
\newblock Valley hall in-plane edge states as building blocks for elastodynamic
  logic circuits.
\newblock {\em Physical Review Applied}, 12(4):044015, 2019.

\bibitem{wang2020valley}
Mudi Wang, Wenyi Zhou, Liya Bi, Chunyin Qiu, Manzhu Ke, and Zhengyou Liu.
\newblock Valley-locked waveguide transport in acoustic heterostructures.
\newblock {\em Nature communications}, 11(1):1--6, 2020.

\bibitem{florijn2014programmable}
Bastiaan Florijn, Corentin Coulais, and Martin van Hecke.
\newblock Programmable mechanical metamaterials.
\newblock {\em Physical review letters}, 113(17):175503, 2014.

\bibitem{wang2021structured}
Yifan Wang, Liuchi Li, Douglas Hofmann, Jos{\'e}~E Andrade, and Chiara Daraio.
\newblock Structured fabrics with tunable mechanical properties.
\newblock {\em Nature}, 596(7871):238--243, 2021.

\bibitem{jenett2020discretely}
Benjamin Jenett, Christopher Cameron, Filippos Tourlomousis, Alfonso~Parra
  Rubio, Megan Ochalek, and Neil Gershenfeld.
\newblock {Discretely assembled mechanical metamaterials}.
\newblock {\em Science advances}, 6(47):eabc9943, 2020.

\bibitem{kochmann2017exploiting}
Dennis~M Kochmann and Katia Bertoldi.
\newblock Exploiting microstructural instabilities in solids and structures:
  from metamaterials to structural transitions.
\newblock {\em Applied mechanics reviews}, 69(5), 2017.

\bibitem{frenzel2017three}
Tobias Frenzel, Muamer Kadic, and Martin Wegener.
\newblock Three-dimensional mechanical metamaterials with a twist.
\newblock {\em Science}, 358(6366):1072--1074, 2017.

\bibitem{jin2020guided}
Lishuai Jin, Romik Khajehtourian, Jochen Mueller, Ahmad Rafsanjani, Vincent
  Tournat, Katia Bertoldi, and Dennis~M Kochmann.
\newblock Guided transition waves in multistable mechanical metamaterials.
\newblock {\em Proceedings of the National Academy of Sciences},
  117(5):2319--2325, 2020.

\bibitem{reid2018auxetic}
Daniel~R Reid, Nidhi Pashine, Justin~M Wozniak, Heinrich~M Jaeger, Andrea~J
  Liu, Sidney~R Nagel, and Juan~J de~Pablo.
\newblock Auxetic metamaterials from disordered networks.
\newblock {\em Proceedings of the National Academy of Sciences},
  115(7):E1384--E1390, 2018.

\bibitem{kadic2012practicability}
Muamer Kadic, Tiemo B{\"u}ckmann, Nicolas Stenger, Michael Thiel, and Martin
  Wegener.
\newblock On the practicability of pentamode mechanical metamaterials.
\newblock {\em Applied Physics Letters}, 100(19):191901, 2012.

\bibitem{fernandes2021mechanically}
Matheus~C Fernandes, Joanna Aizenberg, James~C Weaver, and Katia Bertoldi.
\newblock Mechanically robust lattices inspired by deep-sea glass sponges.
\newblock {\em Nature Materials}, 20(2):237--241, 2021.

\bibitem{bessa2019bayesian}
Miguel~A Bessa, Piotr Glowacki, and Michael Houlder.
\newblock Bayesian machine learning in metamaterial design: Fragile becomes
  supercompressible.
\newblock {\em Advanced Materials}, 31(48):1904845, 2019.

\bibitem{hedayati20213d}
Reza Hedayati, Aysun G{\"u}ven, and Sybrand Van Der~Zwaag.
\newblock 3d gradient auxetic soft mechanical metamaterials fabricated by
  additive manufacturing.
\newblock {\em Applied Physics Letters}, 118(14):141904, 2021.

\bibitem{Wu2021TopologyReview}
Jun Wu, Ole Sigmund, and Jeroen~P. Groen.
\newblock {Topology optimization of multi-scale structures: a review}.
\newblock {\em Structural and Multidisciplinary Optimization},
  63(3):1455--1480, 2021.

\bibitem{Regenwetter2021DeepReview}
Lyle Regenwetter, Amin~Heyrani Nobari, and Faez Ahmed.
\newblock Deep generative models in engineering design: A review.
\newblock {\em Journal of Mechanical Design}, 144(7):071704, 2022.

\bibitem{woldseth2022use}
Rebekka~V Woldseth, Niels Aage, J~Andreas B{\ae}rentzen, and Ole Sigmund.
\newblock On the use of artificial neural networks in topology optimisation.
\newblock {\em Structural and Multidisciplinary Optimization}, 65(10):1--36,
  2022.

\bibitem{Mukherjee2021AcceleratingChallenges}
Sougata Mukherjee, Dongcheng Lu, Balaji Raghavan, Piotr Breitkopf, Subhrajit
  Dutta, Manyu Xiao, and Weihong Zhang.
\newblock {Accelerating Large-scale Topology Optimization: State-of-the-Art and
  Challenges}.
\newblock {\em Archives of Computational Methods in Engineering},
  28(7):4549--4571, 2021.

\bibitem{so2022revisiting}
Sunae So, Jungho Mun, Junghyun Park, and Junsuk Rho.
\newblock Revisiting the design strategies for metasurfaces: Fundamental
  physics, optimization, and beyond.
\newblock {\em Advanced Materials}, page 2206399, 2022.

\bibitem{Wang2020MachinePerspectives}
C.~Wang, X.~P. Tan, S.~B. Tor, and C.~S. Lim.
\newblock {Machine learning in additive manufacturing: State-of-the-art and
  perspectives}.
\newblock {\em Additive Manufacturing}, 36(August):101538, 2020.

\bibitem{Mozaffar2022MechanisticPerspectives}
Mojtaba Mozaffar, Shuheng Liao, Xiaoyu Xie, Sourav Saha, Chanwook Park, Jian
  Cao, Wing~Kam Liu, and Zhengtao Gan.
\newblock {Mechanistic artificial intelligence (mechanistic-AI) for modeling,
  design, and control of advanced manufacturing processes: Current state and
  perspectives}.
\newblock {\em Journal of Materials Processing Technology}, 302(October 2021),
  2022.

\bibitem{So2020DeepNanophotonics}
Sunae So, Trevon Badloe, Jaebum Noh, Junsuk Rho, and Jorge Bravo-Abad.
\newblock {Deep learning enabled inverse design in nanophotonics}.
\newblock {\em Nanophotonics}, 9(5):1041--1057, 2020.

\bibitem{Liu2021HowStructures}
Xin Liu, Su~Tian, Fei Tao, Haodong Du, and Wenbin Yu.
\newblock {How machine learning can help the design and analysis of composite
  materials and structures?}
\newblock {\em AIAA Scitech 2021 Forum}, pages 1--21, 2021.

\bibitem{Kumar2021WhatMechanics}
Siddhant Kumar and Dennis~M Kochmann.
\newblock What machine learning can do for computational solid mechanics.
\newblock In {\em Current trends and open problems in computational mechanics},
  pages 275--285. Springer, 2022.

\bibitem{liu2021tackling}
Zhaocheng Liu, Dayu Zhu, Lakshmi Raju, and Wenshan Cai.
\newblock Tackling photonic inverse design with machine learning.
\newblock {\em Advanced Science}, 8(5):2002923, 2021.

\bibitem{jin2022intelligent}
Yabin Jin, Liangshu He, Zhihui Wen, Bohayra Mortazavi, Hongwei Guo, Daniel
  Torrent, Bahram Djafari-Rouhani, Timon Rabczuk, Xiaoying Zhuang, and Yan Li.
\newblock Intelligent on-demand design of phononic metamaterials.
\newblock {\em Nanophotonics}, 11(3):439--460, 2022.

\bibitem{liu2023deep}
Chen-Xu Liu and Gui-Lan Yu.
\newblock Deep learning for the design of phononic crystals and elastic
  metamaterials.
\newblock {\em Journal of Computational Design and Engineering},
  10(2):602--614, 2023.

\bibitem{li2008introduction}
Shaofan Li and Gang Wang.
\newblock {\em Introduction to micromechanics and nanomechanics}.
\newblock World Scientific Publishing Company, 2008.

\bibitem{Chan2022Yu-ChinDissertation}
Yu-Chin Chan.
\newblock {\em Data-Driven and Diversity-Enhanced Design of Heterogeneous
  Multiscale Structures}.
\newblock PhD thesis, Northwestern University, 2022.

\bibitem{Wang2020}
Liwei Wang, Yu~Chin Chan, Zhao Liu, Ping Zhu, and Wei Chen.
\newblock {Data-driven metamaterial design with Laplace-Beltrami spectrum as
  “shape-DNA”}.
\newblock {\em Structural and Multidisciplinary Optimization},
  61(6):2613--2628, 2020.

\bibitem{Wang2022IH-GAN:Structures}
Jun Wang, Wei~(Wayne) Chen, Daicong Da, Mark Fuge, and Rahul Rai.
\newblock {IH-GAN: A conditional generative model for implicit surface-based
  inverse design of cellular structures}.
\newblock {\em Computer Methods in Applied Mechanics and Engineering},
  396:115060, 2022.

\bibitem{Liu2018}
Zhaocheng Liu, Dayu Zhu, Sean~P. Rodrigues, Kyu~Tae Lee, and Wenshan Cai.
\newblock {Generative Model for the Inverse Design of Metasurfaces}.
\newblock {\em Nano Letters}, 18(10):6570--6576, 2018.

\bibitem{Chan2022RemixingBlending}
Yu~Chin Chan, Daicong Da, Liwei Wang, and Wei Chen.
\newblock {Remixing functionally graded structures: data-driven topology
  optimization with multiclass shape blending}.
\newblock {\em Structural and Multidisciplinary Optimization}, 65(5):1--22,
  2022.

\bibitem{wang2020deep}
Liwei Wang, Yu-Chin Chan, Faez Ahmed, Zhao Liu, Ping Zhu, and Wei Chen.
\newblock Deep generative modeling for mechanistic-based learning and design of
  metamaterial systems.
\newblock {\em Computer Methods in Applied Mechanics and Engineering},
  372:113377, 2020.

\bibitem{kramer1991nonlinear}
Mark~A Kramer.
\newblock Nonlinear principal component analysis using autoassociative neural
  networks.
\newblock {\em AIChE journal}, 37(2):233--243, 1991.

\bibitem{li2020designing}
Xiang Li, Shaowu Ning, Zhanli Liu, Ziming Yan, Chengcheng Luo, and Zhuo Zhuang.
\newblock {Designing phononic crystal with anticipated band gap through a deep
  learning based data-driven method}.
\newblock {\em Computer Methods in Applied Mechanics and Engineering},
  361:112737, 2020.

\bibitem{qiu2019deep}
Tianshuo Qiu, Xin Shi, Jiafu Wang, Yongfeng Li, Shaobo Qu, Qiang Cheng, Tiejun
  Cui, and Sai Sui.
\newblock Deep learning: a rapid and efficient route to automatic metasurface
  design.
\newblock {\em Advanced Science}, 6(12):1900128, 2019.

\bibitem{kingma2013auto}
Diederik~P Kingma and Max Welling.
\newblock Auto-encoding variational bayes.
\newblock {\em arXiv preprint arXiv:1312.6114}, 2013.

\bibitem{Wang2020DeepSystems}
Liwei Wang, Yu~Chin Chan, Faez Ahmed, Zhao Liu, Ping Zhu, and Wei Chen.
\newblock {Deep generative modeling for mechanistic-based learning and design
  of metamaterial systems}.
\newblock {\em Computer Methods in Applied Mechanics and Engineering},
  372:1--41, 2020.

\bibitem{liu2020hybrid}
Zhaocheng Liu, Lakshmi Raju, Dayu Zhu, and Wenshan Cai.
\newblock A hybrid strategy for the discovery and design of photonic
  structures.
\newblock {\em IEEE Journal on Emerging and Selected Topics in Circuits and
  Systems}, 10(1):126--135, 2020.

\bibitem{goodfellow2020generative}
Ian Goodfellow, Jean Pouget-Abadie, Mehdi Mirza, Bing Xu, David Warde-Farley,
  Sherjil Ozair, Aaron Courville, and Yoshua Bengio.
\newblock Generative adversarial networks.
\newblock {\em Communications of the ACM}, 63(11):139--144, 2020.

\bibitem{radford2015unsupervised}
Alec Radford, Luke Metz, and Soumith Chintala.
\newblock Unsupervised representation learning with deep convolutional
  generative adversarial networks.
\newblock {\em arXiv preprint arXiv:1511.06434}, 2015.

\bibitem{gurbuz2021generative}
Caglar Gurbuz, Felix Kronowetter, Christoph Dietz, Martin Eser, Jonas Schmid,
  and Steffen Marburg.
\newblock Generative adversarial networks for the design of acoustic
  metamaterials.
\newblock {\em The Journal of the Acoustical Society of America},
  149(2):1162--1174, 2021.

\bibitem{ma2019probabilistic}
Wei Ma, Feng Cheng, Yihao Xu, Qinlong Wen, and Yongmin Liu.
\newblock Probabilistic representation and inverse design of metamaterials
  based on a deep generative model with semi-supervised learning strategy.
\newblock {\em Advanced Materials}, 31(35):1901111, 2019.

\bibitem{ma2020data}
Wei Ma and Yongmin Liu.
\newblock A data-efficient self-supervised deep learning model for design and
  characterization of nanophotonic structures.
\newblock {\em Science China Physics, Mechanics \& Astronomy}, 63(8):1--8,
  2020.

\bibitem{zhu2005semi}
Xiaojin Zhu.
\newblock Semi-supervised learning literature survey.
\newblock Technical Report 1530, Computer Sciences, University of
  Wisconsin-Madison, 2005.

\bibitem{van2020survey}
Jesper~E Van~Engelen and Holger~H Hoos.
\newblock A survey on semi-supervised learning.
\newblock {\em Machine learning}, 109(2):373--440, 2020.

\bibitem{bellman1957markovian}
Richard Bellman.
\newblock A markovian decision process.
\newblock {\em Journal of mathematics and mechanics}, pages 679--684, 1957.

\bibitem{sajedian2019double}
Iman Sajedian, Heon Lee, and Junsuk Rho.
\newblock Double-deep q-learning to increase the efficiency of metasurface
  holograms.
\newblock {\em Scientific reports}, 9(1):1--8, 2019.

\bibitem{liu2021reinforcement}
Bin Liu, Liujun Xu, and Jiping Huang.
\newblock Reinforcement learning approach to thermal transparency with
  particles in periodic lattices.
\newblock {\em Journal of Applied Physics}, 130(4):045103, 2021.

\bibitem{sui2021deep}
Fanping Sui, Ruiqi Guo, Zhizhou Zhang, Grace~X Gu, and Liwei Lin.
\newblock Deep reinforcement learning for digital materials design.
\newblock {\em ACS Materials Letters}, 3(10):1433--1439, 2021.

\bibitem{wang2021data}
Liwei Wang, Siyu Tao, Ping Zhu, and Wei Chen.
\newblock Data-driven topology optimization with multiclass microstructures
  using latent variable gaussian process.
\newblock {\em Journal of Mechanical Design}, 143(3), 2021.

\bibitem{liu2022growth}
Ke~Liu, Rachel Sun, and Chiara Daraio.
\newblock Growth rules for irregular architected materials with programmable
  properties.
\newblock {\em Science}, 377(6609):975--981, 2022.

\bibitem{Li2019DesignShapes}
Dawei Li, Ning Dai, Yunlong Tang, Guoying Dong, and Yaoyao~Fiona Zhao.
\newblock {Design and Optimization of Graded Cellular Structures with Triply
  Periodic Level Surface-Based Topological Shapes}.
\newblock {\em Journal of Mechanical Design, Transactions of the ASME}, 141(7),
  2019.

\bibitem{kumar2020inverse}
Siddhant Kumar, Stephanie Tan, Li~Zheng, and Dennis~M Kochmann.
\newblock Inverse-designed spinodoid metamaterials.
\newblock {\em npj Computational Materials}, 6(1):1--10, 2020.

\bibitem{Zhu2017}
Bo~Zhu, Mélina Skouras, Desai Chen, and Wojciech Matusik.
\newblock {Two-scale topology optimization with microstructures}.
\newblock {\em ACM Transactions on Graphics}, 2017.

\bibitem{Whiting2020Meta-atomMethod}
Eric~B. Whiting, Sawyer~D. Campbell, Lei Kang, and Douglas~H. Werner.
\newblock {Meta-atom library generation via an efficient multi-objective shape
  optimization method}.
\newblock {\em Optics Express}, 28(16):24229, 2020.

\bibitem{inampudi2018neural}
Sandeep Inampudi and Hossein Mosallaei.
\newblock Neural network based design of metagratings.
\newblock {\em Applied Physics Letters}, 112(24):241102, 2018.

\bibitem{An2021MultifunctionalNetwork}
Sensong An, Bowen Zheng, Hong Tang, Mikhail~Y. Shalaginov, Li~Zhou, Hang Li,
  Myungkoo Kang, Kathleen~A. Richardson, Tian Gu, Juejun Hu, Clayton Fowler,
  and Hualiang Zhang.
\newblock {Multifunctional Metasurface Design with a Generative Adversarial
  Network}.
\newblock {\em Advanced Optical Materials}, 9(5):1--10, 2021.

\bibitem{Malkiel2018PlasmonicLearning}
Itzik Malkiel, Michael Mrejen, Achiya Nagler, Uri Arieli, Lior Wolf, and Haim
  Suchowski.
\newblock {Plasmonic nanostructure design and characterization via Deep
  Learning}.
\newblock {\em Light: Science and Applications}, 7(1), 2018.

\bibitem{ma2018deep}
Wei Ma, Feng Cheng, and Yongmin Liu.
\newblock Deep-learning-enabled on-demand design of chiral metamaterials.
\newblock {\em ACS nano}, 12(6):6326--6334, 2018.

\bibitem{an2019deep}
Sensong An, Clayton Fowler, Bowen Zheng, Mikhail~Y Shalaginov, Hong Tang, Hang
  Li, Li~Zhou, Jun Ding, Anuradha~Murthy Agarwal, Clara Rivero-Baleine, et~al.
\newblock A deep learning approach for objective-driven all-dielectric
  metasurface design.
\newblock {\em ACS Photonics}, 6(12):3196--3207, 2019.

\bibitem{wang2022generalizedOptimization}
Liwei Wang, Zhao Liu, Daicong Da, Yu~Chin Chan, Wei Chen, and Ping Zhu.
\newblock {Generalized de-homogenization via sawtooth-function-based mapping
  and its demonstration on data-driven frequency response optimization}.
\newblock {\em Computer Methods in Applied Mechanics and Engineering},
  395:114967, 2022.

\bibitem{wilt2020accelerating}
Jackson~K Wilt, Charles Yang, and Grace~X Gu.
\newblock Accelerating auxetic metamaterial design with deep learning.
\newblock {\em Advanced Engineering Materials}, 22(5):1901266, 2020.

\bibitem{Wang2022ScalableFactors}
Liwei Wang, Suraj Yerramilli, Akshay Iyer, Daniel Apley, Ping Zhu, and Wei
  Chen.
\newblock {Scalable Gaussian Processes for Data-Driven Design Using Big Data
  With Categorical Factors}.
\newblock {\em Journal of Mechanical Design}, 144(2):1--13, 2022.

\bibitem{choi2019design}
Haejoon Choi, Adrian~Matias Chung~Baek, and Namhum Kim.
\newblock Design of non-periodic lattice structures by allocating pre-optimized
  building blocks.
\newblock In {\em International Design Engineering Technical Conferences and
  Computers and Information in Engineering Conference}, volume 59179, page
  V001T02A037. American Society of Mechanical Engineers, 2019.

\bibitem{Chan2020METASET:Design}
Yu-Chin Chan, Faez Ahmed, Liwei Wang, and Wei Chen.
\newblock Metaset: Exploring shape and property spaces for data-driven
  metamaterials design.
\newblock {\em Journal of Mechanical Design}, 143(3), 2021.

\bibitem{boddapati2023single}
Jagannadh Boddapati, Moritz Flaschel, Siddhant Kumar, Laura De~Lorenzis, and
  Chiara Daraio.
\newblock Single-test evaluation of directional elastic properties of
  anisotropic structured materials.
\newblock {\em arXiv preprint arXiv:2304.09112}, 2023.

\bibitem{LiuADesign}
Zhaocheng Liu, Zhaoming Zhu, and Wenshan Cai.
\newblock Topological encoding method for data-driven photonics inverse design.
\newblock {\em Optics express}, 28(4):4825--4835, 2020.

\bibitem{li2022digital}
Weichen Li, Fengwen Wang, Ole Sigmund, and Xiaojia~Shelly Zhang.
\newblock Digital synthesis of free-form multimaterial structures for
  realization of arbitrary programmed mechanical responses.
\newblock {\em Proceedings of the National Academy of Sciences},
  119(10):e2120563119, 2022.

\bibitem{Kudyshev2020Machine-learning-assistedOptimization}
Zhaxylyk~A. Kudyshev, Alexander~V. Kildishev, Vladimir~M. Shalaev, and
  Alexandra Boltasseva.
\newblock {Machine-learning-assisted metasurface design for high-efficiency
  thermal emitter optimization}.
\newblock {\em Applied Physics Reviews}, 7(2), 2020.

\bibitem{chen2022see}
Zhi Chen, Alexander Ogren, Chiara Daraio, L~Catherine Brinson, and Cynthia
  Rudin.
\newblock How to see hidden patterns in metamaterials with interpretable
  machine learning.
\newblock {\em Extreme Mechanics Letters}, page 101895, 2022.

\bibitem{Ma2020AStructures}
Wei Ma and Yongmin Liu.
\newblock {A data-efficient self-supervised deep learning model for design and
  characterization of nanophotonic structures}.
\newblock {\em Science China: Physics, Mechanics and Astronomy}, 63(8), 8 2020.

\bibitem{so2019designing}
Sunae So and Junsuk Rho.
\newblock Designing nanophotonic structures using conditional deep
  convolutional generative adversarial networks.
\newblock {\em Nanophotonics}, 8(7):1255--1261, 2019.

\bibitem{kazmi2013survey}
Ismail~Khalid Kazmi, Lihua You, and Jian~Jun Zhang.
\newblock A survey of 2d and 3d shape descriptors.
\newblock In {\em 2013 10th International Conference Computer Graphics, Imaging
  and Visualization}, pages 1--10. IEEE, 2013.

\bibitem{tanriover2022deep}
Ibrahim Tanriover, Doksoo Lee, Wei Chen, and Koray Aydin.
\newblock Deep generative modeling and inverse design of manufacturable
  free-form dielectric metasurfaces.
\newblock {\em ACS Photonics}, 2022.

\bibitem{wang2021novel}
Hui Wang, Si-Hang Xiao, and Chong Zhang.
\newblock Novel planar auxetic metamaterial perforated with orthogonally
  aligned oval-shaped holes and machine learning solutions.
\newblock {\em Advanced Engineering Materials}, 23(7):2100102, 2021.

\bibitem{requicha1977constructive}
Aristides~AG Requicha and Herbert~B Voelcker.
\newblock Constructive solid geometry.
\newblock Technical report, 1977.

\bibitem{Guo2014DoingFramework}
Xu~Guo, Weisheng Zhang, and Wenliang Zhong.
\newblock {Doing topology optimization explicitly and geometrically-a new
  moving morphable components based framework}.
\newblock {\em Journal of Applied Mechanics, Transactions ASME}, 81(8):1--12,
  2014.

\bibitem{Lei2019MachineFramework}
Xin Lei, Chang Liu, Zongliang Du, Weisheng Zhang, and Xu~Guo.
\newblock {Machine learning-driven real-time topology optimization under moving
  morphable component-based framework}.
\newblock {\em Journal of Applied Mechanics, Transactions ASME}, 86(1):1--9,
  2019.

\bibitem{lee2023t}
Doksoo Lee, Yu-Chin Chan, Wei Chen, Liwei Wang, Anton van Beek, and Wei Chen.
\newblock t-metaset: Task-aware acquisition of metamaterial datasets through
  diversity-based active learning.
\newblock {\em Journal of Mechanical Design}, 145(3):031704, 2023.

\bibitem{lin2001sampling}
Dennis~KJ Lin, Timothy~W Simpson, and Wei Chen.
\newblock Sampling strategies for computer experiments: design and analysis.
\newblock {\em International Journal of Reliability and applications},
  2(3):209--240, 2001.

\bibitem{jin2002sequential}
Ruichen Jin, Wei Chen, and Agus Sudjianto.
\newblock On sequential sampling for global metamodeling in engineering design.
\newblock In {\em International design engineering technical conferences and
  computers and information in engineering conference}, volume 36223, pages
  539--548, 2002.

\bibitem{so2019simultaneous}
Sunae So, Jungho Mun, and Junsuk Rho.
\newblock Simultaneous inverse design of materials and structures via deep
  learning: demonstration of dipole resonance engineering using core--shell
  nanoparticles.
\newblock {\em ACS applied materials \& interfaces}, 11(27):24264--24268, 2019.

\bibitem{bishop2006pattern}
Christopher~M Bishop and Nasser~M Nasrabadi.
\newblock {\em Pattern recognition and machine learning}, volume~4.
\newblock Springer, 2006.

\bibitem{sambasivan2021everyone}
Nithya Sambasivan, Shivani Kapania, Hannah Highfill, Diana Akrong, Praveen
  Paritosh, and Lora~M Aroyo.
\newblock “everyone wants to do the model work, not the data work”: Data
  cascades in high-stakes ai.
\newblock In {\em proceedings of the 2021 CHI Conference on Human Factors in
  Computing Systems}, pages 1--15, 2021.

\bibitem{fu2014bio}
Katherine Fu, Diana Moreno, Maria Yang, and Kristin~L Wood.
\newblock Bio-inspired design: an overview investigating open questions from
  the broader field of design-by-analogy.
\newblock {\em Journal of Mechanical Design}, 136(11):111102, 2014.

\bibitem{jiang2022data}
Shuo Jiang, Jie Hu, Kristin~L Wood, and Jianxi Luo.
\newblock Data-driven design-by-analogy: state-of-the-art and future
  directions.
\newblock {\em Journal of Mechanical Design}, 144(2), 2022.

\bibitem{miniaci2016spider}
Marco Miniaci, Anastasiia Krushynska, Alexander~B Movchan, Federico Bosia, and
  Nicola~M Pugno.
\newblock Spider web-inspired acoustic metamaterials.
\newblock {\em Applied Physics Letters}, 109(7):071905, 2016.

\bibitem{liu2018fractal}
Jian Liu, Liping Li, Baizhan Xia, and Xianfeng Man.
\newblock Fractal labyrinthine acoustic metamaterial in planar lattices.
\newblock {\em International Journal of Solids and Structures}, 132:20--30,
  2018.

\bibitem{hamzehei20223d}
Ramin Hamzehei, Ali Zolfagharian, Soheil Dariushi, and Mahdi Bodaghi.
\newblock 3d-printed bio-inspired zero poisson’s ratio graded metamaterials
  with high energy absorption performance.
\newblock {\em Smart Materials and Structures}, 31(3):035001, 2022.

\bibitem{li20214d}
Bing Li, Chao Zhang, Fang Peng, Wenzhi Wang, Bryan~D Vogt, and KT~Tan.
\newblock 4d printed shape memory metamaterial for vibration bandgap switching
  and active elastic-wave guiding.
\newblock {\em Journal of Materials Chemistry C}, 9(4):1164--1173, 2021.

\bibitem{nagel2010function}
Jacquelyn~KS Nagel, Robert~L Nagel, Robert~B Stone, and Daniel~A McAdams.
\newblock Function-based, biologically inspired concept generation.
\newblock {\em Ai Edam}, 24(4):521--535, 2010.

\bibitem{goel2015biologically}
Ashok~K Goel, Daniel~A McAdams, and Robert~B Stone.
\newblock {\em Biologically inspired design}.
\newblock Springer, 2015.

\bibitem{farooq2018spectral}
Umar Farooq~Ghumman, Akshay Iyer, Rabindra Dulal, Joydeep Munshi, Aaron Wang,
  TeYu Chien, Ganesh Balasubramanian, and Wei Chen.
\newblock A spectral density function approach for active layer design of
  organic photovoltaic cells.
\newblock {\em Journal of Mechanical Design}, 140(11), 2018.

\bibitem{iyer2020designing}
Akshay Iyer, Rabindra Dulal, Yichi Zhang, Umar~Farooq Ghumman, TeYu Chien,
  Ganesh Balasubramanian, and Wei Chen.
\newblock Designing anisotropic microstructures with spectral density function.
\newblock {\em Computational Materials Science}, 179:109559, 2020.

\bibitem{loh1996latin}
Wei-Liem Loh.
\newblock On latin hypercube sampling.
\newblock {\em The annals of statistics}, 24(5):2058--2080, 1996.

\bibitem{zhang2023entropy}
Hengrui Zhang, Wei Chen, James~M Rondinelli, and Wei Chen.
\newblock Et-al: Entropy-targeted active learning for bias mitigation in
  materials data.
\newblock {\em Applied Physics Reviews}, 10(2):021403, 2023.

\bibitem{choudhary2020joint}
Kamal Choudhary, Kevin~F Garrity, Andrew~CE Reid, Brian DeCost, Adam~J Biacchi,
  Angela~R Hight~Walker, Zachary Trautt, Jason Hattrick-Simpers, A~Gilad Kusne,
  Andrea Centrone, et~al.
\newblock The joint automated repository for various integrated simulations
  (jarvis) for data-driven materials design.
\newblock {\em npj computational materials}, 6(1):173, 2020.

\bibitem{settles2011theories}
Burr Settles.
\newblock From theories to queries: Active learning in practice.
\newblock In {\em Active learning and experimental design workshop in
  conjunction with AISTATS 2010}, pages 1--18. JMLR Workshop and Conference
  Proceedings, 2011.

\bibitem{gal2017deep}
Yarin Gal, Riashat Islam, and Zoubin Ghahramani.
\newblock Deep bayesian active learning with image data.
\newblock In {\em International conference on machine learning}, pages
  1183--1192. PMLR, 2017.

\bibitem{ren2021survey}
Pengzhen Ren, Yun Xiao, Xiaojun Chang, Po-Yao Huang, Zhihui Li, Brij~B Gupta,
  Xiaojiang Chen, and Xin Wang.
\newblock A survey of deep active learning.
\newblock {\em ACM computing surveys (CSUR)}, 54(9):1--40, 2021.

\bibitem{monarch2021human}
Robert~Munro Monarch.
\newblock {\em Human-in-the-Loop Machine Learning: Active learning and
  annotation for human-centered AI}.
\newblock Simon and Schuster, 2021.

\bibitem{farahani2021brief}
Abolfazl Farahani, Sahar Voghoei, Khaled Rasheed, and Hamid~R Arabnia.
\newblock A brief review of domain adaptation.
\newblock {\em Advances in Data Science and Information Engineering:
  Proceedings from ICDATA 2020 and IKE 2020}, pages 877--894, 2021.

\bibitem{Branco2016ADomains}
Paula Branco, Luís Torgo, and Rita~P. Ribeiro.
\newblock {A survey of predictive modeling on imbalanced domains}.
\newblock {\em ACM Computing Surveys}, 49(2):1--56, 2016.

\bibitem{kulesza2012determinantal}
Alex Kulesza, Ben Taskar, et~al.
\newblock Determinantal point processes for machine learning.
\newblock {\em Foundations and Trends{\textregistered} in Machine Learning},
  5(2--3):123--286, 2012.

\bibitem{rasmussen2004gaussian}
Carl~Edward Rasmussen.
\newblock Gaussian processes in machine learning.
\newblock In {\em Summer school on machine learning}, pages 63--71. Springer,
  2004.

\bibitem{rahimi2007random}
Ali Rahimi and Benjamin Recht.
\newblock Random features for large-scale kernel machines.
\newblock {\em Advances in neural information processing systems}, 20, 2007.

\bibitem{gillenwater2012near}
Jennifer Gillenwater, Alex Kulesza, and Ben Taskar.
\newblock Near-optimal map inference for determinantal point processes.
\newblock {\em Advances in Neural Information Processing Systems}, 25, 2012.

\bibitem{affandi2013approximate}
Raja~Hafiz Affandi, Emily Fox, and Ben Taskar.
\newblock Approximate inference in continuous determinantal processes.
\newblock {\em Advances in Neural Information Processing Systems}, 26, 2013.

\bibitem{lee2021dynamic}
Doksoo Lee, Shizhou Jiang, Oluwaseyi Balogun, and Wei Chen.
\newblock Dynamic control of plasmonic localization by inverse optimization of
  spatial phase modulation.
\newblock {\em ACS Photonics}, 9(2):351--359, 2021.

\bibitem{zhang2022uncertainty}
Hengrui Zhang, Wei Chen, Akshay Iyer, Daniel~W Apley, and Wei Chen.
\newblock Uncertainty-aware mixed-variable machine learning for materials
  design.
\newblock {\em Scientific reports}, 12(1):19760, 2022.

\bibitem{saal2013materials}
James~E Saal, Scott Kirklin, Muratahan Aykol, Bryce Meredig, and Christopher
  Wolverton.
\newblock Materials design and discovery with high-throughput density
  functional theory: the open quantum materials database (oqmd).
\newblock {\em Jom}, 65:1501--1509, 2013.

\bibitem{jin2003efficient}
Ruichen Jin, Wei Chen, and Agus Sudjianto.
\newblock An efficient algorithm for constructing optimal design of computer
  experiments.
\newblock In {\em International design engineering technical conferences and
  computers and information in engineering conference}, volume 37009, pages
  545--554, 2003.

\bibitem{du2004sequential}
Xiaoping Du and Wei Chen.
\newblock Sequential optimization and reliability assessment method for
  efficient probabilistic design.
\newblock {\em J. Mech. Des.}, 126(2):225--233, 2004.

\bibitem{pronzato2012design}
Luc Pronzato and Werner~G M{\"u}ller.
\newblock Design of computer experiments: space filling and beyond.
\newblock {\em Statistics and Computing}, 22:681--701, 2012.

\bibitem{shang2021fully}
Boyang Shang and Daniel~W Apley.
\newblock Fully-sequential space-filling design algorithms for computer
  experiments.
\newblock {\em Journal of Quality Technology}, 53(2):173--196, 2021.

\bibitem{cover2006elements}
Thomas~M Cover and Joy~A Thomas.
\newblock Elements of information theory second edition solutions to problems.
\newblock {\em Internet Access}, pages 19--20, 2006.

\bibitem{ashby2013designing}
Mike Ashby.
\newblock Designing architectured materials.
\newblock {\em Scripta Materialia}, 68(1):4--7, 2013.

\bibitem{wang2022data}
Liwei Wang, Anton van Beek, Daicong Da, Yu-Chin Chan, Ping Zhu, and Wei Chen.
\newblock Data-driven multiscale design of cellular composites with multiclass
  microstructures for natural frequency maximization.
\newblock {\em Composite Structures}, 280:114949, 2022.

\bibitem{xia2015design}
Liang Xia and Piotr Breitkopf.
\newblock Design of materials using topology optimization and energy-based
  homogenization approach in matlab.
\newblock {\em Structural and multidisciplinary optimization},
  52(6):1229--1241, 2015.

\bibitem{zandehshahvar2022manifold}
Mohammadreza Zandehshahvar, Yashar Kiarashinejad, Muliang Zhu, Hossein Maleki,
  Tyler Brown, and Ali Adibi.
\newblock Manifold learning for knowledge discovery and intelligent inverse
  design of photonic nanostructures: breaking the geometric complexity.
\newblock {\em ACS Photonics}, 9(2):714--721, 2022.

\bibitem{abdi2010principal}
Herv{\'e} Abdi and Lynne~J Williams.
\newblock Principal component analysis.
\newblock {\em Wiley interdisciplinary reviews: computational statistics},
  2(4):433--459, 2010.

\bibitem{van2008visualizing}
Laurens Van~der Maaten and Geoffrey Hinton.
\newblock Visualizing data using t-sne.
\newblock {\em Journal of machine learning research}, 9(11), 2008.

\bibitem{scholkopf2001estimating}
Bernhard Sch{\"o}lkopf, John~C Platt, John Shawe-Taylor, Alex~J Smola, and
  Robert~C Williamson.
\newblock Estimating the support of a high-dimensional distribution.
\newblock {\em Neural computation}, 13(7):1443--1471, 2001.

\bibitem{mcinnes2018umap}
Leland McInnes, John Healy, and James Melville.
\newblock Umap: Uniform manifold approximation and projection for dimension
  reduction.
\newblock {\em arXiv preprint arXiv:1802.03426}, 2018.

\bibitem{picard2023dated}
Cyril Picard, J{\"u}rg Schiffmann, and Faez Ahmed.
\newblock Dated: Guidelines for creating synthetic datasets for engineering
  design applications.
\newblock {\em arXiv preprint arXiv:2305.09018}, 2023.

\bibitem{ning2020low}
Shaowu Ning, Fengyuan Yang, Chengcheng Luo, Zhanli Liu, and Zhuo Zhuang.
\newblock Low-frequency tunable locally resonant band gaps in acoustic
  metamaterials through large deformation.
\newblock {\em Extreme Mechanics Letters}, 35:100623, 2020.

\bibitem{ma2022deep}
Chunping Ma, Yilong Chang, Shuai Wu, and Ruike~Renee Zhao.
\newblock Deep learning-accelerated designs of tunable magneto-mechanical
  metamaterials.
\newblock {\em ACS Applied Materials \& Interfaces}, 14(29):33892--33902, 2022.

\bibitem{zhelyeznyakov2021deep}
Maksym~V Zhelyeznyakov, Steve Brunton, and Arka Majumdar.
\newblock Deep learning to accelerate scatterer-to-field mapping for inverse
  design of dielectric metasurfaces.
\newblock {\em ACS Photonics}, 8(2):481--488, 2021.

\bibitem{an2022deep}
Sensong An, Bowen Zheng, Mikhail~Y Shalaginov, Hong Tang, Hang Li, Li~Zhou,
  Yunxi Dong, Mohammad Haerinia, Anuradha~Murthy Agarwal, Clara Rivero-Baleine,
  et~al.
\newblock Deep convolutional neural networks to predict mutual coupling effects
  in metasurfaces.
\newblock {\em Advanced Optical Materials}, 10(3):2102113, 2022.

\bibitem{ghavanloo2019wave}
Esmaeal Ghavanloo and S~Ahmad Fazelzadeh.
\newblock Wave propagation in one-dimensional infinite acoustic metamaterials
  with long-range interactions.
\newblock {\em Acta Mechanica}, 230:4453--4461, 2019.

\bibitem{zhu2020nonlocal}
Hongfei Zhu, Sansit Patnaik, Timothy~F Walsh, Bradley~H Jared, and Fabio
  Semperlotti.
\newblock Nonlocal elastic metasurfaces: Enabling broadband wave control via
  intentional nonlocality.
\newblock {\em Proceedings of the National Academy of Sciences},
  117(42):26099--26108, 2020.

\bibitem{ertsgaard2014dynamic}
Christopher~T Ertsgaard, Rachel~M McKoskey, Isabel~S Rich, and Nathan~C
  Lindquist.
\newblock Dynamic placement of plasmonic hotspots for super-resolution
  surface-enhanced raman scattering.
\newblock {\em ACS nano}, 8(10):10941--10946, 2014.

\bibitem{buijs2021programming}
Robin~D Buijs, Tom~AW Wolterink, Giampiero Gerini, Ewold Verhagen, and A~Femius
  Koenderink.
\newblock Programming metasurface near-fields for nano-optical sensing.
\newblock {\em Advanced Optical Materials}, 9(15):2100435, 2021.

\bibitem{spagele2021multifunctional}
Christina Sp{\"a}gele, Michele Tamagnone, Dmitry Kazakov, Marcus Ossiander,
  Marco Piccardo, and Federico Capasso.
\newblock Multifunctional wide-angle optics and lasing based on supercell
  metasurfaces.
\newblock {\em Nature Communications}, 12(1):1--10, 2021.

\bibitem{liu2018generative}
Zhaocheng Liu, Dayu Zhu, Sean~P Rodrigues, Kyu-Tae Lee, and Wenshan Cai.
\newblock Generative model for the inverse design of metasurfaces.
\newblock {\em Nano letters}, 18(10):6570--6576, 2018.

\bibitem{brunton2016discovering}
Steven~L Brunton, Joshua~L Proctor, and J~Nathan Kutz.
\newblock Discovering governing equations from data by sparse identification of
  nonlinear dynamical systems.
\newblock {\em Proceedings of the national academy of sciences},
  113(15):3932--3937, 2016.

\bibitem{raissi2019physics}
Maziar Raissi, Paris Perdikaris, and George~E Karniadakis.
\newblock Physics-informed neural networks: A deep learning framework for
  solving forward and inverse problems involving nonlinear partial differential
  equations.
\newblock {\em Journal of Computational physics}, 378:686--707, 2019.

\bibitem{goswami2022physics}
Somdatta Goswami, Minglang Yin, Yue Yu, and George~Em Karniadakis.
\newblock A physics-informed variational deeponet for predicting crack path in
  quasi-brittle materials.
\newblock {\em Computer Methods in Applied Mechanics and Engineering},
  391:114587, 2022.

\bibitem{zhang2022metanor}
Lu~Zhang, Huaiqian You, and Yue Yu.
\newblock Metanor: A meta-learnt nonlocal operator regression approach for
  metamaterial modeling.
\newblock {\em arXiv preprint arXiv:2206.02040}, 2022.

\bibitem{rixner2022self}
Maximilian Rixner and Phaedon-Stelios Koutsourelakis.
\newblock Self-supervised optimization of random material microstructures in
  the small-data regime.
\newblock {\em npj Computational Materials}, 8(1):46, 2022.

\bibitem{motamedi2021data}
Mohammad Motamedi, Nikolay Sakharnykh, and Tim Kaldewey.
\newblock A data-centric approach for training deep neural networks with less
  data.
\newblock {\em arXiv preprint arXiv:2110.03613}, 2021.

\bibitem{whang2023data}
Steven~Euijong Whang, Yuji Roh, Hwanjun Song, and Jae-Gil Lee.
\newblock Data collection and quality challenges in deep learning: A
  data-centric ai perspective.
\newblock {\em The VLDB Journal}, pages 1--23, 2023.

\bibitem{mazumder2022dataperf}
Mark Mazumder, Colby Banbury, Xiaozhe Yao, Bojan Karla{\v{s}}, William~Gaviria
  Rojas, Sudnya Diamos, Greg Diamos, Lynn He, Douwe Kiela, David Jurado, et~al.
\newblock Dataperf: Benchmarks for data-centric ai development.
\newblock {\em arXiv preprint arXiv:2207.10062}, 2022.

\bibitem{strickland2022andrew}
Eliza Strickland.
\newblock Andrew ng, ai minimalist: The machine-learning pioneer says small is
  the new big.
\newblock {\em IEEE Spectrum}, 59(4):22--50, 2022.

\bibitem{zhao2016perspective}
He~Zhao, Xiaolin Li, Yichi Zhang, Linda~S Schadler, Wei Chen, and L~Catherine
  Brinson.
\newblock Perspective: Nanomine: A material genome approach for polymer
  nanocomposites analysis and design.
\newblock {\em APL Materials}, 4(5):053204, 2016.

\bibitem{mccusker2020nanomine}
Jamie~P McCusker, Neha Keshan, Sabbir Rashid, Michael Deagen, Cate Brinson, and
  Deborah~L McGuinness.
\newblock Nanomine: A knowledge graph for nanocomposite materials science.
\newblock In {\em The Semantic Web--ISWC 2020: 19th International Semantic Web
  Conference, Athens, Greece, November 2--6, 2020, Proceedings, Part II}, pages
  144--159. Springer, 2020.

\bibitem{brinson2020polymer}
L~Catherine Brinson, Michael Deagen, Wei Chen, James McCusker, Deborah~L
  McGuinness, Linda~S Schadler, Marc Palmeri, Umar Ghumman, Anqi Lin, and
  Bingyin Hu.
\newblock Polymer nanocomposite data: curation, frameworks, access, and
  potential for discovery and design.
\newblock {\em ACS Macro Letters}, 9(8):1086--1094, 2020.

\bibitem{wilkinson2016fair}
Mark~D Wilkinson, Michel Dumontier, IJsbrand~Jan Aalbersberg, Gabrielle
  Appleton, Myles Axton, Arie Baak, Niklas Blomberg, Jan-Willem Boiten,
  Luiz~Bonino da~Silva~Santos, Philip~E Bourne, et~al.
\newblock The fair guiding principles for scientific data management and
  stewardship.
\newblock {\em Scientific data}, 3(1):1--9, 2016.

\bibitem{boeckhout2018fair}
Martin Boeckhout, Gerhard~A Zielhuis, and Annelien~L Bredenoord.
\newblock The fair guiding principles for data stewardship: fair enough?
\newblock {\em European journal of human genetics}, 26(7):931--936, 2018.

\bibitem{smajic2009comparison}
Jasmin Smajic, Christian Hafner, Ludmila Raguin, Kakhaber Tavzarashvili, and
  Matthew Mishrikey.
\newblock Comparison of numerical methods for the analysis of plasmonic
  structures.
\newblock {\em Journal of Computational and Theoretical Nanoscience},
  6(3):763--774, 2009.

\bibitem{poole2015metric}
Daniel~J Poole, Christian~B Allen, and Thomas~CS Rendall.
\newblock Metric-based mathematical derivation of efficient airfoil design
  variables.
\newblock {\em AIAA Journal}, 53(5):1349--1361, 2015.

\bibitem{allen2018wing}
Christian~B Allen, Daniel~J Poole, and Thomas Rendall.
\newblock Wing aerodynamic optimization using efficient
  mathematically-extracted modal design variables.
\newblock {\em Optimization and Engineering}, 19(2):453--477, 2018.

\bibitem{li2019surrogate}
Jichao Li, Jinsheng Cai, and Kun Qu.
\newblock Surrogate-based aerodynamic shape optimization with the active
  subspace method.
\newblock {\em Structural and Multidisciplinary Optimization}, 59(2):403--419,
  2019.

\bibitem{chen2020airfoil}
Wei Chen, Kevin Chiu, and Mark~D Fuge.
\newblock Airfoil design parameterization and optimization using b{\'e}zier
  generative adversarial networks.
\newblock {\em AIAA journal}, 58(11):4723--4735, 2020.

\bibitem{li2020efficient}
Jichao Li, Mengqi Zhang, Joaquim~RRA Martins, and Chang Shu.
\newblock Efficient aerodynamic shape optimization with deep-learning-based
  geometric filtering.
\newblock {\em AIAA Journal}, 58(10):4243--4259, 2020.

\bibitem{sekar2019inverse}
Vinothkumar Sekar, Mengqi Zhang, Chang Shu, and Boo~Cheong Khoo.
\newblock Inverse design of airfoil using a deep convolutional neural network.
\newblock {\em Aiaa Journal}, 57(3):993--1003, 2019.

\bibitem{chen2022inverse}
Qiuyi Chen, Jun Wang, Phillip Pope, Mark Fuge, et~al.
\newblock Inverse design of two-dimensional airfoils using conditional
  generative models and surrogate log-likelihoods.
\newblock {\em Journal of Mechanical Design}, 144(2), 2022.

\bibitem{glaws2022invertible}
Andrew Glaws, Ryan~N King, Ganesh Vijayakumar, and Shreyas Ananthan.
\newblock Invertible neural networks for airfoil design.
\newblock {\em AIAA Journal}, 60(5):3035--3047, 2022.

\bibitem{azad2016metasurface}
Abul~K Azad, Wilton~JM Kort-Kamp, Milan Sykora, Nina~R Weisse-Bernstein, Ting~S
  Luk, Antoinette~J Taylor, Diego~AR Dalvit, and Hou-Tong Chen.
\newblock Metasurface broadband solar absorber.
\newblock {\em Scientific reports}, 6(1):20347, 2016.

\bibitem{jiang2022dispersion}
Weifeng Jiang, Yangyang Zhu, Guofu Yin, Houhong Lu, Luofeng Xie, and Ming Yin.
\newblock Dispersion relation prediction and structure inverse design of
  elastic metamaterials via deep learning.
\newblock {\em Materials Today Physics}, 22:100616, 2022.

\bibitem{gu2018bioinspired}
Grace~X Gu, Chun-Teh Chen, Deon~J Richmond, and Markus~J Buehler.
\newblock Bioinspired hierarchical composite design using machine learning:
  simulation, additive manufacturing, and experiment.
\newblock {\em Materials Horizons}, 5(5):939--945, 2018.

\bibitem{sajedian2019finding}
Iman Sajedian, Jeonghyun Kim, and Junsuk Rho.
\newblock Finding the optical properties of plasmonic structures by image
  processing using a combination of convolutional neural networks and recurrent
  neural networks.
\newblock {\em Microsystems \& nanoengineering}, 5(1):1--8, 2019.

\bibitem{wiecha2019deep}
Peter~R Wiecha and Otto~L Muskens.
\newblock Deep learning meets nanophotonics: a generalized accurate predictor
  for near fields and far fields of arbitrary 3d nanostructures.
\newblock {\em Nano letters}, 20(1):329--338, 2019.

\bibitem{an2020deep}
Sensong An, Bowen Zheng, Mikhail~Y Shalaginov, Hong Tang, Hang Li, Li~Zhou, Jun
  Ding, Anuradha~Murthy Agarwal, Clara Rivero-Baleine, Myungkoo Kang, et~al.
\newblock Deep learning modeling approach for metasurfaces with high degrees of
  freedom.
\newblock {\em Optics Express}, 28(21):31932--31942, 2020.

\bibitem{donda2021ultrathin}
Krupali Donda, Yifan Zhu, Aur{\'e}lien Merkel, Shi-Wang Fan, Liyun Cao, Sheng
  Wan, and Badreddine Assouar.
\newblock Ultrathin acoustic absorbing metasurface based on deep learning
  approach.
\newblock {\em Smart Materials and Structures}, 30(8):085003, 2021.

\bibitem{garland2021pragmatic}
Anthony~P Garland, Benjamin~C White, Scott~C Jensen, and Brad~L Boyce.
\newblock Pragmatic generative optimization of novel structural lattice
  metamaterials with machine learning.
\newblock {\em Materials \& Design}, 203:109632, 2021.

\bibitem{zhang2021genetic}
Junming Zhang, Guowu Wang, Tao Wang, and Fashen Li.
\newblock Genetic algorithms to automate the design of metasurfaces for
  absorption bandwidth broadening.
\newblock {\em ACS Applied Materials \& Interfaces}, 13(6):7792--7800, 2021.

\bibitem{ji2022design}
Qingxiang Ji, Yunchao Qi, Chenwei Liu, Songhe Meng, Jun Liang, Muamer Kadic,
  and Guodong Fang.
\newblock Design of thermal cloaks with isotropic materials based on machine
  learning.
\newblock {\em International Journal of Heat and Mass Transfer}, 189:122716,
  2022.

\bibitem{pahlavani2022deep}
Helda Pahlavani, Muhamad Amani, Mauricio~Cruz Sald{\'\i}var, Jie Zhou,
  Mohammad~J Mirzaali, and Amir~A Zadpoor.
\newblock Deep learning for the rare-event rational design of 3d printed
  multi-material mechanical metamaterials.
\newblock {\em Communications Materials}, 3(1):1--11, 2022.

\bibitem{lee2022generative}
Sangryun Lee, Zhizhou Zhang, and Grace~X Gu.
\newblock Generative machine learning algorithm for lattice structures with
  superior mechanical properties.
\newblock {\em Materials Horizons}, 9(3):952--960, 2022.

\bibitem{wang2022design}
Zihan Wang, Weikang Xian, M~Ridha Baccouche, Horst Lanzerath, Ying Li, and
  Hongyi Xu.
\newblock Design of phononic bandgap metamaterials based on gaussian mixture
  beta variational autoencoder and iterative model updating.
\newblock {\em Journal of Mechanical Design}, 144(4):041705, 2022.

\bibitem{hensman2013gaussian}
James Hensman, Nicolo Fusi, and Neil~D Lawrence.
\newblock Gaussian processes for big data.
\newblock {\em arXiv preprint arXiv:1309.6835}, 2013.

\bibitem{evans2018scalable}
Trefor Evans and Prasanth Nair.
\newblock Scalable gaussian processes with grid-structured eigenfunctions
  (gp-grief).
\newblock In {\em International Conference on Machine Learning}, pages
  1417--1426. PMLR, 2018.

\bibitem{bostanabad2019globally}
Ramin Bostanabad, Yu-Chin Chan, Liwei Wang, Ping Zhu, and Wei Chen.
\newblock {Globally approximate gaussian processes for big data with
  application to data-driven metamaterials design}.
\newblock {\em Journal of Mechanical Design}, 141(11), 2019.

\bibitem{wang2019exact}
Ke~Wang, Geoff Pleiss, Jacob Gardner, Stephen Tyree, Kilian~Q Weinberger, and
  Andrew~Gordon Wilson.
\newblock Exact gaussian processes on a million data points.
\newblock {\em Advances in Neural Information Processing Systems}, 32, 2019.

\bibitem{liu2020gaussian}
Haitao Liu, Yew-Soon Ong, Xiaobo Shen, and Jianfei Cai.
\newblock When gaussian process meets big data: A review of scalable gps.
\newblock {\em IEEE transactions on neural networks and learning systems},
  31(11):4405--4423, 2020.

\bibitem{Chen2018ComputationalFamilies}
Desai Chen, Mélina Skouras, Bo~Zhu, and Wojciech Matusik.
\newblock {Computational discovery of extremal microstructure families}.
\newblock {\em Science Advances}, 4(1):1--8, 2018.

\bibitem{elzouka2020interpretable}
Mahmoud Elzouka, Charles Yang, Adrian Albert, Ravi~S Prasher, and Sean~D
  Lubner.
\newblock Interpretable forward and inverse design of particle spectral
  emissivity using common machine-learning models.
\newblock {\em Cell Reports Physical Science}, 1(12):100259, 2020.

\bibitem{lin2020generalized}
Jimmy Lin, Chudi Zhong, Diane Hu, Cynthia Rudin, and Margo Seltzer.
\newblock Generalized and scalable optimal sparse decision trees.
\newblock In {\em International Conference on Machine Learning}, pages
  6150--6160. PMLR, 2020.

\bibitem{yang2022high}
Zhenze Yang and Markus~J Buehler.
\newblock High-throughput generation of 3d graphene metamaterials and property
  quantification using machine learning.
\newblock {\em Small Methods}, 6(9):2200537, 2022.

\bibitem{chen2022gan}
Wei Chen, Doksoo Lee, Oluwaseyi Balogun, and Wei Chen.
\newblock Gan-duf: Hierarchical deep generative models for design under
  free-form geometric uncertainty.
\newblock {\em Journal of Mechanical Design}, 145(1):011703, 2022.

\bibitem{shen2022nature}
Sabrina Chin-yun Shen and Markus~J Buehler.
\newblock Nature-inspired architected materials using unsupervised deep
  learning.
\newblock {\em Communications Engineering}, 1(1):1--15, 2022.

\bibitem{karras2019style}
Tero Karras, Samuli Laine, and Timo Aila.
\newblock A style-based generator architecture for generative adversarial
  networks.
\newblock In {\em Proceedings of the IEEE/CVF conference on computer vision and
  pattern recognition}, pages 4401--4410, 2019.

\bibitem{liu2021thermal}
Bin Liu, Liujun Xu, and Jiping Huang.
\newblock Thermal transparency with periodic particle distribution: A machine
  learning approach.
\newblock {\em Journal of Applied Physics}, 129(6):065101, 2021.

\bibitem{van2016deep}
Hado Van~Hasselt, Arthur Guez, and David Silver.
\newblock Deep reinforcement learning with double q-learning.
\newblock In {\em Proceedings of the AAAI conference on artificial
  intelligence}, volume~30, 2016.

\bibitem{raina2022learning}
Ayush Raina, Jonathan Cagan, and Christopher Mccomb.
\newblock Learning to design without prior data: Discovering generalizable
  design strategies using deep learning and tree search.
\newblock {\em Journal of Mechanical Design}, pages 1--38, 2022.

\bibitem{jiang2019global}
Jiaqi Jiang and Jonathan~A Fan.
\newblock Global optimization of dielectric metasurfaces using a physics-driven
  neural network.
\newblock {\em Nano letters}, 19(8):5366--5372, 2019.

\bibitem{jiang2020simulator}
Jiaqi Jiang and Jonathan~A Fan.
\newblock Simulator-based training of generative neural networks for the
  inverse design of metasurfaces.
\newblock {\em Nanophotonics}, 9(5):1059--1069, 2020.

\bibitem{lu2021physics}
Lu~Lu, Raphael Pestourie, Wenjie Yao, Zhicheng Wang, Francesc Verdugo, and
  Steven~G Johnson.
\newblock Physics-informed neural networks with hard constraints for inverse
  design.
\newblock {\em SIAM Journal on Scientific Computing}, 43(6):B1105--B1132, 2021.

\bibitem{kollmann2020deep}
Hunter~T Kollmann, Diab~W Abueidda, Seid Koric, Erman Guleryuz, and Nahil~A
  Sobh.
\newblock Deep learning for topology optimization of 2d metamaterials.
\newblock {\em Materials \& Design}, 196:109098, 2020.

\bibitem{zhao2021machine}
Tianyu Zhao, Yiwen Li, Lei Zuo, and Kai Zhang.
\newblock Machine-learning optimized method for regional control of sound
  fields.
\newblock {\em Extreme Mechanics Letters}, 45:101297, 2021.

\bibitem{liu2021intelligent}
Che Liu, Wen~Ming Yu, Qian Ma, Lianlin Li, and Tie~Jun Cui.
\newblock Intelligent coding metasurface holograms by physics-assisted
  unsupervised generative adversarial network.
\newblock {\em Photonics Research}, 9(4):B159--B167, 2021.

\bibitem{liu2018training}
Dianjing Liu, Yixuan Tan, Erfan Khoram, and Zongfu Yu.
\newblock Training deep neural networks for the inverse design of nanophotonic
  structures.
\newblock {\em Acs Photonics}, 5(4):1365--1369, 2018.

\bibitem{so2021demand}
Sunae So, Younghwan Yang, Taejun Lee, and Junsuk Rho.
\newblock On-demand design of spectrally sensitive multiband absorbers using an
  artificial neural network.
\newblock {\em Photonics Research}, 9(4):B153--B158, 2021.

\bibitem{Yeung2021MultiplexedNetworks}
Christopher Yeung, Ju~Ming Tsai, Brian King, Benjamin Pham, David Ho, Julia
  Liang, Mark~W. Knight, and Aaswath~P. Raman.
\newblock {Multiplexed supercell metasurface design and optimization with
  tandem residual networks}.
\newblock {\em Nanophotonics}, 10(3):1133--1143, 1 2021.

\bibitem{zhen2021realizing}
Zheng Zhen, Chao Qian, Yuetian Jia, Zhixiang Fan, Ran Hao, Tong Cai, Bin Zheng,
  Hongsheng Chen, and Erping Li.
\newblock Realizing transmitted metasurface cloak by a tandem neural network.
\newblock {\em Photonics Research}, 9(5):B229--B235, 2021.

\bibitem{mirza2014conditional}
Mehdi Mirza and Simon Osindero.
\newblock Conditional generative adversarial nets.
\newblock {\em arXiv preprint arXiv:1411.1784}, 2014.

\bibitem{sohn2015learning}
Kihyuk Sohn, Honglak Lee, and Xinchen Yan.
\newblock Learning structured output representation using deep conditional
  generative models.
\newblock {\em Advances in neural information processing systems}, 28, 2015.

\bibitem{jiang2019free}
Jiaqi Jiang, David Sell, Stephan Hoyer, Jason Hickey, Jianji Yang, and
  Jonathan~A Fan.
\newblock Free-form diffractive metagrating design based on generative
  adversarial networks.
\newblock {\em ACS nano}, 13(8):8872--8878, 2019.

\bibitem{wang2020deeplearning}
Hai~Peng Wang, Yun~Bo Li, He~Li, Shu~Yue Dong, Che Liu, Shi Jin, and Tie~Jun
  Cui.
\newblock Deep learning designs of anisotropic metasurfaces in ultrawideband
  based on generative adversarial networks.
\newblock {\em Advanced Intelligent Systems}, 2(9):2000068, 2020.

\bibitem{wen2020robust}
Fufang Wen, Jiaqi Jiang, and Jonathan~A Fan.
\newblock Robust freeform metasurface design based on progressively growing
  generative networks.
\newblock {\em ACS Photonics}, 7(8):2098--2104, 2020.

\bibitem{luo2020probability}
Ying-Tao Luo, Peng-Qi Li, Dong-Ting Li, Yu-Gui Peng, Zhi-Guo Geng, Shu-Huan
  Xie, Yong Li, Andrea Al{\`u}, Jie Zhu, and Xue-Feng Zhu.
\newblock Probability-density-based deep learning paradigm for the fuzzy design
  of functional metastructures.
\newblock {\em Research}, 2020, 2020.

\bibitem{gu2018novo}
Grace~X Gu, Chun-Teh Chen, and Markus~J Buehler.
\newblock De novo composite design based on machine learning algorithm.
\newblock {\em Extreme Mechanics Letters}, 18:19--28, 2018.

\bibitem{barri2021multifunctional}
Kaveh Barri, Pengcheng Jiao, Qianyun Zhang, Jun Chen, Zhong~Lin Wang, and
  Amir~H Alavi.
\newblock Multifunctional meta-tribomaterial nanogenerators for energy
  harvesting and active sensing.
\newblock {\em Nano Energy}, 86:106074, 2021.

\bibitem{wang2019robust}
Evan~W Wang, David Sell, Thaibao Phan, and Jonathan~A Fan.
\newblock Robust design of topology-optimized metasurfaces.
\newblock {\em Optical Materials Express}, 9(2):469--482, 2019.

\bibitem{roxworthy2014reconfigurable}
Brian~J Roxworthy, Abdul~M Bhuiya, Xin Yu, Edmond~KC Chow, and Kimani~C
  Toussaint~Jr.
\newblock Reconfigurable nanoantennas using electron-beam manipulation.
\newblock {\em Nature communications}, 5(1):4427, 2014.

\bibitem{chen2019looks}
Chaofan Chen, Oscar Li, Daniel Tao, Alina Barnett, Cynthia Rudin, and
  Jonathan~K Su.
\newblock This looks like that: deep learning for interpretable image
  recognition.
\newblock {\em Advances in neural information processing systems}, 32, 2019.

\bibitem{yang2021gami}
Zebin Yang, Aijun Zhang, and Agus Sudjianto.
\newblock Gami-net: An explainable neural network based on generalized additive
  models with structured interactions.
\newblock {\em Pattern Recognition}, 120:108192, 2021.

\bibitem{milton2006cloaking}
Graeme~W Milton, Marc Briane, and John~R Willis.
\newblock On cloaking for elasticity and physical equations with a
  transformation invariant form.
\newblock {\em New journal of physics}, 8(10):248, 2006.

\bibitem{zuo2017multi}
Wenjie Zuo and Kazuhiro Saitou.
\newblock Multi-material topology optimization using ordered simp
  interpolation.
\newblock {\em Structural and Multidisciplinary Optimization}, 55(2):477--491,
  2017.

\bibitem{gu2021material}
Dongdong Gu, Xinyu Shi, Reinhart Poprawe, David~L Bourell, Rossitza Setchi, and
  Jihong Zhu.
\newblock Material-structure-performance integrated laser-metal additive
  manufacturing.
\newblock {\em Science}, 372(6545):eabg1487, 2021.

\bibitem{garner2019compatibility}
Eric Garner, Helena~MA Kolken, Charlie~CL Wang, Amir~A Zadpoor, and Jun Wu.
\newblock Compatibility in microstructural optimization for additive
  manufacturing.
\newblock {\em Additive Manufacturing}, 26:65--75, 2019.

\bibitem{sydney2016biomimetic}
A~Sydney~Gladman, Elisabetta~A Matsumoto, Ralph~G Nuzzo, Lakshminarayanan
  Mahadevan, and Jennifer~A Lewis.
\newblock Biomimetic 4d printing.
\newblock {\em Nature materials}, 15(4):413--418, 2016.

\bibitem{li2022empowering}
Zhaoyi Li, Rapha{\"e}l Pestourie, Zin Lin, Steven~G Johnson, and Federico
  Capasso.
\newblock Empowering metasurfaces with inverse design: principles and
  applications.
\newblock {\em ACS Photonics}, 9(7):2178--2192, 2022.

\bibitem{wang2022mechanical}
Liwei Wang, Jagannadh Boddapati, Ke~Liu, Ping Zhu, Chiara Daraio, and Wei Chen.
\newblock Mechanical cloak via data-driven aperiodic metamaterial design.
\newblock {\em Proceedings of the National Academy of Sciences},
  119(13):e2122185119, 2022.

\bibitem{buckmann2014elasto}
Tiemo B{\"u}ckmann, Michael Thiel, Muamer Kadic, Robert Schittny, and Martin
  Wegener.
\newblock An elasto-mechanical unfeelability cloak made of pentamode
  metamaterials.
\newblock {\em Nature communications}, 5(1):1--6, 2014.

\bibitem{boley2019shape}
J~William Boley, Wim~M Van~Rees, Charles Lissandrello, Mark~N Horenstein,
  Ryan~L Truby, Arda Kotikian, Jennifer~A Lewis, and L~Mahadevan.
\newblock Shape-shifting structured lattices via multimaterial 4d printing.
\newblock {\em Proceedings of the National Academy of Sciences},
  116(42):20856--20862, 2019.

\bibitem{Panetta2015ElasticFabrication}
Julian Panetta, Qingnan Zhou, Luigi Malomo, Nico Pietroni, Paolo Cignoni, and
  Denis Zorin.
\newblock {Elastic textures for additive fabrication}.
\newblock {\em ACM Transactions on Graphics}, 34(4):1--12, 2015.

\bibitem{schumacher2015microstructures}
Christian Schumacher, Bernd Bickel, Jan Rys, Steve Marschner, Chiara Daraio,
  and Markus Gross.
\newblock {Microstructures to control elasticity in 3D printing}.
\newblock {\em ACM Transactions on Graphics (TOG)}, 34(4):1--13, 2015.

\bibitem{wu2018multiscale}
Tong Wu and Andres Tovar.
\newblock Multiscale, thermomechanical topology optimization of self-supporting
  cellular structures for porous injection molds.
\newblock {\em Rapid Prototyping Journal}, 2018.

\bibitem{deng2013multi}
Jiadong Deng, Jun Yan, and Gengdong Cheng.
\newblock Multi-objective concurrent topology optimization of thermoelastic
  structures composed of homogeneous porous material.
\newblock {\em Structural and Multidisciplinary Optimization}, 47(4):583--597,
  2013.

\bibitem{yan2016multi}
Jun Yan, Xu~Guo, and Gengdong Cheng.
\newblock Multi-scale concurrent material and structural design under
  mechanical and thermal loads.
\newblock {\em Computational Mechanics}, 57(3):437--446, 2016.

\bibitem{kadic2020elastodynamic}
Muamer Kadic, Martin Wegener, Andr{\'e} Nicolet, Fr{\'e}d{\'e}ric Zolla,
  S{\'e}bastien Guenneau, and Andr{\'e} Diatta.
\newblock Elastodynamic behavior of mechanical cloaks designed by direct
  lattice transformations.
\newblock {\em Wave Motion}, 92:102419, 2020.

\bibitem{zhang2020multiscale}
Yan Zhang, Mi~Xiao, Liang Gao, Jie Gao, and Hao Li.
\newblock Multiscale topology optimization for minimizing frequency responses
  of cellular composites with connectable graded microstructures.
\newblock {\em Mechanical Systems and Signal Processing}, 135:106369, 2020.

\bibitem{zhao2019efficient}
Junpeng Zhao, Heonjun Yoon, and Byeng~D Youn.
\newblock An efficient concurrent topology optimization approach for frequency
  response problems.
\newblock {\em Computer Methods in Applied Mechanics and Engineering},
  347:700--734, 2019.

\bibitem{thomsen2018buckling}
Christian~Rye Thomsen, Fengwen Wang, and Ole Sigmund.
\newblock Buckling strength topology optimization of 2d periodic materials
  based on linearized bifurcation analysis.
\newblock {\em Computer Methods in Applied Mechanics and Engineering},
  339:115--136, 2018.

\bibitem{wang20213d}
Fengwen Wang and Ole Sigmund.
\newblock 3d architected isotropic materials with tunable stiffness and
  buckling strength.
\newblock {\em Journal of the Mechanics and Physics of Solids}, 152:104415,
  2021.

\bibitem{pham2019damage}
Minh-Son Pham, Chen Liu, Iain Todd, and Jedsada Lertthanasarn.
\newblock Damage-tolerant architected materials inspired by crystal
  microstructure.
\newblock {\em Nature}, 565(7739):305--311, 2019.

\bibitem{yin2021strong}
Sha Yin, Weihua Guo, Huitian Wang, Yao Huang, Ruiheng Yang, Zihan Hu, Dianhao
  Chen, Jun Xu, and Robert~O Ritchie.
\newblock Strong and tough bioinspired additive-manufactured dual-phase
  mechanical metamaterial composites.
\newblock {\em Journal of the Mechanics and Physics of Solids}, 149:104341,
  2021.

\bibitem{jiang2021tailoring}
Huan Jiang, Aaron Coomes, Zhennan Zhang, Hannah Ziegler, and Yanyu Chen.
\newblock Tailoring 3d printed graded architected polymer foams for enhanced
  energy absorption.
\newblock {\em Composites Part B: Engineering}, 224:109183, 2021.

\bibitem{cheng2019functionally}
Lin Cheng, Jiaxi Bai, and Albert~C To.
\newblock {Functionally graded lattice structure topology optimization for the
  design of additive manufactured components with stress constraints}.
\newblock {\em Computer Methods in Applied Mechanics and Engineering},
  344:334--359, 2019.

\bibitem{liu2022kriging}
Xiliang Liu, Liang Gao, Mi~Xiao, and Yan Zhang.
\newblock Kriging-assisted design of functionally graded cellular structures
  with smoothly-varying lattice unit cells.
\newblock {\em Computer Methods in Applied Mechanics and Engineering},
  390:114466, 2022.

\bibitem{li2019design}
Dawei Li, Ning Dai, Yunlong Tang, Guoying Dong, and Yaoyao~Fiona Zhao.
\newblock Design and optimization of graded cellular structures with triply
  periodic level surface-based topological shapes.
\newblock {\em Journal of Mechanical Design}, 141(7), 2019.

\bibitem{wang2017multiscale}
Yingjun Wang, Hang Xu, and Damiano Pasini.
\newblock Multiscale isogeometric topology optimization for lattice materials.
\newblock {\em Computer Methods in Applied Mechanics and Engineering},
  316:568--585, 2017.

\bibitem{wu2019topology}
Zijun Wu, Liang Xia, Shuting Wang, and Tielin Shi.
\newblock Topology optimization of hierarchical lattice structures with
  substructuring.
\newblock {\em Computer Methods in Applied Mechanics and Engineering},
  345:602--617, 2019.

\bibitem{white2019multiscale}
Daniel~A White, William~J Arrighi, Jun Kudo, and Seth~E Watts.
\newblock Multiscale topology optimization using neural network surrogate
  models.
\newblock {\em Computer Methods in Applied Mechanics and Engineering},
  346:1118--1135, 2019.

\bibitem{groen2018multi}
Jeroen~Peter Groen.
\newblock {\em Multi-scale design methods for topology optimization}.
\newblock DTU Mechanical Engineering, 2018.

\bibitem{avellaneda1987optimal}
Marco Avellaneda.
\newblock Optimal bounds and microgeometries for elastic two-phase composites.
\newblock {\em SIAM Journal on Applied Mathematics}, 47(6):1216--1228, 1987.

\bibitem{WuZijun2020Tsto}
Zijun Wu, Fei Fan, Renbin Xiao, and Lianqing Yu.
\newblock {The substructuring‐based topology optimization for maximizing the
  first eigenvalue of hierarchical lattice structure}.
\newblock {\em International journal for numerical methods in engineering},
  121(13):2964--2978, 2020.

\bibitem{vogiatzis2018computational}
Panagiotis Vogiatzis, Ming Ma, Shikui Chen, and Xianfeng~David Gu.
\newblock Computational design and additive manufacturing of periodic conformal
  metasurfaces by synthesizing topology optimization with conformal mapping.
\newblock {\em Computer Methods in Applied Mechanics and Engineering},
  328:477--497, 2018.

\bibitem{jiang2021generative}
Long Jiang, Xianfeng~David Gu, and Shikui Chen.
\newblock Generative design of bionic structures via concurrent multiscale
  topology optimization and conformal geometry method.
\newblock {\em Journal of Mechanical Design}, 143(1), 2021.

\bibitem{ma2022compliance}
Chuang Ma, Dingchuan Xue, Shaoshuai Li, Zhengcheng Zhou, Yichao Zhu, and
  Xu~Guo.
\newblock Compliance minimisation of smoothly varying multiscale structures
  using asymptotic analysis and machine learning.
\newblock {\em Computer Methods in Applied Mechanics and Engineering},
  395:114861, 2022.

\bibitem{pantz2008post}
Olivier Pantz and Karim Trabelsi.
\newblock {A post-treatment of the homogenization method for shape
  optimization}.
\newblock {\em SIAM Journal on Control and Optimization}, 47(3):1380--1398,
  2008.

\bibitem{groen2018homogenization}
Jeroen~P Groen and Ole Sigmund.
\newblock Homogenization-based topology optimization for high-resolution
  manufacturable microstructures.
\newblock {\em International Journal for Numerical Methods in Engineering},
  113(8):1148--1163, 2018.

\bibitem{Groen2020De-homogenizationTopologies}
Jeroen~P. Groen, Florian~C. Stutz, Niels Aage, Jakob~A. B{\ae}rentzen, and Ole
  Sigmund.
\newblock {De-homogenization of optimal multi-scale 3D topologies}.
\newblock {\em Computer Methods in Applied Mechanics and Engineering},
  364:112979, 2020.

\bibitem{Groen2021Multi-scaleModeling}
J.~P. Groen, C.~R. Thomsen, and O.~Sigmund.
\newblock {Multi-scale topology optimization for stiffness and
  de-homogenization using implicit geometry modeling}.
\newblock {\em Structural and Multidisciplinary Optimization},
  63(6):2919--2934, 2021.

\bibitem{elingaard2022homogenization}
Martin~Ohrt Elingaard, Niels Aage, Jakob~Andreas B{\ae}rentzen, and Ole
  Sigmund.
\newblock De-homogenization using convolutional neural networks.
\newblock {\em Computer Methods in Applied Mechanics and Engineering},
  388:114197, 2022.

\bibitem{Geoffroy-DondersPerle2020Coom}
Perle Geoffroy-Donders, Gregoire Allaire, Georgios Michailidis, and Olivier
  Pantz.
\newblock {Coupled optimization of macroscopic structures and lattice infill}.
\newblock {\em International journal for numerical methods in engineering},
  2020.

\bibitem{Geoffroy-DondersPerle20203too}
Perle Geoffroy-Donders, Grégoire Allaire, and Olivier Pantz.
\newblock {3-d topology optimization of modulated and oriented periodic
  microstructures by the homogenization method}.
\newblock {\em Journal of computational physics}, 401:108994--, 2020.

\bibitem{TamijaniAliY2020Tamd}
Ali~Y Tamijani, Shajayra~Patricia Velasco, and Lee Alacoque.
\newblock {Topological and morphological design of additively-manufacturable
  spatially-varying periodic cellular solids}.
\newblock {\em Materials {\&} design}, 196:109155--, 2020.

\bibitem{jensen2022homogenization}
Peter D{\o}rffler~Ladegaard Jensen, Ole Sigmund, and Jeroen~P Groen.
\newblock De-homogenization of optimal 2d topologies for multiple loading
  cases.
\newblock {\em Computer Methods in Applied Mechanics and Engineering},
  399:115426, 2022.

\bibitem{KumarTej2019AdKa}
Tej Kumar and Krishnan Suresh.
\newblock {A density-and-strain-based K-clustering approach to microstructural
  topology optimization}.
\newblock {\em Structural and multidisciplinary optimization},
  61(4):1399--1415, 2019.

\bibitem{liu2020data}
Zhen Liu, Liang Xia, Qi~Xia, and Tielin Shi.
\newblock {Data-driven design approach to hierarchical hybrid structures with
  multiple lattice configurations}.
\newblock {\em Structural and Multidisciplinary Optimization},
  61(6):2227--2235, 2020.

\bibitem{wang2018concurrent}
Chuang Wang, Ji~Hong Zhu, Wei~Hong Zhang, Shao~Ying Li, and Jie Kong.
\newblock {Concurrent topology optimization design of structures and
  non-uniform parameterized lattice microstructures}.
\newblock {\em Structural and Multidisciplinary Optimization}, 58(1):35--50,
  2018.

\bibitem{Chan2021RemixingBlending}
Yu-Chin Chan, Daicong Da, Liwei Wang, and Wei Chen.
\newblock {Remixing Functionally Graded Structures: Data-Driven Topology
  Optimization with Multiclass Shape Blending}.
\newblock {\em Structural and Multidisciplinary Optimization}, 2021.

\bibitem{ZhengLi2021Dtoo}
Li~Zheng, Siddhant Kumar, and Dennis~M Kochmann.
\newblock {Data-driven topology optimization of spinodoid metamaterials with
  seamlessly tunable anisotropy}.
\newblock {\em Computer methods in applied mechanics and engineering},
  383:113894--, 2021.

\bibitem{senhora2022optimally}
Fernando~V Senhora, Emily~D Sanders, and Glaucio~H Paulino.
\newblock Optimally-tailored spinodal architected materials for multiscale
  design and manufacturing.
\newblock {\em Advanced Materials}, page 2109304, 2022.

\bibitem{wang2020concurrent}
Chuang Wang, Xiaojun Gu, Jihong Zhu, Han Zhou, Shaoying Li, and Weihong Zhang.
\newblock Concurrent design of hierarchical structures with three-dimensional
  parameterized lattice microstructures for additive manufacturing.
\newblock {\em Structural and Multidisciplinary Optimization}, 61(3):869--894,
  2020.

\bibitem{zhang2019concurrent}
Yan Zhang, Hao Li, Mi~Xiao, Liang Gao, Sheng Chu, and Jinhao Zhang.
\newblock Concurrent topology optimization for cellular structures with
  nonuniform microstructures based on the kriging metamodel.
\newblock {\em Structural and Multidisciplinary Optimization},
  59(4):1273--1299, 2019.

\bibitem{zhang2020maximizing}
Yan Zhang, Liang Gao, and Mi~Xiao.
\newblock {Maximizing natural frequencies of inhomogeneous cellular structures
  by Kriging-assisted multiscale topology optimization}.
\newblock {\em Computers {\textbackslash}{\&} Structures}, 230:106197, 2020.

\bibitem{soyarslan20183d}
Celal Soyarslan, Swantje Bargmann, Marc Pradas, and J{\"o}rg Weissm{\"u}ller.
\newblock 3d stochastic bicontinuous microstructures: Generation, topology and
  elasticity.
\newblock {\em Acta materialia}, 149:326--340, 2018.

\bibitem{buckmann2015mechanical}
Tiemo B{\"u}ckmann, Muamer Kadic, Robert Schittny, and Martin Wegener.
\newblock Mechanical cloak design by direct lattice transformation.
\newblock {\em Proceedings of the National Academy of Sciences},
  112(16):4930--4934, 2015.

\bibitem{BickelBernd2010Dafo}
Bernd Bickel, Moritz B{\"{a}}cher, Miguel Otaduy, Hyunho Lee, Hanspeter
  Pfister, Markus Gross, and Wojciech Matusik.
\newblock {Design and fabrication of materials with desired deformation
  behavior}.
\newblock {\em ACM transactions on graphics}, 29(4):1--10, 2010.

\bibitem{ChuChen2008DfAM}
Chen Chu, Greg Graf, and David~W Rosen.
\newblock {Design for Additive Manufacturing of Cellular Structures}.
\newblock {\em Computer-aided design and applications}, 5(5):686--696, 2008.

\bibitem{CramerAndrewD2015Mifm}
Andrew~D Cramer, Vivien~J Challis, and Anthony~P Roberts.
\newblock {Microstructure interpolation for macroscopic design}.
\newblock {\em Structural and multidisciplinary optimization}, 53(3):489--500,
  2015.

\bibitem{HanYafeng2018ANDM}
Yafeng Han and Wen~Feng Lu.
\newblock {A Novel Design Method for Nonuniform Lattice Structures Based on
  Topology Optimization}.
\newblock {\em Journal of mechanical design (1990)}, 140(9), 2018.

\bibitem{MironovVladimir2009OpTs}
Vladimir Mironov, Richard~P Visconti, Vladimir Kasyanov, Gabor Forgacs,
  Christopher~J Drake, and Roger~R Markwald.
\newblock {Organ printing: Tissue spheroids as building blocks}.
\newblock {\em Biomaterials}, 30(12):2164--2174, 2009.

\bibitem{chen2018computational}
Desai Chen, M{\'e}lina Skouras, Bo~Zhu, and Wojciech Matusik.
\newblock Computational discovery of extremal microstructure families.
\newblock {\em Science advances}, 4(1):eaao7005, 2018.

\bibitem{DaDaicong2022Datd}
Daicong Da, Yu-Chin Chan, Liwei Wang, and Wei Chen.
\newblock {Data-driven and topological design of structural metamaterials for
  fracture resistance}.
\newblock {\em Extreme Mechanics Letters}, 50:101528--, 2022.

\bibitem{wolfram1984cellular}
Stephen Wolfram.
\newblock Cellular automata as models of complexity.
\newblock {\em Nature}, 311(5985):419--424, 1984.

\bibitem{nguyen2011multiscale}
Vinh~Phu Nguyen, Martijn Stroeven, and Lambertus~Johannes Sluys.
\newblock Multiscale continuous and discontinuous modeling of heterogeneous
  materials: a review on recent developments.
\newblock {\em Journal of Multiscale Modelling}, 3(04):229--270, 2011.

\bibitem{wallin2020nonlinear}
Mathias Wallin and Daniel~A Tortorelli.
\newblock Nonlinear homogenization for topology optimization.
\newblock {\em Mechanics of Materials}, 145:103324, 2020.

\bibitem{nakshatrala2013nonlinear}
Praveen~Babu Nakshatrala, DA~Tortorelli, and KB3069875 Nakshatrala.
\newblock Nonlinear structural design using multiscale topology optimization.
  part i: Static formulation.
\newblock {\em Computer Methods in Applied Mechanics and Engineering},
  261:167--176, 2013.

\bibitem{hassani1998review}
Behrooz Hassani and Ernest Hinton.
\newblock A review of homogenization and topology optimization
  i—homogenization theory for media with periodic structure.
\newblock {\em Computers \& Structures}, 69(6):707--717, 1998.

\bibitem{da2021design}
Daicong Da and Liang Xia.
\newblock Design of heterogeneous mesostructures for nonseparated scales and
  analysis of size effects.
\newblock {\em International Journal for Numerical Methods in Engineering},
  122(5):1333--1351, 2021.

\bibitem{fu2019topology}
Junjian Fu, Liang Xia, Liang Gao, Mi~Xiao, and Hao Li.
\newblock Topology optimization of periodic structures with substructuring.
\newblock {\em Journal of Mechanical Design}, 141(7), 2019.

\bibitem{yao2020fea}
Houpu Yao, Yi~Gao, and Yongming Liu.
\newblock {FEA-Net: A physics-guided data-driven model for efficient mechanical
  response prediction}.
\newblock {\em Computer Methods in Applied Mechanics and Engineering},
  363:112892, 2020.

\bibitem{saha2021hierarchical}
Sourav Saha, Zhengtao Gan, Lin Cheng, Jiaying Gao, Orion~L Kafka, Xiaoyu Xie,
  Hengyang Li, Mahsa Tajdari, H~Alicia Kim, and Wing~Kam Liu.
\newblock Hierarchical deep learning neural network (hidenn): An artificial
  intelligence (ai) framework for computational science and engineering.
\newblock {\em Computer Methods in Applied Mechanics and Engineering},
  373:113452, 2021.

\bibitem{efendiev2009multiscale}
Yalchin Efendiev and Thomas~Y Hou.
\newblock {\em Multiscale finite element methods: theory and applications},
  volume~4.
\newblock Springer Science \& Business Media, 2009.

\bibitem{liu2018narrow}
Haixiang Liu, Yuanming Hu, Bo~Zhu, Wojciech Matusik, and Eftychios Sifakis.
\newblock Narrow-band topology optimization on a sparsely populated grid.
\newblock {\em ACM Transactions on Graphics (TOG)}, 37(6):1--14, 2018.

\bibitem{mukherjee2021accelerating}
Sougata Mukherjee, Dongcheng Lu, Balaji Raghavan, Piotr Breitkopf, Subhrajit
  Dutta, Manyu Xiao, and Weihong Zhang.
\newblock Accelerating large-scale topology optimization: state-of-the-art and
  challenges.
\newblock {\em Archives of Computational Methods in Engineering},
  28(7):4549--4571, 2021.

\end{thebibliography}

\end{document}